\documentclass[aps, prb, 10pt, superscriptaddress, preprintnumbers, twocolumn]{revtex4-1}

\usepackage{graphicx}
\usepackage{amsfonts}
\usepackage{amsmath}
\usepackage{hyperref}
\usepackage[mathscr]{urwchancal}
\usepackage{amssymb}
\usepackage{mathrsfs} 

\newcommand{\la}{\label}
\newcommand{\be}{\begin{equation}}
\newcommand{\ee}{\end{equation}}
\newcommand{\bea}{\begin{eqnarray}}
\newcommand{\eea}{\end{eqnarray}}
\newcommand{\p}{\partial}
\newcommand{\1}{\frac{1}{2}}

\newcommand{\comment}[1]{}
\newcommand{\cp} {\mathbb{CP}^1}
\newcommand{\zn} {\mathbb{Z}_n}

\usepackage{color}

\begin{document}

\title{Geometric Defects in Quantum Hall States}
\preprint{EFI-16-11}
\author{Andrey~Gromov}
\affiliation{Kadanoff Center for Theoretical Physics and Enrico Fermi Institute, University of Chicago, Chicago, Illinois 60637}
\email{gromovand@uchicago.edu}

\begin{abstract}
We describe a geometric (or gravitational) analogue of the Laughlin quasiholes in the fractional quantum Hall states. Analogously to the quasiholes these defects can be constructed by an insertion of an appropriate vertex operator into the conformal block representation of a trial wavefunction, however, unlike the quasiholes these defects are extrinsic and do not correspond to true excitations of the quantum fluid. We construct a wavefunction in the presence of such defects and explain how to assign an electric charge and a spin to each defect, and calculate the adiabatic, non-abelian statistics of the defects. The defects turn out to be equivalent to the genons in that their adiabatic exchange statistics can be described in terms of representations of the mapping class group of an appropriate higher genus Riemann surface. We present a general construction that, in principle, allows to calculate the statistics of $\mathbb Z_n$ genons for any ``parent'' topological phase. We illustrate the construction on the example of the Laughlin state and perform an explicit calculation of the braiding matrices. In addition to non-abelian statistics geometric defects possess a universal abelian overall phase, determined by the gravitational anomaly.
\end{abstract}

\maketitle



\section{Introduction}
The last two decades brought the rise of interest in topological properties of materials. These properties (in two spatial dimensions) manifest themselves in a number of ways: fractionalized excitations, protected gapless edge modes, anyonic statistics, degeneracy on higher genus surfaces, quantized linear response functions and many more. After the work of  [\onlinecite{moore1991nonabelions}] on non-abelian anyons in fractional quantum Hall (FQH) states the exotic FQH states took the spotlight. In the older approach the non-abelian statistics is encoded into the properties (more concretely, monodromy) of conformal blocks in a rational conformal field theory (RCFT). Nowadays more abstract methods are used to describe the non-abelian statistics\cite{kitaev2006anyons}.

More recently we have learned that it is illuminating to subject a topological phase of matter to a geometric background. One completely academic way to accomplish this is to formally couple the physical degrees of freedom to the curvature of space\cite{WenZeeShiftPaper, 2011-Haldane-FQHE, hughes2011torsional, hughes2012torsion, Abanov-2014, GCYFA, klevtsov2014random, can2014geometry, CLW, laskin2015collective, Klevtsov-fields, klevtsov2015quantum, gromov2015boundary, cappelli2015multipole}. There are many physical ways to think about the geometry, for example shears and stresses in the material \cite{1995-AvronSeilerZograf, LandauLifshitz-7} or inhomogeneous band curvature\cite{2011-Haldane-FQHE, johri2015probing}, geometric defects such as dislocations and disclinations\cite{barkeshli2012topological, cho2015condensation, laskin2016emergent}, temperature gradients\cite{2000-ReadGreen, Stone-Gravitational, gromov-thermal, bradlyn2014low} - all can be modeled by either homogeneous or singular perturbations in geometry. Geometry provides new parameter (or {\it moduli}) spaces to study Berry phases\cite{ bradlyn2015topological,  klevtsov2015precise}, new tools to compute the linear response functions of stress, energy current and momentum\cite{bradlyn-read-2012kubo, Abanov-2014}, induces a new type of gravitational Aharonov-Bohm effect\cite{WenZeeShiftPaper, einarsson1991fractional, einarsson1995fractional}, allows to incorporate extra symmetry requirements such as non-relativistic diffeomorphism invariance \cite{2012-HoyosSon, son2013newton, Gromov-galilean, jensen2014coupling, geracie2015fields}, provides a way to determine the central charge beyond the ${\mbox \rm mod} \,\,8$ restriction of the topological quantum field theory (TQFT)\cite{GCYFA, bradlyn2015topological,  klevtsov2015precise}, mimics the order parameter of nematic phase transition\cite{maciejko2013field, you2013field} and, possibly, describes otherwise invisible, neutral degrees of freedom\cite{2011-Haldane-FQHE, cappelli2015multipole}. Geometric background unveils the universal features of topological phases of matter that are hidden in flat space.  

Laughlin introduced the quasiholes \cite{1983-Laughlin} as charge depletions induced via adiabaticly threading magnetic flux through an infinitely thin solenoid perpendicular to the surface of FQH sample. When magnetic flux is arbitrary a defect is created, however when the magnetic flux is integer the defect can be removed by a gauge transformation. Thus the states with and without defect are gauge equivalent, therefore the defect is an eigenstate of the Hamiltonian and its energy does not depend on the position of the flux insertion as long as it is far away from the boundary and other defects. In other words the defect is mobile and, yet, in other words there is no Dirac string connecting the defect to infinity. The wavefunction is regular and single-valued in electron coordinates. There is an effective theory that encodes charge, spin and statistics of Laughlin quasiholes - a $U(1)$ Chern-Simons theory, where the quasiholes correspond to Wilson lines in some representation of $U(1)$. 

\begin{figure}
  \includegraphics[width=\linewidth]{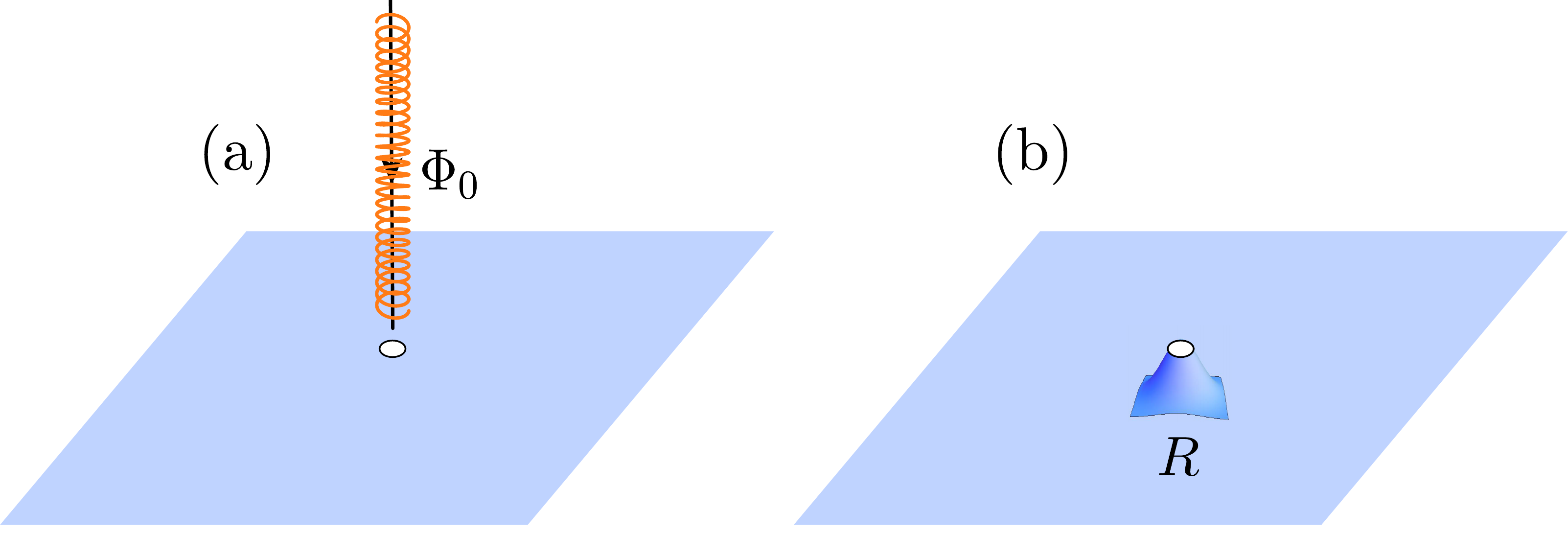}
  \caption{(a). A quasihole created by a unit flux adiabatically threaded through an infinitesimal solenoid, perpendicular to the sample. (b) A genon created by adiabatically ``threading'' a unit curvature flux through the sample. }
  \label{curvaturehole}
\end{figure}

In the present paper we wish to study the behavior of analogous defects created by the fluxes of curvature. It is not hard to imagine threading a ``unit'' flux of curvature through the quantum Hall system and determine conditions under which such defect behaves similar to a quasihole (see FIG. \ref{curvaturehole}). These defects, as we will learn, are fundamentally different from the quasiholes no matter how the curvature flux is quantized. Our goal is to determine charge, spin and statistics of such defects. While, it was not our original intention, these defects will turn out to be equivalent to the genons\cite{barkeshli2013twist} when appropriate quantization of the curvature flux is imposed. Thus for the rest of the paper we will refer to these defects as {\it genons}. We will find that the relevant fluxes of curvature are negative integers in the units of $4\pi$. Such fluxes naturally occur on branched coverings. The branch points of a covering correspond to the genons. It is possible to assign a charge, a spin and even a primary field to a branch point. The branch points have non-abelian statistics determined by a representation of the mapping class group (MPG) acting on the space of groundsates. The representation is fixed by the topological phase of matter ``placed'' on a branched covering. There is an explicit representation for the action of the MPG on the moduli space of any Riemann surface given in terms of $Sp(2g,\mathbb Z)$ matrices which we will utilize to determine the braid matrices of genons. In addition to the universal non-abelian statistics the genons appear to possess a universal abelian $U(1)$ phase that is determined by the central charge. This universal phase arises because, contrary to the common wisdom, the partition function and correlation functions in a topological phase of matter depend on geometry in a controlled way fixed by the Weyl anomaly. Any branched covering is topologically equivalent to a smooth Riemann surface, but the geometry (metric, curvature distribution and, perhaps, group of automorphisms) is different. When a topological phase of matter is constructed with the tools of conformal field theory (CFT) it ``feels'' the variations of geometry through the Weyl and gravitational anomalies\cite{bradlyn2015topological,Klevtsov-fields,CLW,klevtsov2015precise}. When it is constructed via Chern-Simons theory it feels the geometry through the framing anomaly\cite{witten1989quantum}(see [\onlinecite{GCYFA}] for FQH application). The universal $U(1)$ phase that appears when two genons are braided is a manifestation of these effects. This universal $U(1)$ phase depends on the ``parent'' quantum Hall state (or, more generally, topological phase of matter) only through the central charge.  

Previously, the genons were introduced in a system of $n$ layers of an arbitrary topological phase of matter $\mathfrak C$. The symmetry that interchanges the copies is $\mathbb Z_n$ and genons are introduced as twist defects of this symmetry \cite{barkeshli2013twist}. It was also realized that such system can be mapped to a one copy of $\mathfrak C$ on a  Riemann surface which genus that scales with the number of genons, hence the name. In the present work we start from the other end - we consider a gapped quantum system on a higher genus surface with a $\mathbb Z_n$ automorphism and study the properties of the branch points.  This correspondence is not too surprising since in the simplest case the degeneracy of, say, a Laughlin state at filling $\frac{1}{q}$ grows with genus as $q^g$, which can also be interpreted as $g$ copies of Laughlin state on a torus.

This paper is organized as follows.  In Section $2$ we will review different approaches to quasiholes on the example of the Laughlin state. While the material is standard we will pay extra attention to how the traditional constructions generalize to the curved space. The Section 2 is organized into three parts. Each part reviews an independent approach: plasma analogy, conformal field theory (CFT) and topological quantum field theory (TQFT). The Section $3$ is devoted to genons and is organized the same way as Section $2$ to help the reader to see the parallel in the construction and to pinpoint the key aspects in which the genons differ from the quasiholes. In Section $4$ we present discussions and conclusions. Various Appendices are devoted to either computational details or to the material that did not logically fit into the main presentation. 

\section{Quasiholes}

In this Section we review the standard approaches to quasiholes. The intuition we obtain in this Section will guide us in the next Section when we discuss the genons. There are several standard approaches to quasiholes, all of these approaches allow to calculate quantum numbers and statistics leading, of course, to the same results. These approaches are, however, quite different at first sight as they emphasize different physical and mathematical ideas. In our presentation we will allow the physical space to be curved, the effects of curvature cannot be found in classic reviews, thus we feel that this review Section is of some value.

\subsection{Coulomb Plasma}

In this Subsection we will study quasiholes as impurities in the Coulomb plasma. We will restrict our attention to the Laughlin state. The plasma description of more sophisticated trial states has been explored \cite{gurarie1997plasma, bonderson2011plasma, 2009-Read-HallViscosity}, but we will not need it in what follows.

\subsubsection{Plasma in curved space}

In flat space the modulus squared of the Laughlin function reads \cite{1983-Laughlin}
\be\la{Lfunction}
|\Psi(\{z_i\})|^2 = \prod_{i<j}|z_i - z_j|^{2q} \exp-\sum_{i=1}^N \frac{|z_i|^2}{2\ell^2} = \exp \Big(-\beta U\Big)\,,
\ee
where $q$ determines the inverse filling $q = \nu^{-1}$, $\ell$ is the magnetic length fixed by the background magnetic field $\ell^{-2} = \bar B$ and
\be\la{plasmaflat}
U = - 2q^2 \sum_{i<j}\ln|z_i - z_j| + q \sum_{i} \frac{|z_i|^2}{2\ell^2}\, , \qquad \beta = \frac{1}{q}
\ee
is the energy of the Coulomb plasma in external potential. When the plasma is in the screening phase (for $q < 70$) the electron density is homogeneous and can be found from the Poisson equation \cite{1983-Laughlin}
\be\la{rhoflat}
\bar \rho =  \frac{1}{4\pi q} \Delta  \left( \frac{|z|^2}{2\ell^2}\right) = \frac{1}{q}\frac{1}{2\pi \ell^2}\,,
\ee  
where $\Delta = 4 \p \bar \p$ is the Laplace operator.
 
The generalization to curved space (and constant magnetic field) is straightforward \cite{klevtsov2014random, Klevtsov-fields, CLW}. First, we choose the conformal coordinates so that the metric is diagonal, which is always possible in $2$D,
\be\la{metricconf}
ds^2 = \sqrt{g} dz d\bar z
\ee
Second, we replace the background charge (last term in  Eq. \eqref{plasmaflat}) by a function that depends on the geometry
\be
q\sum_{i} \frac{|z_i|^2}{2\ell^2} \longrightarrow  q \sum_{i} \frac{\mathcal K(z_i,\bar z_i)}{2\ell^2}\,.
\ee
Third, we demand the ``generalized screening''
\be \la{GS}
\bar \rho = \frac{1}{4\pi q} \Delta_{g} \left(\frac{\mathcal K(z,\bar z)}{2\ell^2}\right) = \frac{1}{q} \frac{1}{2\pi \ell^2}\,,
\ee
where $\Delta_{g} = \frac{4}{\sqrt{g}}\p \bar\p$ is the Laplace operator in conformal coordinates.
This condition means that the electron density is still ``constant'', but transforms as a {\it scalar density} under a coordinate transformation.
From \eqref{GS} where we can read off 
\be\la{Kahlerdef}
\frac{1}{4}\Delta_{g} \mathcal K(z,\bar z) =1\,,\quad {\mbox or}\quad  \p \bar \p \mathcal K(z,\bar z) =  \sqrt{g}\,.
\ee
A function satisfying (\ref{Kahlerdef}) is known as a K\"{a}hler potential. 

To summarize, the (unnormalized) absolute value squared of the Laughlin function in curved space and constant magnetic field is given by\cite{}\footnote{In fact, there is extra freedom in introducing the coupling to the curved space. One can always multiply the wavefunction by a factor $\prod_{i} \sqrt{g(z_i)}^j$. In this paper we take $j=0$, but in general $j$ will change the geometric spin.}
\be\la{Lcurved}
|\Psi(\{z_i\})|^2 = \prod_{i<j}(z_i - z_j)^{2q} \exp\sum_{i=1}^N -\frac{\mathcal K(z_i,\bar z_i)}{2\ell^2}\,.
\ee

\subsubsection{Defects in the plasma}
When smooth deviations of magnetic field $B$ and curvature $R$ are introduced on top of a fixed background the density is given by (we have divided by $\sqrt{g}$, so that both magnetic field and curvature are defined appropriately)\cite{WenZeeShiftPaper, douglas2010bergman,klevtsov2014random,  CLW, Abanov-2014}
\be\la{density}
\rho = \bar \rho + \nu \frac{ B}{2\pi} + \nu \bar s \frac{ R}{4\pi} + o(\ell^2)\,,
\ee
where we have introduced a new universal quantum number $\bar s$ known as mean orbital spin\cite{WenZeeShiftPaper}. Analogously to the filling fraction $\nu$, the mean orbital spin is related to a universal transport coefficient known as Hall viscosity $\eta_H$ \cite{2009-Read-HallViscosity} (see also \cite{2011-HoyosSon} for effective theory explanation of the relation) and to the topological shift $\mathcal S$ \cite{1983-Haldane-hierarchy, WenZeeShiftPaper, 1993-frohlich}. In the Laughlin state the value of mean orbital spin is usually cited as $\bar s = \frac{1}{2\nu}$ \cite{1995-AvronSeilerZograf, 2009-Read-HallViscosity}, however it is possible to tweak the Laughlin state and change $\bar s$ without changing $\nu$ \cite{klevtsov2015precise}. In a general situation $\bar s$ carries extra information about the state \cite{2009-Read-HallViscosity}. This can be easily seen on the example of trial conformal block states, where $\bar s$ equals to the conformal weight (and conformal spin) of the electron operator. Mean orbital spin $\bar s$ has recently been measured in an integer QH system of photonic Landau levels \cite{schine2015synthetic} as a fractional charge trapped on a conical singularity. We will also need an integrated version of \eqref{density}
\be\la{shift}
N = \nu N_\phi + \chi \nu \bar s,
\ee
where $\chi$ is the Euler characteristic and $N_\phi$ is total magnetic flux, corresponding to $\bar B$.
From the plasma perspective a quasihole can be viewed as follows. Consider and adiabatic insertion of a singular perturbation of magnetic field (on top of $\bar B$)
\be\la{singularB}
B = - 2\pi p \delta (z-a)\,,
\ee
with $p$ being an arbitrary number for a moment. Then density is inhomogeneous around $z=a$ and there is a charge excess or depletion given given by
\be\la{delN1}
\delta N = \int d^2x \sqrt{g} \Big( \rho-\bar \rho\big) = -\frac{p}{q}\,.
\ee
When $p$ is an {\it integer} the extra magnetic flux is not seen by other particles, therefore the defect can be removed by a singular gauge transformation. This defect is a quasihole. We have not yet fixed the sign of $p$.

Next, we are going to determine the wavefunction describing the quasihole by matching $\delta N$ to a plasma computation.
Consider the Laughlin state \eqref{Lcurved} in the background \eqref{singularB}. Clearly, the particles will repel or attract to the point $z=a$. Thus, we are led to the ansatz for the wavefunction
\be
|\Psi(\{z_i\},a)|^2 = \prod_i|z_i-a|^{2n}\prod_{i<j}|z_i - z_j|^{2q} \exp\sum_{i=1}^N -\frac{\mathcal K(z,\bar z)}{2\ell^2}\,\,,
\ee
where $n$ must be a {\it positive} real number.
We again demand generalized screening
\be
\rho = \frac{1}{4\pi q} \Delta_{LB} \left(\sum_{i} \frac{\mathcal K(z,\bar z)}{2\ell^2} - n\sum_i\ln|z_i-a|^2 \right)\,.
\ee
The Laplacian of the second term is easily evaluated
\be
\frac{n}{4\pi q}\Delta_{LB}\sum_{i}\ln|z_i-a| = -\frac{n}{q} \delta(z-a)
\ee
Thus particle excess around $z=a$ is given by
\be\la{delN2}
\delta N^\prime = -\frac{n}{q}\,.
\ee

Comparison of \eqref{delN1} and \eqref{delN2} shows that $n=p$ and implies that $p$ is a positive integer. Thus the Laughlin function in the presence of one quasihole is
\be\la{quasihole}
\Psi(\{z_i\},a)= \prod_i(z_i-a)^{p}\prod_{i<j}(z_i - z_j)^{q} \exp\sum_{i=1}^N -\frac{\mathcal K(z,\bar z)}{4\ell^2}\,,
\ee
where we have also removed the absolute value and made a gauge choice to fix the overall phase. The electric charge of the quasihole is $-\frac{p}{q}$.

From \eqref{quasihole} we also see that $p$ has to be {\it positive}, otherwise $\Psi$ is singular at the position of the quasihole, and $p$ has to be an integer, otherwise the wavefunction will not be single-valued in electron coordinates. Notice that the plasma computation only allowed us to derive the norm, not the phase of the wavefunction, however the two are tied to each other due to the fact that the wavefunction is holomorphic (up to the  background charge factor). 

Other values of flux are, of course, possible, but these will result in either multivalued or singular (or both) ``wavefunctions''. Alternatively, the wavefunction can be made single-valued, but non-holomorphic when the flux is not quantized. Since quasiholes admit a nice, holomorphic {\it and} single-valued, wavefunction one can think of them as an intrinsic property of the state or ``excitations''.

\subsubsection{Charge and statistics from an Aharonov-Bohm phase}

We have already established the charge of a quasihole in the previous Subsection. We will compute the charge from a Berry phase calculation. Consider a process when a quasihole adiabatically travels (counterclockwise) in a closed loop $\mathcal C$, given parametrically by $z_0(t)$, that encircles a planar region $\Sigma$ of area $A$. It is absolutely {\it crucial} that the region $\Sigma$ is in the plane (or flat torus). In a seminal paper~[\onlinecite{arovas1984fractional}] it was shown that in the end of the process the wavefunction \eqref{quasihole} acquires a Berry phase $e^{2\pi i\gamma_{AB}}$ where $\gamma_{AB}$ satisfies
\be
\frac{d \gamma_{AB}}{ dt} = i \int d^2z \rho(z) \frac{d}{d t} \ln (z-z_0(t))\,,
\ee
where $\rho(z)$ is given by \eqref{density} and $B$ is given by \eqref{singularB} (where the $\delta$-function is slightly smoothed out in a rotationally invariant way). Then writing $\rho(z) = \bar \rho + \delta\rho(z)$ we have 
\bea\nonumber
\gamma_{AB} &&= i\oint dz_0 \int d^2z \rho(z) \frac{1}{z-z_0} \\ \la{delta}
&&= -2\pi \bar \rho + i \oint dz_0 \int d^2 z \frac{\delta \rho(z-z_0)}{z-z_0}\,.
\eea
The last term can easily be shown to vanish in {\it flat }space. The final answer for the AB phase is, then
\be\la{AB}
\gamma_{AB} =   -2\pi \bar \rho= - p\nu \Phi(\Sigma)\,,
\ee
where $\Phi(\Sigma)$ is the total flux of magnetic field piercing the surface $\Sigma$. Eq.\eqref{AB} is simply an Aharonov-Bohm effect that determines the charge of the quasihole to be 
\be
Q= -\nu p = - \frac{p}{q}.
\ee
When the path $\mathcal C$ contains another quasihole of charge $-\nu p^\prime$ inside there is an extra ``statistical phase''
\be\la{qhstat}
2\gamma_{stat} =  -p\nu \delta\Phi(\Sigma) =  p p^\prime \nu\,,
\ee
where the factor of $2$ is put to emphasize that we took one quasihole completely around another. Then $e^{2\pi i\gamma_{stat}}$ gives the exchange statistics.

\subsubsection{Spin of a quasihole}

In curved space we can go one step further than Ref. [\onlinecite{arovas1984fractional}] and calculate the spin of the quasihole. The spin is {\it defined} through the curvature analogue of the Aharonov-Bohm effect. Namely, we consider the same adiabatic process described before, but in curved space. Then on general grounds we have to expect a geometric phase
\be\la{curvatureAB}
\Psi \longrightarrow  e^{2\pi i S N_R(\Sigma)}\Psi\,,
\ee
where $N_R(\Sigma)$ is the curvature flux through $\Sigma$ and the quantum number $S$ is {\it defined} to be the spin of a quasihole.

\begin{figure}
  \includegraphics[width=2in]{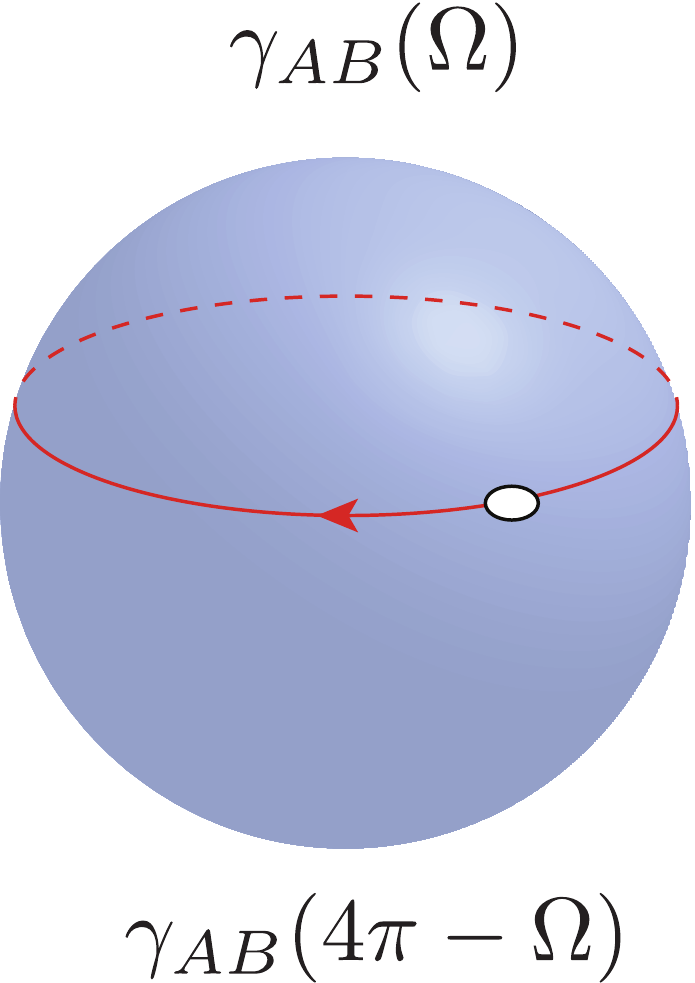}
  \caption{A quasihole is dragged around a loop on a sphere. The Berry phase should not depend on what is considered the inside and outside of the contour $\gamma_{AB}(\Omega) =\gamma_{AB}(4\pi - \Omega)$ mod $2\pi$. This condition alone leads to the presence of an extra intrinsic spin of a quasihole and is used to derive its value.}
  \label{sphere}
\end{figure}

 In fact, the presence of such phase is necessary to ensure that quasihole braiding is self-consistent on a sphere (or any curved surface for that matter). To see this \cite{einarsson1991fractional, einarsson1995fractional} we note that given a closed path on a sphere the notion of the interior of the path is ambiguous (see FIG. \ref{sphere}). The interior can be either to the left or to the right from the boundary of a path. Self-consistency requires that the AB phase must not depend on what is considered to be the interior of the path. To be more precise, consider a sphere of radius $1$. The total solid angle is then $4\pi$. Consider a closed path $\mathcal C$ that cuts out a solid angle $\Omega$ from the sphere. The Aharonov-Bohm phase must satisfy
\be\la{check}
\gamma_{AB}(\Omega) + \gamma_{AB}(4\pi - \Omega) = 2\pi k\,,
\ee
where $k$ is an integer. This relation implies 
\be\la{problem}
-p \nu N_\phi = k\,,
\ee
which can only be satisfied when $N_\phi$ is proportional to $\nu^{-1}$. This, however, contradicts \eqref{density} and \eqref{singularB} since the total magnetic flux is given by
\be
N_\phi = \nu^{-1} N + 2 \bar s - p\,.
\ee
In order to resolve this contradiction we require that there is an {\it extra} AB phase 
\be
\gamma^\prime_{AB} = \frac{\Omega}{4\pi} \left( \frac{\nu p^2}{2} - \nu\bar sp \right)\,.
\ee
Inclusion of this phase allows the condition \eqref{problem} to be satisfied
\be
- p \nu \left(N_\phi - 2\bar s + p\right) = -p \nu \cdot \nu^{-1} N = -p N \in \mathbb Z\,.
\ee 
We conclude that the total AB phase on a sphere is $\gamma_{AB} + \gamma_{AB}^\prime$. When written covariantly the second phase is simply
\be
\gamma^\prime_{AB} = \left( \frac{\nu p^2}{2} - \nu \bar s\right) N_{R}(\Sigma)\,,
\ee
which implies that on general grounds the spin of the quasihole is
\be\la{spinqh}
S = \frac{\nu p^2}{2} - \nu\bar s p\,.
\ee
The first term in this relation is well-known as the topological spin $\theta_p = e^{2\pi i \frac{p^2}{2q}}$. It appears due to a short distance effect - interaction between charge and flux making up the quasihole. The second term appears due to the interaction of the quasihole with the curvature of the sphere, the ``strength'' of this interaction is encoded in the quantum number $\bar s$. Note that the second term is responsible for the violation of the ``spin-statistics theorem''. 

If we were to demand the ``spin-statistics theorem'' in addition to \eqref{check} we would find an extra condition
\be
2 \bar s \nu p \in \mathbb Z\,,
\ee
which holds identically  for the Laughlin state if $\bar s = \frac{1}{2\nu}$. This was probably the case considered in Ref.~[\onlinecite{read2008quasiparticle}]. In the general situation, the quantum number $\bar s$ can be tuned {\it independently}\cite{klevtsov2015precise, laskin2016emergent} of the filling fraction. For example\cite{2007-TokatlyVignale,2009-Read}, consider the electrons filling only the $N$-th Landau level. In this case $\nu =1$, but $\bar s =\frac{2N+1}{2}$, that is $\bar s$ is fixed by the cyclotron orbital angular momentum of the electron. Another example is provided by the Read-Reazyi series \cite{Read1999}, where $\bar s = \frac{1}{2\nu} + h_\psi$. In both of these examples the spin $S$ of the quasihole is incompatible with the spin-statistics theorem. The effect of the mean orbital spin can be seen in the flat space.  For example, the Hall viscosity is sensitive to the mean orbital spin $\eta_H = \frac{\bar s}{2}  \rho$. \cite{2011-ReadRezayi}

Another check of \eqref{spinqh} is provided if one chooses $p = \nu^{-1}$ and $\bar s = \frac{1}{2\nu}$. In this case we find that the spin of a real hole vanishes identically \cite{sondhi1992long} which is the consequence of the sum rule for second moment of density in the Coulomb plasma.

 Direct Berry phase calculation of the spin $S$ is also possible. In fact, the spin-statistics violating second term in \eqref{spinqh} is easy to derive - it comes from \eqref{AB} combined with \eqref{density}. It is much harder to derive the topological spin. It turns out, perhaps surprisingly, that in curved space one cannot disregard the second term in \eqref{delta}. The adiabatic drag of a smoothed out quasihole around a close loop induces a $2\pi$ rotation of the quasihole ``around itself'', which is reflected in the Berry phase, we refer the interested reader to a computation of [\onlinecite{einarsson1995fractional}] that carefully regulates the quasihole's finite size. An independent computation that involves functional integration can also be found in [\onlinecite{can2014field},\onlinecite{can2014geometry}]. We will rederive the relation \eqref{spinqh} two more times in this Section, using the effective approaches: Moore-Read construction, generalized to curved space, and the Wen-Zee construction. It seems to be a general theme for the topological spin $\theta$ - it {\it can} be seen in curved space, but only after short distance manipulations.

\subsection{Conformal Field Theory}

Laughlin state as well as many other states (but not all known states) can be constructed as certain correlation functions or conformal blocks in a CFT \cite{moore1991nonabelions}. We, again, will focus on the Laughlin state. Our formulation will slightly differ from the original Moore-Read construction, but all of the results can be obtained from either point of view.

\subsubsection{Conformal field theory data}
The relevant CFT for the Laughlin state is $c=1$ boson. Below we briefly list the objects of interest. We fix the topology of a sphere with constant magnetic field $\bar B = \ell^{-2}$ and round metric that gives rise to constant curvature $R$. We will consider a theory in the presence of a background that breaks the scale symmetry.

The ``CFT''  has a Lagrangian description given by \cite{kvorning2013quantum, Klevtsov-fields}
\be\la{boson}
S[\varphi]= \frac{1}{\pi} \int \p\varphi\bar \p \varphi +\frac{i}{2\sqrt{q}} \bar B\varphi + i\frac{\bar s}{4\sqrt{q}} R\varphi\,.
\ee
Strictly speaking, \eqref{boson} is not a CFT since the scale $\ell$ is explicitly in the action, but some of the CFT terminology and ideas will hold for this very special ``perturbation'' (it is not a conformal perturbation in the usual sense since $\varphi$ is not a primary field) . We also note that the perturbation is equivalent to the neutralizing background operator introduced in [\onlinecite{moore1991nonabelions}] since the action can be re-written as
\be\la{MRlink}
S[\varphi] = \frac{1}{\pi} \int \p\varphi\bar \p \varphi +i\sqrt{q}\int \rho \varphi\,,
\ee
where the density $ \rho$ is given by
\be\la{rho}
 \rho = \frac{\nu}{2\pi \ell^2} + \frac{\nu\bar s}{4\pi} R\,.
\ee
The holomorphic stress tensor is (without background charge) \footnote{Notice that it happens to be Sugawara stress tensor of simple currents $J$.}
\be
T = - \frac{1}{2} : \p \varphi(z) \p \varphi(z) :\,.
\ee
There are interesting primary fields in the ``CFT'' \eqref{boson}. Define the vertex operators
\be
V_p(z,\bar z) = : e^{i\frac{p}{\sqrt{q}} \varphi(z,\bar z)}:\,.
\ee
The correlation function of the vertex operators is given by
\be\la{corf}
\Big\langle \prod_i V_{p_i}(z_i,\bar z_i) \Big\rangle = \prod_{i<j}|z_i-z_j|^{2\frac{p_ip_j}{q}}
\ee
The theory \eqref{boson} has a (broken by the background charge) $U(1)$ shift symmetry $\varphi \rightarrow \varphi + \alpha$. This symmetry imposes the neutrality condition on the vertex operator correlation functions 
\be\la{neutr}
\sum_i p_i = q\int\rho\,,
\ee
so that when \eqref{neutr} does not hold the correlation function vanishes.

The field $\varphi$ is chosen to be a {\it compact} boson with compactification radius $r=\sqrt{q}$
\be
\varphi \sim \varphi + 2\pi\sqrt{q}\,.
\ee
Then condition $V_p \sim e^{2\pi i p}V_p$ implies that $p$ is an integer. With these choices the vertex operator is well-defined. We also define the electron operator setting $p=q$
\be
V_{q} = e^{i\sqrt{q} \varphi(z)}\,.
\ee
The electron operator has trivial monodromy with other operators. This property ensures that the conformal block wavefunction is single-valued in the electron coordinates. There is a {\it finite} number of well-defined primary vertex operators $V_p$ since $p$ can always be shifted by $q$ at the expense of multiplying by the ``trivial'' operator. A CFT with a finite number of primary fields is called {\it rational}. We must emphasize that multiplication by an electron operator does not change either braiding properties or topological spin, it does, however, change the quasihole spin \eqref{spinqh}. It was suggested in [\onlinecite{moore1991nonabelions}] that other primary operators will describe quasiholes. Since there is only a finite number of primary fields there will be only a finite number of quasihole types (labeled by their fractional charge).

The last term in \eqref{boson} is known as the background charge \cite{CFT-book}. This term modifies the stress tensor by an additive term
\be\la{stressnew}
T_{new} = - \frac{1}{2} : \p \varphi(z) \p \varphi(z):  - i \frac{\bar s}{\sqrt{q}} \p^2 \varphi \,.
\ee
This modification leads to the change in both conformal dimensions of primary fields and the central charge (defined from either 2-point function of stress tensor $T$ or, more generally, through the trace anomaly). The conformal dimension of the vertex operator $V_p$ is given by
\be
h_p = \frac{p^2}{2q} - \frac{\bar s p}{q} = \frac{\nu p^2}{2} - \nu\bar s p\,,
\ee
which agrees with \eqref{spinqh}. This is probably the easiest way to derive the spin of a quasihole and it follows directly from the Moore-Read construction, provided that the neutralizing background is interpreted as part of the action.

The central charge is (dubbed ``Hall central charge'' and denoted $c_H$ in [\onlinecite{klevtsov2015precise}, \onlinecite{laskin2016emergent}]; dubbed ``apparent central charge'' and denoted $c_{app}$ in [\onlinecite{bradlyn2015topological}] )
\be\la{capp}
c_\mathrm{w} = 1 - 12\nu \bar s^2\,.
\ee
This quantity appears in the Ward identity for the Weyl symmetry of \eqref{boson}. Alternatively, it can be derived from a two-point function of the stress tensor \eqref{stressnew}. The first term can be understood as a genuine Weyl anomaly of the functional integration measure, whereas the second term is induced by the neutralizing background. To be more precise, the Weyl Ward identity takes form
\be\la{Weylll}
\langle T_{z\bar z} \rangle =  \frac{c_\mathrm{w}}{24\pi} R\,,
\ee
Eq.\eqref{Weylll} motivates the notation $c_\mathrm{w}$.

\subsubsection{Laughlin function}
The (absolute value) of the Laughlin function is given by the correlation function of the electron operators\cite{moore1991nonabelions, Klevtsov-fields}
\be
\Big\langle\prod_{i=1}^NV_q(z_i,\bar z_i) \Big\rangle = \prod_{i<j}|z_i - z_j|^{2q} \exp\sum_{i=1}^N -\frac{\mathcal K(z_i,\bar z_i)}{2\ell^2}
\ee
The neutrality condition \eqref{neutr} takes form
\be
N = \nu N_\phi + 2\nu \bar s = \nu N_\phi + 1
\ee
giving the correct relation between the number of magnetic flux quanta, number of electrons and the shift.

Quasiholes of electric charge $-p/q$ are generated by extra insertions primary fields $V_{p}(a)$, giving the norm \eqref{quasihole}. Quasihole wavefunctions can also be understood as correlation functions of {\it only} electron operators evaluated on a singular magnetic field background \eqref{singularB}. Clearly, shifting the magnetic field in the action \eqref{boson} by a $\delta$-function inserts precisely the operator $V_p(a)$ into the correlation function.

The spin of a quasihole equals to the scaling dimension of the operator $V_p(a)$. Due to the background charge (last term in \eqref{boson}) the spin does not equal to the statistical spin, but is given by \eqref{spinqh}, where the last term comes precisely from the background charge.

The computation of statistics can be done in a very elegant way\cite{moore1991nonabelions}. We can separate the vertex operators into chiral and anti-chiral parts
\be
V_p(z,\bar z) = V_p(z)\otimes V_p(\bar z)
\ee
and calculate only the holomorphic part of the correlator with two quasihole insertions $V_{p_J}(a_J)$. This yields the expression
\bea\nonumber
\Psi(\{z_i\},\{a_J\})= && \frac{1}{\sqrt{\mathcal{N}(a_J)}}(a_1-a_2)^{\frac{p_1p_2}{q}}\prod_{i,J}(z_i-a_J)^{p_J}\cdot \\
&&\prod_{i<j}(z_i - z_j)^{q} e^{-\sum_{i=1}^N \frac{\mathcal K(z_i,\bar z_i)}{4\ell^2}}\,,
\eea
where $\mathcal N(a_J)$ is an appropriate normalization factor (single-valued in $a_J$ and exponentially saturating to a constant as $|a_1-a_2|$ increases).
The statistics can be read off from the monodromy of the wavefunction under analytic continuation of $a_1$ around $a_2$. Of course, the monodromy result agrees with the previous computations. Miraculously, the CFT representation of the Laughlin wavefunction selects a nice gauge (in the space of Berry connections) so that the Berry gauge field vanishes along the quasihole trajectory and monodromy of the wavefunction completely accounts for the adiabatic statistics. This fact was first used in [\onlinecite{moore1991nonabelions}]. The detailed discussion of conditions that ensure equality between the Berry phase and the monodromy can be found in [\onlinecite{2009-Read-HallViscosity}].

There are three important insights that the CFT construction gave us. First, there is a relation between primary fields and fractional anyonic excitations. Second, the statistics of quasiholes can be read off from the monodromy of the wavefunction. Third, the action \eqref{boson} hints us that it is also possible to produce a ``vertex operator'' insertions via choosing a singular configuration of curvature $R$. These insertions will be discussed in the next Section.

\subsubsection{Moduli spaces on a torus}
The previous construction can also be done on a torus geometry. In writing the action \eqref{boson} we were slightly imprecise, because we have integrated by parts the last two terms. On torus we must be more careful. First, we simplify the action by choosing a flat torus, so that $R=0$, however the stress tensor is still given by \eqref{stressnew}. The action takes form
\be\la{bosontor}
S[\varphi]= \frac{1}{\pi} \int \p\varphi\bar \p \varphi - \frac{i}{2\sqrt{q}} \bar A^{(0)} d\varphi +\frac{i}{2\sqrt{q}}\bar B\varphi\,,
\ee
where we have kept the fluxes of the vector potential $\bar A^{(0)}$. More concretely we can break the vector potential into two pieces $\bar A$ and $\bar A^{(0)}$ such that $d \bar A = \bar B$ and $d \bar A^{(0)}=0$. On a sphere the last condition would imply that $\bar A^{(0)}$ contains no information, however on a torus $\bar A^{(0)}$ parametrizes the fluxes through the cycles of the torus as 
\be
\Phi_i = \frac{1}{2\pi}\oint_{c_i} \bar A^{(0)}\,,
\ee
where $c_i$ is either $a$ or $b$ cycle of the torus. Thus, the correlation functions of electron operators will parametrically depend on the {\it moduli} $\Phi_i$. The space of $\Phi_i$ is also known under the name Jacobian variety and flux torus, and is topologically a torus with $\Phi_i \in [0,1)$. The Berry phase in the space of $\Phi_i$ computes the Hall conductance \cite{TKNN}. 

There is another parameter space in the game. Fixing the torus to be flat leaves an infinite number of inequivalent tori, parametrized by a complex modular parameter $\tau$ defined in the complex upper half plane $\mathbb H$. The simplest way to understand where the modulus $\tau$ enters the equations is to notice that there are infinitely many flat metrics parametrized as
\be
ds^2 = |dz + \tau d\bar z|^2\,.
\ee
There is an $SL(2,\mathbb Z)$ redundancy in the definition of $\tau$. Thus, the space of $\tau$ is an orbifold $\mathbb H/SL(2,\mathbb Z)$. The Berry phase in the space of $\tau$ computes the Hall viscosity\cite{1995-AvronSeilerZograf, levay1995berry}. To fix the terminology we note that $\mathbb H$ (in the general genus $g$ case) becomes the Teichm\"{u}ller space $\mathscr T_g$, whereas the factor $\mathscr T_1(\Sigma)/SL(2,\mathbb Z)$ is known as the moduli space $\mathscr M_1$.

When a quantum Hall system is placed on a surface of higher genus $g>1$ there is an extra novelty: the curvature cannot be chosen to be $0$ everywhere, instead the best one can do is to choose it to be $R=-1$, alternatively the Euler characteristic does not vanish. This leads to an extra term in the Berry curvature on the space of $3g-3$ moduli. This extra term computes the central charge $c_\mathrm{w}$\cite{klevtsov2015precise}. 

The correlation functions of electron operators turn into {\it finite} sums over the extended conformal blocks \cite{CFT-book}. Each conformal block corresponds to a good wave-function, thus the space of ``Laughlin states'' is not one-dimensional like it was on a sphere. In fact, there are precisely $q$ independent extended conformal blocks \cite{verlinde1988fusion} and thus, the degeneracy of the Laughlin state is $q$ [\onlinecite{wen1990ground}]. A convenient choice of basis in the space of the unnormalized degenerate ground states is \cite{haldane1985periodic, 2009-Read-HallViscosity}
\bea \nonumber
\Psi_p =\mathcal N_0&& \Big(({\mbox \rm Im}\,\tau)^\1\eta(\tau)^2\Big)^{N\frac{q}{2}} \frac{1}{\eta(\tau)}F_q\begin{bmatrix}
    \frac{\Phi_1 + p}{q}  \\
    \Phi_2 \\
\end{bmatrix}\Big(Z\Big|\tau\Big)\\ \la{LaughlinT}
&&\cdot \prod_{i<j}\frac{\theta_1(z_i-z_j|\tau)^q}{\eta(\tau)^q}  e^{-\sum_i\frac{ ({\mbox \rm Im} \,z_i)^2}{4\ell^2}}\,,
\eea
where $Z = \sum_i z_i$ and $\mathcal N_0$ is the normalization constant that depends on $\tau$ only through the area of the torus, which is held fixed in all computations. The factors of the Dedekind function $\eta(\tau)$ are needed to insure that the right transformation properties under the $\mathcal S$ generator of $SL(2,\mathbb Z)$, {\it i.e.} under $\tau \longrightarrow -\frac{1}{\tau}$. In particular, the ratio $\theta_1 / \eta(\tau)$ is a modular form of weight $0$. The factors of $({\mbox \rm Im}\, \tau)^\frac{q}{2}$ come from every insertion of the vertex operator and there are $N$ such insertions. The combination $({\mbox \rm Im}\, \tau)^\frac{1}{2}\eta^2(\tau)$ is again a modular form with weight $0$ and so is the wavefunction. This condition is necessary since the norm should not transform when going between two equivalent (in $SL(2,\mathbb Z$) sense) choices of $\tau$.

The center-of-mass factor $F_q$ expressed in terms of $\theta$-function with characteristics \cite{fay1973theta} as
\be
F_q\begin{bmatrix}
    a  \\
    b \\
\end{bmatrix}\Big(z\Big|\tau\Big) =\theta\begin{bmatrix}
    a  \\
    b \\
\end{bmatrix}\Big(q z\Big|q\tau\Big)\,.
\ee
The only information about the degeneracy is contained in the center-of-mass factor.
The $\theta$-function with characteristics is defined as
\be\la{thetadef}
\theta\begin{bmatrix}
    a  \\
    b \\
\end{bmatrix}\Big(z|\tau\Big) = \sum_{n=-\infty}^\infty e^{\pi i \tau (n+ a)^2 + 2\pi i(n+a)(z+b)}\,.
\ee
Finally, $\theta_1(z_i-z_j|\tau) = \theta\begin{bmatrix}
    1/2  \\
    1/2 \\
\end{bmatrix}\Big(z_i-z_j|\tau\Big) $ is the odd $\theta$-function - merely a doubly-periodic generalization of the Jastrow factor $(z_i - z_j)$.

It is possible to calculate the charge of a quasihole by performing a large gauge transformation that affects only $\Phi_2$. Consider a basis state $\Psi_p$ with $\Phi_1=0$ and perform an adiabatic change $\Phi_2 \longrightarrow \Phi_2 + 1$. Then
\be\la{gauge}
\Psi_p \longrightarrow e^{2\pi i \frac{p}{q}} \Psi_p \,.
\ee
Since there are as many ground states as there are types of quasiholes we can restore the entire charge lattice by performing the ``flux insertions'' in different ground states. Since the charge is determined from a phase we can only obtain it up to an integer.

It is also possible to calculate the topological part of the spin of a quasihole \eqref{spinqh}. For simplicity we assume $\Phi_1=0$. We will perform a large coordinate transformation known as Dehn twist $\mathcal T_a$. This coordinate transformation is equivalent to an operation on the Teichm\"{u}ller  space $\tau \longrightarrow \tau + 1$. At this point we have to be careful. When $q$ is {\it even} ({\it i.e.} we are dealing with the bosonic Laughlin state) $\mathcal T_a$ is diagonal in the basis $\Psi_p$. Then we have
\be\la{Dehntwist}
\mathcal T_a \Psi_p = e^{2\pi i \frac{p^2}{2q}}\Psi_p
\ee
or
\be\la{Tmatrix}
(\mathcal T_a)_{pp^\prime}  = \delta_{pp^\prime}e^{2\pi i \frac{pp^\prime}{2q}}
\ee
and we read off $s_{stat} = \frac{p^2}{2q} \,\,\,{\mbox \rm mod } \,\,\,1$.

However, when $q$ is {\it odd} the Dehn twist is not diagonal anymore. This happens because the second characteristic of the $\theta$ function is shifted by the ``spin'' of the electron operator $\frac{q}{2}$ which is half integer. There are two ways to avoid this problem. The first way is to reduce the $SL(2,\mathbb Z)$ to a normal subgroup $\Gamma$ generated by $\mathcal S$ and $\mathcal T_a^2$. Then, it is easy to see that that $\mathcal T_a^2$ is diagonal since the problematic shift becomes $2 \cdot \frac{q}{2}$ which is now an integer\cite{cappelli1997modular}. Another way out is to make a large gauge transformation shifting $\Phi_2$ to $\Phi_2 + \frac{1}{2}$ together with $\mathcal T_a$. The combined transformation is diagonal in the $\Psi_p$ basis. Then \eqref{Dehntwist} holds for the combined transformation, however there is an extra minus sign.

The states $\Psi_p$ transform non-trivially under the $\mathcal S$ generator of $SL(2,\mathbb Z)$ according to
\be
\Psi_p = \sum_{p^\prime} \mathcal S_{pp^\prime} \Psi_{p^\prime}\,.
\ee
It is not hard to show that the basis functions \eqref{LaughlinT} transform by a unitary $\mathcal S$-matrix given by
\be\la{Smatrix}
\mathcal S_{pp^\prime} =  \frac{1}{\sqrt q}e^{-2\pi i\frac{pp^\prime}{q} }\,,
\ee
where we have dropped the overall $U(1)$ phase which depends on positions of the particles $z_i$ and their number $N$.

The central charge $c$ (and not $c_\mathrm{w}$!) can be determined ${\mbox \rm mod} \,\,\, 8$ from the general relation (we have specified it for the bosonic Laughlin state) \cite{kitaev2006anyons}
\be
e^{\frac{2\pi i}{8} c} = \frac{1}{\sqrt{q}} \sum_{p=0}^{q-1} e^{2\pi i \frac{p^2}{q}}\,.
\ee
In the next Section we will find that $\mathcal S$ and $\mathcal T$ (as well as their generalization to higher genus) can also be related to the braid matrices of genons.

\subsection{Chern-Simons Theory}
In this Subsection we also briefly review the Chern-Simons effective approach to FQHE states, emphasizing the role of curved space. We will restrict our discussion to one-component states. 
\subsubsection{The action}
The effective action reads
\be\la{Seff}
S_{eff} = \int -\frac{q}{4\pi} ada + \frac{1}{2\pi} adA + \frac{s}{2\pi}ad\omega + a_\mu j^\mu\,,
\ee
where $a$ is the $U(1)$ ``statistical'' gauge field, $q$ is the inverse filling and the {\it level} of Chern-Simons theory, and $s$ is the spin quantum number. The last term describes the quasihole current. The action \eqref{Seff} is quadratic and the partition function can be calculated exactly with ease. Omitting the details we have
\bea\nonumber
Z[A,j^\mu] &&=Z_0 \exp\Big({i S_{\rm CS}[A,g]}\Big) \cdot \exp{2\pi i \Big( \frac{1}{2q}\int j^\mu \Delta^{-1}_{\mu\nu} j^\nu} \Big)\\ \la{PF}
&&\cdot \exp\Big(-i\frac{1}{q}\int A_\mu j^\mu\Big)\cdot \exp\Big(-i\frac{s}{q}\int \omega_\mu j^\mu\Big)\,,
\eea
where
\bea\nonumber
 S_{\rm CS}[A, g]= \frac{1}{q} \frac{1}{4\pi}&& \int AdA + \frac{1}{2\pi}\frac{s}{q}\int Ad\omega + \frac{1}{4\pi}\frac{s^2}{q} \int \omega d \omega \\ \la{bulkS}
 &&- \frac{{\mbox \rm sign}(q)}{96\pi}\int {\mbox \rm Tr} \left[\Gamma d \Gamma + \frac{2}{3} \Gamma^3\right]\,.
\eea
$Z_0$ is the topological invariant known as Reidemeister torsion, $\Delta^{-1}_{\mu\rho}$ is the propagator of the Chern-Simons theory
\be
\Delta^{\mu\rho} = \epsilon^{\mu\rho\nu}\p_\nu\,, \qquad \Delta^{\mu\nu} \Delta^{-1}_{\nu\rho} = \delta^{\mu}_\rho\,.
\ee

\subsubsection{Charge, spin and statistics}
Notice that apart from the constant $Z_0$ the partition function is a phase. Different factors in this phase describe different quantum numbers discussed before. We will start with charge and spin first. In order to study one quasihole we choose the quasihole current to be
\be
j^0 = p \delta^{(2)}(x-x(t)) \qquad j^{i} = p\dot x^i \delta^{(2)}(x-x(t))\,, 
\ee
where $x(t)$ is the trajectory of a quasihole. We choose the trajectory to be a closed curve $\mathcal C$ so that region $\Sigma$ is bounded by $\mathcal C$. Then the factor
\be
\exp -i\frac{1}{q}\int A_\mu j^\mu = \exp {-i \frac{p}{q} \oint_{\mathcal C} A} =  \exp{-2\pi i \frac{p}{q} \Phi(\Sigma)}
\ee
allows one to extract the charge of the quasihole $-p/q$.

\begin{figure}
  \includegraphics[width=2in]{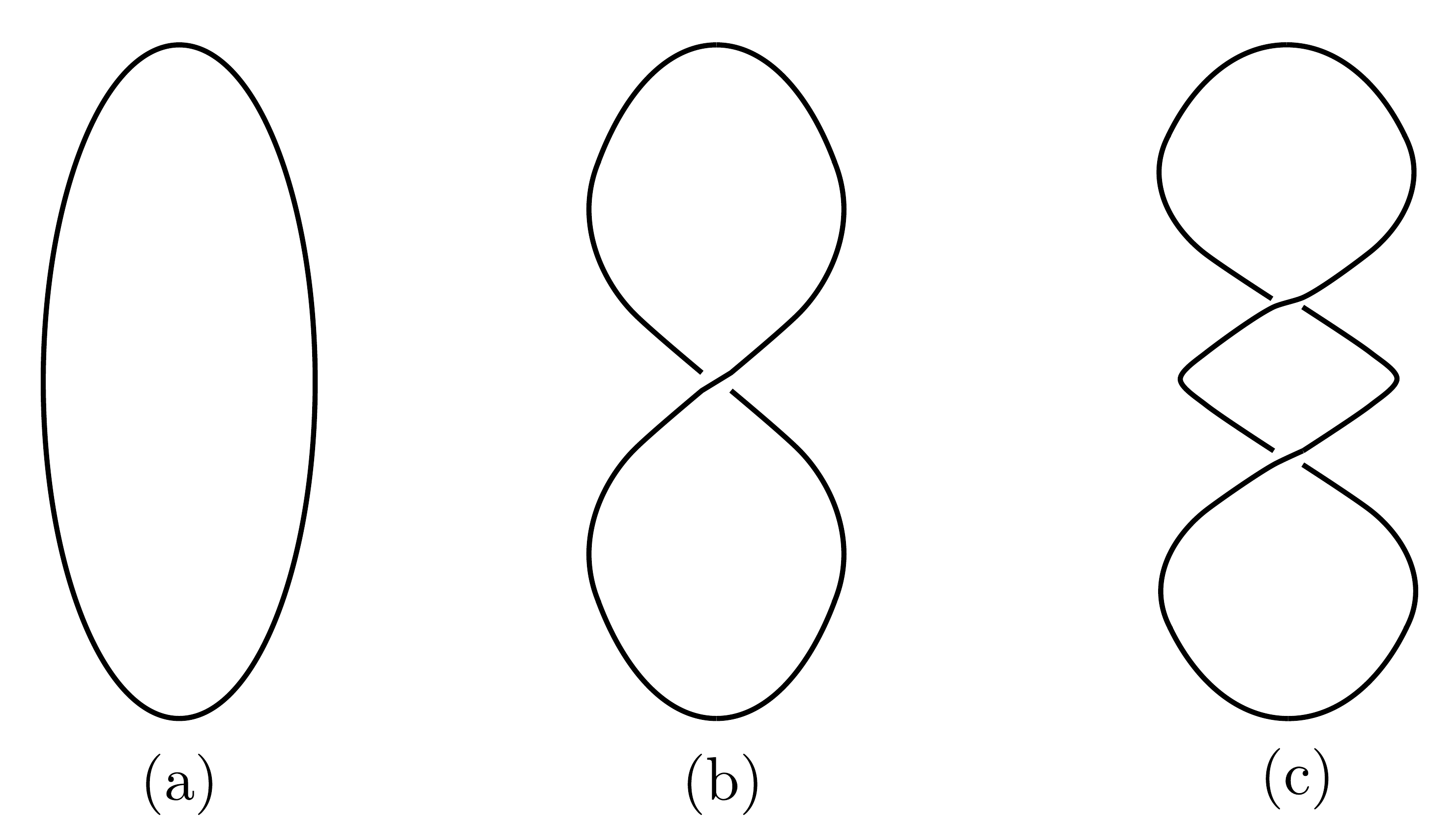}
  \caption{(a) A wilson loop that evaluates to the Aharonov-Bohm phases times the geometric factor $\exp{-2\pi i T[\mathcal C]}$. (b) Framing of the curve is shifted by one unit, relative to (a). This results in an extra phase given by $\theta_p=e^{2\pi i \frac{p^2}{2q}}$. (c) Framing of the curve is shifted by two units, relative to (a). This results in an extra phase given by $\theta_p^2$.}
  \label{Twist}
\end{figure}

The computation of spin is somewhat more sophisticated. First, there is and obvious Aharonov-Bohm term
\be
\exp \Big(-i\frac{s}{q}\int \omega_\mu j^\mu\Big) =\exp \Big(-2\pi i\frac{sp}{q}N_R(\Sigma)\Big)\,,
\ee
notice that the phenomenological coefficient $s$ matches to $\bar s$.
However, it turns out that it is not the whole story since the factor
\be
 \exp{2\pi i \Big( \frac{1}{2q}\int j^\mu \Delta^{-1}_{\mu\nu} j^\nu} \Big)
\ee
also contributes to the phase. This term can be written as a limit of the Gauss linking number of a thin ribbon with edges $\mathcal C$ and $\mathcal C_{\epsilon}$, where the latter is defined using a framing of the curve $\mathcal C$. The curve $\mathcal C_\epsilon$ is defined as follows. If the curve $\mathcal C$ is described by $\vec r = \vec r(t)$ then $\mathcal C_{\epsilon}$ is described by $\vec r_{\epsilon} = \vec r (t) + \epsilon \vec n(t)$. The vector field $\vec n(t)$ is the framing. The Gauss linking number is given by
\be
I[\mathcal C, \mathcal C_{\epsilon}] = \frac{1}{4\pi}\oint_{\mathcal C} \oint_{\mathcal C_{\epsilon}} dx^\mu dy^\nu \epsilon_{\mu\nu\lambda} \frac{x^\lambda - y^\lambda}{|x-y|^3}\, ,
\ee
where one has to first evaluate the integral and then take $\epsilon \longrightarrow 0$ limit. Careful analysis shows \cite{polyakov1988fermi, tze1988manifold} that this limit is given by the {\it writhe} of the curve $\mathcal C$, which, in its turn, is given by
\be
W[\mathcal C] = L[\mathcal C]-\frac{1}{2\pi}\oint_{\mathcal C} d\vec x \cdot [\vec n \times \dot {\vec n}] =( L[\mathcal C] - T[\mathcal C])\,,
\ee
where the first term is the ``self-linking'' number and is a topological invariant. The second term is a geometric invariant (it depends on the choice of framing, however transforms in a controlled way under the change of framing), known as twist of a curve. When the framing is changed the twist changes by an additive constant (see FIG. \ref{Twist} for clarification of this statement). We can choose the framing to be induced by the framing of the ambient space. Then (up to an additive constant) the twist is proportional to the curvature flux \cite{lee1994orbital, 2014-ChoYouFradkin}
\be
T[\mathcal C] = -\frac{1}{2\pi} \oint_{\mathcal C} dx^\mu \omega_\mu = -\frac{1}{4\pi} \int_{\Sigma} d^2x \sqrt{g} R =  -N_R(\Sigma)\,.
\ee
Putting things together we get the phase factor (we have dropped the phases that do not depend on $N_R(\Sigma)$)
\be
\exp\Big(2\pi i \frac{p^2}{2q} W[\mathcal C]\Big)=\exp \Big(2\pi i\cdot \frac{p^2}{2q} N_R(\Sigma)\Big)
\ee
The total phase factor proportional to the curvature flux is
\be
e^{2\pi i S N_R(\Sigma)} =  \exp\Big[2\pi i  \Big(\frac{\nu p^2}{2} - \nu s p\Big) N_R(\Sigma)\Big]\, ,
\ee
thus we again obtain \eqref{spinqh}. 

To calculate the statistics we choose the quasihole current to be
\bea
j^0 &&= p \delta^{(2)}(x-x(t)) + p^\prime \delta^{(2)}(x-x_0),  \\
 j^{i} &&= p\dot x^i \delta^{(2)}(x-x(t))\,. 
\eea
The mutual statistics comes from the factor
\be
 \exp{2\pi i \Big( \frac{1}{q}\int j^\mu \Delta^{-1}_{\mu\nu} j^\nu} \Big) = e^{2\pi i \frac{pp^\prime}{q}}\,,
\ee
which agrees with previous Subsections.

Some clarification is required on the relation between the spin and statistics, which is a delicate subject in quantum Hall physics, due to the apparent absence of Lorentz invariance. In this paper we took a straightforward perspective. We {\it define} spin of a quasihole $S$ through the ``gravitational Aharonov-Bohm effect'' \eqref{curvatureAB}. When the effective Chern-Simons theory is Lorentz invariant (which is not the case for a realistic QH system), {\it i.e.} when $s=0$ in \eqref{Seff} the spin $S$ satisfies the spin-statistics relation as can explicitly be seen from \eqref{qhstat} and \eqref{spinqh}. However, when the Lorentz symmetry is {\it manifestly} broken by the Wen-Zee coupling (the third term in \eqref{Seff}) the spin-statistics relation does not hold. Now, the topological spin is defined to be an eigenvalue of the Dehn twist \eqref{Dehntwist}. The topological spin is insensitive to the value of $s$ and it does not couple to curvature, so the ``topological spin-statistics relation'' holds. The validity of the spin-statistics relation depends on which object is called spin. We choose to call $S$ the spin since it is (i) the quantum number that appears in the gravitational Aharonov-Bohm phase, and (ii) it is the conformal spin (which equals to the $SO(2)$ spin) of the vertex operator that creates a quasihole when the coupling to curvature is included into the Moore-Read construction. Finally, the spin can be defined for a particle-antiparticle pair. In the quasihole case such spin equals to $\frac{p^2}{q}$ and it satisfies spin-statistics theorem. The terms linear in charge cancel for particle and anti-particle cancel between each other. There is an extensive literature on the spin of quasiholes and anyons. We refer the interested reader to [\onlinecite{sondhi1992long, einarsson1991fractional, einarsson1995fractional, leinaas2002spin, li1993intrinsic, li1993anyons, lee1994orbital, read2008quasiparticle, laskin2015collective, can2014geometry, bradlyn2015topological}].

\section{Genons}

This Section contains the new results obtained in the present paper. We will introduce the curvature defect in a way that closely resembles the previous section. We will construct the wavefunction in the presence of the defects, determine the quantization conditions on the curvature flux and calculate charge, spin and statistics. We will discover that these defects are the genons \cite{barkeshli2013twist}, however our approach allows us to obtain extra information about the genons such as charge and spin. Since we restrict our attention to the Laughlin state we will also make progress in explicitly deriving the braiding properties of genons that correspond to higher genus surface and outline the general procedure that allows us to obtain (at least in principle) the braiding matrices for any ``parent'' topological phase. 

\subsection{Coulomb Plasma}
In this Section we will study the local behavior of the Laughlin state in the vicinity of a curvature defect. This will allow us to make a ansatz for the wavefunction and to compute the electric charge of a single genon.

\subsubsection{One defect}
We will start with a blunt brute force approach to the Coulomb plasma. We have already learned that the singular configurations of magnetic field \eqref{singularB} with quantized flux behave as particle-like local excitations. Now we consider a singular configuration of curvature $R$ on a sphere.
\be
R = \bar R - 4\pi \alpha \delta(z-a)\,.
\ee
Following the logic of Section 2 we calculate the electric charge depletion near $z=a$. We use \eqref{density} and find
\be
\delta N = \int \sqrt{g} (\rho - \bar \rho) = - \frac{\bar s\alpha}{q}\,.
\ee
 The unnormalized wavefunction with this concentration of charge in the vicinity of $z=a$ is
\be
\Psi(\{z_i\}, a) = \prod_i(z_i-a)^{\bar s\alpha}\prod_{i<j}(z_i - z_j)^{q} e^{-\sum_{i=1}^N \frac{\mathcal K(z_i,\bar z_i)}{4\ell^2}}\,.
\ee
The wavefunction is regular in the electron coordinates when $\bar s\alpha$ is a positive integer, and consequently, curvature flux is a negative integer in the units of $4\pi$. This flux is invisible to the electrons since it leads to an Aharonov-Bohm phase $e^{-2\pi i \bar s\alpha}$, provided that $\bar s$ is an integer. This holds, for example, for a bosonic Laughlin state. In the more general case the flux quantization is affected to ensure that this Aharonov-Bohm phase is trivial. We choose to parametrize $\alpha = n-1$ for the reasons that will become clear shortly. Thus the charge of the curvature defect is
\be\la{genoncharge}
Q = -\nu \bar s(n-1)\,.
\ee
This is as far as we can go with Coulomb plasma. For the remainder of the Subsection we will explore the geometry of the defects.
\subsubsection{Geometry}
The geometric singularity that corresponds to a genon is a conical singularity of degree $n$. Close to the singularity the metric in conformal coordinates is given by
\be
ds^2 = |z-a|^{2n-2}dz d\bar z\,.
\ee
Curvature is found from
\be
\sqrt{g}R = - 4 \p \bar \p \ln \sqrt{g} = -4\pi(n-1)\delta(z-a) \,.
\ee
The K\"ahler potential is given by
\be
\mathcal K(z,\bar z) = \frac{1}{n^2}|z-a|^{2n}\,.
\ee
Points of negative, quantized curvature usually appear on branched coverings of Riemann surfaces and come in pairs connected by a branch cut. In the present case $z=a$ is a branch point of degree $n$ and there is another branch point at infinity. First, we calculate the Euler characteristic of a surface with many genons. Denote the surface with $2M$ branch points of degree $n$ as $\Sigma_{n,2M}$ then Riemann-Hurwitz theorem allows to calculate the Euler characteristic
\be
\chi(\Sigma_{n,2M}) = 2 - 2(n-1)(M-1)\,
\ee
and the genus is
\be
g(\Sigma_{n,2M}) = (n-1)(M-1)\,.
\ee
The first non-trivial case that we will study in great detail is $g=0$ which implies $M=1$ ($n=1$ corresponds to the absence of a singular point). In this case for any $n$ the surface retains the topology of the sphere, in other words one pair of genons of any charge does not change the topology, however they do change the geometry.

The next simplest case, $4$ genons, has the Euler characteristic $\chi(\Sigma_{n,4}) = 2 - 2(n-1)$ and the genus $g(\Sigma_{n,4}) = n-1$. For $M>1$ the genus increases with the charge of the genon.

\begin{figure}
  \includegraphics[width=\linewidth]{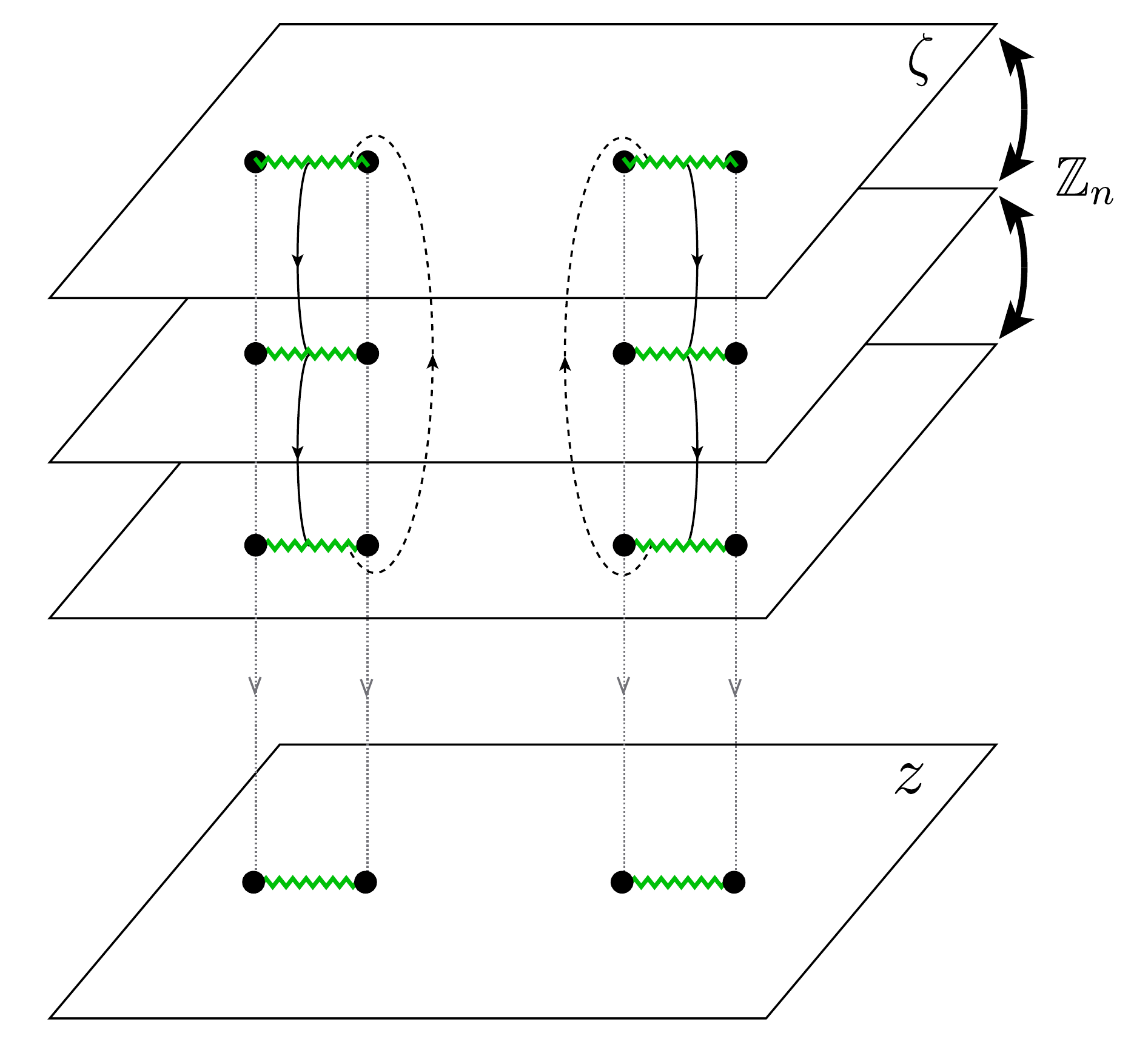}
  \caption{An example of the genon surface $\Sigma_{3,4}$. In $\zeta$ variables there is an explicit $\zn$ automorphism of the surface. Every marked point is a branch point of degree $3$. Also, every sheet has $8\pi$ units of curvature at infinity. The $n$ to $1$ map $\zeta \mapsto z$ is a projection that creates conical defects of degree $n$ in the $z$-plane. This projection can be understood as a factor of the genon surface by the action of $\zn$.}
  \label{ZETA}
\end{figure}

A simpler description of the $\Sigma_{n,4M}$ surface can be given in a different set of coordinates that we call $\zeta$ (see FIG. \ref{ZETA}). Locally, the mapping between $\zeta \longrightarrow z(\zeta)$ is $n$ to $1$ except at the position of the genon and given by
\be\la{zeta}
\zeta = (z-a_i)^n\,.
\ee
In these coordinates the metric is regular at the origin and takes the form 
\be
ds^2 = d\zeta d\bar \zeta = \left|\frac{\p \zeta}{\p z}\right|^{2}dzd\bar z = \frac{1}{n^2} |z-a_i|^{2n-2}dz d\bar z\,.
\ee
the geometry is encoded in the boundary conditions: when $z$ is analytically continued around $a$, $\zeta$ travels around the origin $n$ times. In this representation it is clear that we are working with a very special surface with an extra $\mathbb Z_n$ automorphism. This automorphism will play an important role in the derivation of braiding statistics of genons. 
\subsection{CFT on a Singular Surface}

Conformal field theory proved to be useful as an alternative way to derive the wavefunction and in calculation of quantum numbers of the quasiholes. In this Subsection we will use the CFT on the singular geometry described above to derive some properties of genons. We will  interpret the singular points as primary fields in an orbifold CFT. This, in turn, will allow us to calculate spin and statistics of the genons. We wish to interpret the correlation function of electron operators evaluated on the $\Sigma_{n,2M}$ surface as the wavefunction in the presence of genons. Then the full power of CFT  can be used to determine the correlation functions including the dependance on the positions of the genons $a_i$. This method conceptually generalizes to any number of genons, but the computations quickly escalate in difficulty due to the topology change induced by the presence of genons. We will take the Moore-Read point of view and interpret the neutralizing background as an extra insertion in the correlation functions. The non-holomorphic factors are dropped for brevity. In this Section $V_q(\zeta,\bar \zeta)$ denotes an arbitrary electron operator in a conformal block trial state, however all of the explicit calculations will be carried out for the compact boson at $R=2\pi \sqrt{q}$, {\it i.e.} for the Laughlin state.

\subsubsection{Two genons}
To make things simpler we start with the surface $\Sigma_{n,2}$. In this case for any $n$ the surface is topologically a sphere.

We consider the correlation function on the surface $\Sigma_{n,2}$ in coordinates $\zeta$ understood as a function of the coordinates $z$
\be
\Psi_{2g}(\{z_i\},a_i) \equiv \Big\langle \prod_i V_q(\zeta_i,\bar\zeta_i) \Big\rangle_{\Sigma_{n,2},\zeta(z)}\,.
\ee
To evaluate the correlation function we make a coordinate transformation from coordinates $\zeta$ to coordinates $z$. There is no good global formula for the transformation, however in a vicinity of each branch point $a_i$ the transformation takes form \eqref{zeta}.
Despite the lack of global formula for the transformation law we can still write down a ``global'' formula for the induced metric. It is given by
\be
ds^2 = |z-a_1|^{2n-2}|z-a_2|^{2n-2}dz d\bar z\,.
\ee
This formula works in a chart that does not include a small vicinity of infitnity. 
The curvature density is then given by
\be\la{2gr}
\sqrt{g}R = - 4\pi(n-1) \Big( \delta(z-a_1) + \delta(z-a_2) \Big)\,.
\ee
This equation is written in the chart that does not include infinity. There is extra curvature at infinity that ensures that the Euler characteristic comes out correctly \footnote{To be more precise Eq. \eqref{2gr} is somewhat symbolic as every term in the sum is written in a chart that covers vicinity of $z=a$. Indeed when singularities $a_i$ move around they induce the perturbations of metric outside of their location}
\bea\nonumber
\chi(\Sigma_{n,2}) &&= \int_{U(a_1)} \sqrt{g} R + \int_{U(a_2)} \sqrt{g} R  +\int_{U(\infty)} \sqrt{g} R  \\
&&= -(n-1) - (n-1) +2n\,,
\eea
where $U(a)$ a small neighborhood of $z=a$.

In the $z$-coordinates we have
\be
 \Psi_{2g} = \prod_{i=1}^N\left( \frac{\p z}{\p\zeta}\right)^{h}\Big|_{z=z_i}\left( \frac{\p \bar z}{\p\bar\zeta}\right)^{\bar h}\Big|_{\bar z=\bar z_i}\cdot\Big \langle  \prod_{i=1}^N V_q(z_i,\bar z_i) \Big\rangle_{\Sigma_{n,2},z} \, .
\ee
The first factor comes from the transformation of the electron operator insertions. The correlation function is still difficult to evaluate since the geometry of the space is both non-trivial and singular. The next step is to remove the metric singularity by a Weyl transformation
\be\la{metricWeyl}
g^\prime = e^{-2\sigma(z)} g\,,
\ee
where $\sigma(z) = (n-1) (\ln|z-a_1| +\ln|z-a_2|)$ so that $ds^2 = g^\prime_{z\bar z} dz d \bar z = dz d \bar z$. After the Weyl transformation the correlation function acquires an extra factor given by the integrated Weyl anomaly
\be\la{GWF}
  \Psi_{2g} =e^{ \frac{c}{48\pi} S_L[\sigma]}\cdot\prod_{i=1}^N\Big( \frac{\p z}{\p\zeta}\Big)^{h}\cdot\Big \langle  \prod_{i=1}^N e^{i\sqrt{q}\phi(z_i)} \Big\rangle_{\cp,z}\,,
\ee
where $S_L[\sigma]$ is the Liouville action given by
\be 
S_L[\sigma] = \int \p \sigma \bar\p \sigma + R[g^\prime]\sigma\,,
\ee
where $R[g^\prime] = 0$ is the curvature of the metric $g^\prime$. The wavefunction \eqref{GWF} is equivalent to the one used in [\onlinecite{bradlyn2015topological}] in the presence of arbitrary smooth deformation of the metric. The factors $\left(\frac{\p z}{\p \zeta}\right)^h = \sqrt{g}^h = e^{h\sigma}$ can be regarded as ``gravitational dressing'' of the electron operators $V_q$.

The Weyl transformation acts only on the metric and not on the coordinates. The functional integration measure is not Weyl invariant due to the Weyl anomaly which leads to the Liouville factor. The neutralizing background is explicitly {\it not} Weyl invariant. Indeed, under a Weyl transformation the neutralizing background (second term in \eqref{rho}) transforms as 
\be\la{densityWeyl}
 \rho^\prime =   \rho - \frac{\nu \bar s}{4\pi} \Delta \sigma\,.
\ee
There are two ways to do the Weyl rescaling. One (more traditional) is to simply allow the metric to change according to \eqref{metricWeyl} and the background density will transform according to \eqref{densityWeyl}. Another way is to make a Weyl transformation {\it keeping the density $\rho$ fixed}. This can be achieved via accompanying the Weyl transformation with a {\it simultaneous} transformation of magnetic field $B \rightarrow B + \frac{\bar s}{2} \Delta \sigma$. Then $ \rho ^\prime =  \rho$. Different choices will correspond to slightly different prescription for the braiding of the genons. In the first case the genons are braided ``as is'', while in the second choice the adiabatic dragging of genons is accompanied by adiabatic variations of magnetic field so that the combination $B + \bar s R$ is kept constant\cite{bradlyn2015topological}. These braiding processes are equivalent in a sense that knowing the result of one braiding experiment one can reconstruct the result of the other. We choose to accompany Weyl variations with variations of magnetic field for aesthetic reasons. In the case of CFT trial states this choice will result in replacing $ c_{\mathrm w}$ with $c$.

Finally, the correlation function has to be evaluated on $\cp$ with metric $g^\prime$, which we have done in the previous Section. We are now left with the problem of evaluating the Liouville action on the singular metric induced by the map $z(\zeta)$. Fortunately, the Liouville action has been evaluated on precisely this metric in the study of orbifold CFTs \cite{lunin2001correlation}. See also the Appendix \ref{Laction}. We have
\be\la{2gfactor}
e^{\frac{c}{48\pi}S_{L}[\sigma]} = |a_1-a_2|^{-\frac{c}{6} \left(n-\frac{1}{n}\right)}\,.
\ee

 Putting things together and taking only the holomorphic part we present the two genon ``wavefunction'' on top of the Laughlin state
\bea\nonumber
\Psi_{2g} = \mathcal N_0(a_i) &&(a_1 - a_2)^{-\frac{c}{12}(n -\frac{1}{n})}\prod_{i,k}(z_i - a_k)^{\bar s(n-1)}\\ \la{2gwf}
&&\cdot \prod_{i<j}(z_i - z_j)^q e^{-\sum_i \frac{\mathcal K(z_i, \bar z_i) }{4\ell^2}}\,.
\eea
Following Ref.~[\onlinecite{bradlyn2015topological}] we assume that the state \eqref{GWF} is normalized when integrated with $\sqrt{g}$ the monodromy of conformal block will equal the Berry phase acquired by the wavefunction when metric, ({\it i.e.} $a_i$) is varied adiabatically.

Adiabatic exchange of $a_1$ with $a_2$ produces a phase
\be\la{genonstat}
 \gamma_{stat}=  \frac{c}{24}\Big(n-\frac{1}{n}\Big)\,.
\ee 
This phase allows us to extract the central charge $c$ (and $c_\mathrm{w}$) from a braiding experiment. Notice, that the braiding phase is universal and depends on the ``parent state'' only through the central charge $c$. There is an identical effect in the evaluation of the entanglement entropy (EE) in $2$D CFT - for a single interval EE is completely determined by the central charge.

We can also calculate the spin of the genon by evaluating the conformal dimension of a branch point. This is a classic computation that can be found in [\onlinecite{knizhnik1987analytic}], see also [\onlinecite{bershadsky1987conformal, bershadsky1987g, bershadsky1988fermionic, dixon1987conformal}]. It is done by evaluating the expectation value of the stress tensor $\langle T(\zeta)\rangle_{\zeta}$ on the $\Sigma_{n,2}$ surface, using the same method we described before, and comparing it to a general form of a the two point function of stress tensor with a primary field on the plane. The result is
\be\la{spingenon}
h_n  =\frac{c}{24}\left(n - \frac{1}{n}\right)\equiv S\,.
\ee
Alternatively, this result can be deduced by inspecting \eqref{2gfactor} and observing that it looks exactly like a two-point function of primary fields with conformal dimension (and conformal spin) given by \eqref{spingenon}. Thus we have calculated charge, spin and statistics of genons associated to $\Sigma_{n,2}$. Unlike the quasiholes, the genons do satisfy the spin-statistics theorem.

Two genons are qualitatively different from any other number of genons in that for any $n$ they are abelian, in other words, the state in the presence of two genons is {\it non-degenerate}. Similar effect happens when one considers a Moore-Read state with only two quasiholes. It is possible to consider genons on a non-compact manifold such as pseudo-sphere. In this case it should be possible to have more than two ``abelian genons''. We will not pursue that route in the present paper. 
\subsubsection{Four genons}

Next, we consider the case of four genons. The genus of the surface $\Sigma_{n,4}$ is $n-1$. For simplicity we choose $n=2$ to get the topology of the torus (however, with singular metric). When more than two genons are present the genus is increased and, consequently, the Laughlin state becomes degenerate and braiding can, in principle, induce non-abelian monodromy among the ground states. We will find this to be the case.

Before proceeding with the computation of the correlation functions we will warm up with an evaluation the partition function for a compact $c=1$ boson at rational compactification radius $r=\sqrt{q}$ on $\Sigma_{2,4}$ and compare it to the torus partition function. We think of a partition function as an unnormalized expectation value of the identity operator $Z_{\Sigma_{2,4}} = \langle 1 \rangle_{\Sigma_{2,4}, \zeta}$. We have
\bea\nonumber
 Z_{\Sigma_{2,4}}  &&= \langle 1 \rangle_{\Sigma_{2,4}, \zeta} =  \langle 1 \rangle_{\Sigma_{2,4}, z} = e^{\frac{c}{48\pi} S_L[\sigma]} \langle 1\rangle_{T,z} \\
 &&= \prod_{i<j}|a_i-a_j|^{-\frac{c}{12}} Z_T\,,
\eea
where we have used \cite{lunin2001correlation}
\be
e^{\frac{c}{96\pi}S_L[\sigma]} = \prod_{i<j}^4 |a_i - a_j|^{-\frac{c}{12}}
\ee
and $Z_T$ is the standard (diagonal) partition function on a torus given by
\be
Z_T = \sum_{p=0}^{q-1}\chi_p(\tau) \bar \chi_p(\tau)\,,
\ee
where $\chi_p$ are the $\widehat{u(1)}_q$ characters given essentially by the center of mass functions introduced in Section 2 \be
\chi_p(\tau)\sim \frac{1}{\eta(\tau)}F_q\begin{bmatrix}
    \frac{p}{q}  \\
    0 \\
\end{bmatrix}\,.
\ee
The final expression,
\be
Z_{\Sigma_{2,4}}  =  \prod_{i<j}|a_i-a_j|^{-\frac{c}{12}} \sum_{p=0}^{q-1}\chi_p(\tau) \bar \chi_p(\bar\tau)\,,
\ee
must be understood as an implicit function of $a_i$ with $\tau$ expressed in terms of $a_i$  according to \cite{dixon1987conformal, lunin2001correlation}
\be\la{supereq}
x \equiv \frac{(a_1 - a_3)(a_2 - a_4)}{(a_1 - a_4)(a_2 - a_3)}  = \left(\frac{\theta_3(\tau)}{\theta_4(\tau)}\right)^4\,.
\ee

We emphasize that the partition functions (and, consequently, correlation functions) on $\Sigma_{2,4}$ and $T$ are {\it not equal to each other}, but instead differ by a factor, fixed by the conformal anomaly.

Next, we wish to compute the correlation function
\be
\Big\langle \prod_i V_q(\zeta_i,\bar \zeta_i)\Big\rangle_{\Sigma_{2,4},\zeta}\,.
\ee
Going through the same steps as in the case of two genons we find the wavefunction as
\be\la{WF4g}
\Psi_{4g}(a_i) = e^{ \frac{c}{48\pi} S_L[\sigma]}\prod_{i=1}^N\Big( \frac{\p z}{\p\zeta}\Big)^{h}\Big|_{z=z_i}\Big \langle  \prod_{i=1}^N V_q(z_i)\Big\rangle_{T,z}\,,
\ee
where the last expectation value has to be computed on a torus and is precisely the torus wavefunction we studied in the previous Section. Any conformal block in the last factor is a good choice for the four genon wavefunction. Thus the state with four genons at $n=2$ is $q$-fold degenerate. Increasing  the number of genons to $2M$ will lead to $q^{M-1}$-fold degeneracy (recall that the genus of $\Sigma_{2,2M}$ is $M-1)$  which implies quantum dimension $\sqrt{q}$ for each genon \cite{barkeshli2013twist}. 

When $q=2$ the quantum dimension is $\sqrt{2}$ and scaling dimension is $h_2=\frac{1}{16}$ (when $c=1$). The state with two More-Read quasiholes is abelian and the (abelian) exchange phase is $e^{-2\pi i \frac{1}{8}}$, which appears to be different from \eqref{genonstat}. Thus genons are similar to non-abelian quasiholes that appear in the Moore-Read state, since they have the same scaling dimension and the same quantum dimension, but different overall braiding phase. We will later show that they have the same braid matrices, up to a (fixed) phase.

The final expression for the degenerate four genon wavefunction is
\bea\nonumber
\Psi_{4g,p} = \mathcal N(a)\prod_{i<j}^4 (a_i - a_j)^{-\frac{c}{24}}&& \cdot \prod_{i,k}(z_i - a_k)^{(n-1)  \bar s} \\\la{4genon}
&& \cdot \Psi_p(\{z_i\}|\tau)\,,
\eea
where $\Psi_p$ is given by \eqref{LaughlinT} and $\tau$ is expressed in terms of anharmonic ratio of $a_i$ through \eqref{supereq}.

It is important to understand that $a_i$ do {\it not} live on a torus, instead, they live on the $z$-plane with singular points that is related to the smooth torus by a Weyl transformation. The first and second factors in \eqref{4genon} describe the local behavior of the wavefunction when either two genons or  an electron and a genon come close to each other. The third factor must be understood as a function of $z_i$ and $a_k$. This re-writing has an advantage that now we can understand the braiding of genons (action of the braid group on $a_i$) in terms of the action of the modular group on $\tau$. To understand this correspondence we will use the crucial Eq.\,\eqref{supereq}. First, recall the properties of the $\theta$-constants $\theta_i(\tau)=\theta_i(z=0,\tau)$ \cite{CFT-book}. We will need the transformation laws under the Dehn twist
\bea
\theta_1(\tau + 1 ) &&= e^{\frac{2\pi i}{8}} \theta_1(\tau)\,,\\
\theta_2(\tau + 1) &&= e^{\frac{2\pi i}{8}}\theta_2(\tau)\,,\\
\theta_3(\tau +1) &&= \theta_4(\tau)\,,\\
\theta_4(\tau + 1) &&= \theta_3(\tau)\,,
\eea
and the transformation laws under $\mathcal S$ transformation
\bea
\theta_2 \left(- \frac{1}{\tau}\right) &&= \sqrt{-i\tau}\theta_4(\tau)\,,\\
\theta_3\left(-\frac{1}{\tau}\right) &&= \sqrt{-i\tau} \theta_3(\tau)\,,\\
\theta_1\left(-\frac{1}{\tau}\right) &&= - i\sqrt{-i\tau} \theta_1(\tau)\,,\\
\theta_4\left(-\frac{1}{\tau}\right) && = \sqrt{-i\tau}\theta_2(\tau)\,.
\eea
We will also need the Jacobi identity
\bea
\theta_3^4(\tau) - \theta_4^4(\tau)-\theta_2^4(\tau)&&=0\,,\\
\theta_1(\tau)=0\,.
\eea
Using these identities it is not hard to derive the action of $\mathcal S$ and $\mathcal T_a$ on the anharmonic ratio
\be
\mathcal T_a \circ x = \frac{1}{x}\,, \qquad \mathcal S \circ x = \frac{x}{x-1}\,.
\ee
It is now a matter of simple algebra to derive the braiding matrices. First, consider $\mathcal B_{23}$ - the braiding of $a_2$ with $a_3$. We have
\be
\mathcal B_{23} \circ x = x(a_2\rightarrow a_3, a_3 \rightarrow a_2) = 1-x\,.
\ee
In terms of the modular transformations we have
\be
\mathcal B_{23} \circ x = (\mathcal S \mathcal T_a \mathcal S^{-1}) \circ x \equiv  \mathcal T_b\circ x
\ee
We conclude that the braid $\mathcal B_{23}$  induces a Dehn twist around the $b$-cycle $\mathcal T_b$. The $\mathcal S$ transformation induces an overall phase, however this phase cancels since $\mathcal B_{23} =\mathcal S \mathcal T \mathcal S^{-1} $. The only contribution to an overall phase comes from the Liouville action that is given by \eqref{genonstat}.

Next, we consider $\mathcal B_{12}$ - the braiding of $a_1$ and $a_2$. We have
\be
\mathcal B_{12} \circ x = x(a_1\rightarrow a_2, a_2 \rightarrow a_1) = \frac{1}{x}\,,
\ee
which implies
\bea
\mathcal B_{12}\circ x = \mathcal T_a \circ x\,.
\eea
We conclude that the braid $\mathcal B_{12}$ induces a Dehn twist around an $a$-cycle $\mathcal T_a$. It is well-known that transformations $\mathcal T_a$ and $\mathcal T_b$ generate the full modular group and, consequently, do $\mathcal B_{12}$ and $\mathcal B_{23}$. These relations completely fix the non-abelian part of the transformation. Finally, we note that $\mathcal B_{34} = \mathcal B_{12} = \mathcal T_a$. 

To summarize the braid matrices act on the space of ground states as follows
\bea\la{B23}
(\mathcal B_{12}\Psi_{4g})_p = (\mathcal T_a)_{p p^\prime} \Psi_{4g,p^\prime}\, ,\\ \la{B12}
(\mathcal B_{23}\Psi_{4g})_p = (\mathcal T_b)_{p p^\prime} \Psi_{4g,p^\prime}\,.
\eea
These relations, together with \eqref{Tmatrix}-\eqref{Smatrix}, give explicit braid matrices for $4$ genons on top of the Laughlin state. Note that Eq.\,\eqref{supereq} does not care about the quantum Hall state in question, thus mapping between the genons and modular group is going to hold for any ``parent'' topological phase. Notice that relations \eqref{B23}-\eqref{B12} are general in that they will work whenever the action of the modular group on the ground state space is known. This observation hints that the homomorphism between the braid group and mapping class group is universal in that it is independent of the topological phase. The only input from the topological phase comes in the explicit form and size of the braid matrices.

We are led to conclude that braid matrices of genons form $q$-dimensional representation of $SL(2,\mathbb Z)$. When $q=2$ the genon braid matrix calculated from \eqref{Tmatrix}-\eqref{Smatrix} agrees with the braid matrices for the Moore-Read quasiholes\cite{barkeshli2013twist}. Curiously, when $q=1$ (IQH state) genons are abelian for any $n$ and $M$. The braiding induces only a universal $U(1)$ phase coming from the Liouville action.

The reason we were able to be very explicit in this Subsection is due to the existence of Eq.\eqref{supereq}. Regrettably, the situation is different for $g\geq2$. It does not appear to be possible to directly derive a higher genus analogue of \eqref{supereq} however, it is possible to develop a ``routine'' procedure that derives the inverse of \eqref{supereq}, meaning the expression of the moduli in terms of the cross-ratios (as opposed to cross-ratios in terms of moduli). This procedure leads to very complicated expressions and is explained in the Appendix \ref{ZnSect}, where the $g=1$ case is explicitly worked out and higher genus general (but quite implicit) expressions are presented.

\subsection{CFT on Higher Genus Surfaces}
In this Subsection we will explain how to calculate the adiabatic statistics of genons beyond the toric geometry. To do so we are inevitably led to a study conformal block trial state on higher genus surfaces $\Sigma_{n,2M}$ with genus $g(\Sigma_{n,2M})=(M-1)(n-1)$.

Before going into any details we informaly outline the general strategy. A smooth, compact Riemann surface has $3g-3$ dimensional moduli space $\mathscr M_g$ which can be parametrized by a complex, symmetric period matrix $\Omega_{ij}$ which size is $g\times g$. Similarly to how $\tau\equiv \Omega_{11}$ is acted on by the group of large diffeomorphisms of a torus $SL(2,\mathbb Z)$, the period matrix $\Omega_{ij}$ is acted on by the group of large diffeomorphisms of a higher genus Riemann surface, which can be expressed as $Sp(2g,\mathbb Z)$ matrices. The period matrix can be related to the positions of the genons $a_i$ which are simply a different way to parametrize the {\it same} moduli space (to be more precise, they parametrize the moduli of Riemann surfaces with $\zn$ automorphism, which is a small corner in the entire moduli space). This relation can be put in more or less explicit form and provides a generalization of \eqref{supereq} which was crucial in deriving the braiding matrices. With this relation at hand it is, in principle, possible to translate braiding of genons into the action of large diffeomorphisms. There are two ways to derive this relation. One is based on algebraic geometry and is presented below. Another one, based on calculus of cut abelian differentials, is presented in the Appendix \ref{ZnSect}. The latter is much less transparent and involves complicated calculations, whereas the former is very intuitive.

\subsubsection{Some algebraic geometry}
We start with an elementary introduction to algebraic geometry of Riemann surfaces. An accessible review of these issues can be found in [\onlinecite{alvarez1986theta}], [{\onlinecite{farb2011primer}]. Below we will provide an absolute minimum, mostly to fix the terminology and notations. Let $\Sigma$ be a Riemann surface of genus $g$. Every such surface has a non-trivial $2g$-dimensional homology group $H_1(\Sigma)$. Loosely speaking, homology group describes the ``independent'' closed curves (referred to as cycles) on a Riemann surface and equips them with an intuitive multiplication law. The simple illustration is a torus, which has $2$ non-trivial cycles traditionally denoted $a$ and $b$. On a higher genus surface there are $g$ $a$-cycles $a_i$, and $g$ $b$-cycles $b_i$ (see FIG. \ref{HomB} for an example). These cycles form a basis in the homology group $H_1(\Sigma)$. Every curve can be decomposed in the basis of these cycles. Now consider two curves $\mathcal C_1$ and $\mathcal C_2$. Let $J(\mathcal C_1,\mathcal C_2)$ be an intersection number of these curves, {\it i.e.} a number of times the curve $\mathcal C_1$ intersects the curve $\mathcal C_2$ with the sign assignment that depends on the orientation (to clarify, at the intersection point tangent vectors to the curves can form either left or right pair; this determines the sign assignment). The intersection number $J$ depends only on the homology class of curves $\mathcal C_i$ and not on the details of their shape. When evaluated on the basis $(a_i, b_i)$ $J$ becomes
\be
J(a_i,a_j) = J(b_i,b_j) = 0, \quad J(a_i,b_j) = -J(b_i,a_j) = \delta_{ij}\,,
\ee
that is $J$ is a block diagonal $2g \times 2g$ matrix
\be
J = \begin{bmatrix}
    0 && I_g \\
    -I_g && 0 \\
\end{bmatrix}\,,
\ee
where $I_g$ is $g\times g$ unit matrix.

\begin{figure}
  \includegraphics[width=\linewidth]{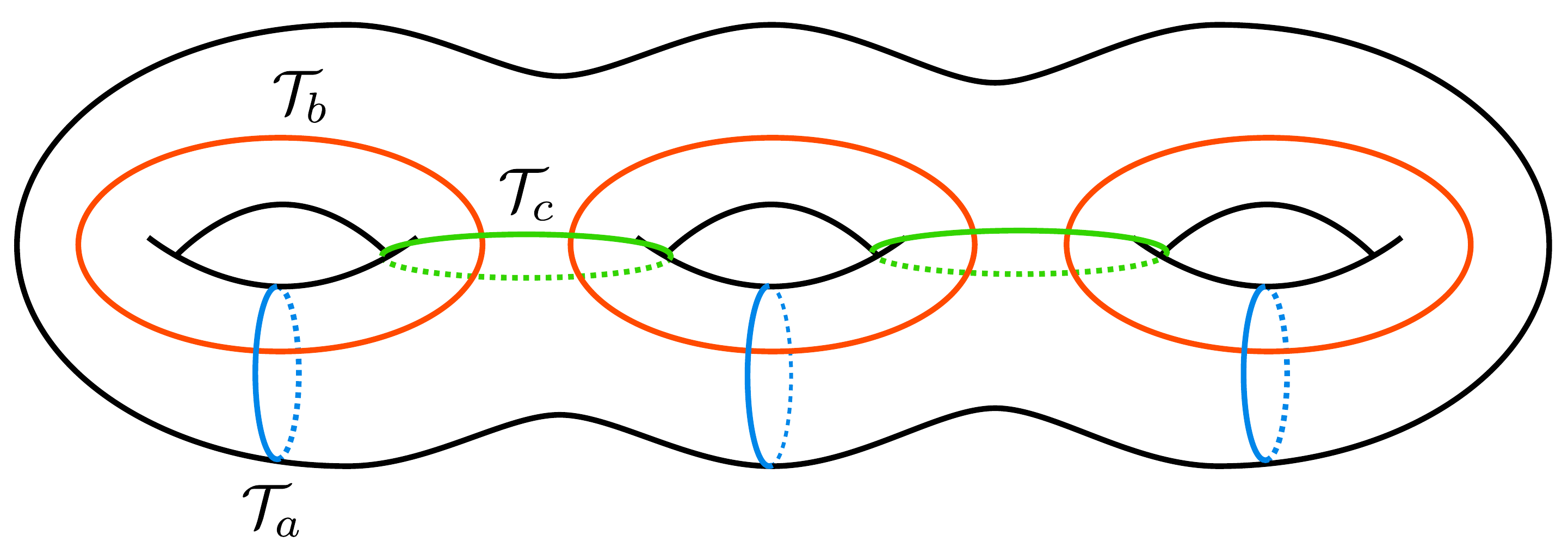}
  \caption{The homology basis consists of $a$ and $b$ cycles. The mapping class group is generated by $3g-1$ Dehn twists $\mathcal T_a, \mathcal T_b$ and $\mathcal T_c$ around $a$, $b$ and $c$ cycles correspondingly. Notice that there is no $\mathcal T_c$ generator at genus $1$.}
  \label{HomB}
\end{figure}

Intersection numbers are topological invariants of a pair of curves and cannot depend on the choice of coordinates. There are two types of (orientation preserving) coordinate transformations: small and large. Small coordinate transformations are homotopic (can be smoothly deformed) to an identity, whereas the large ones are not. Large coordinate transformations form the mapping class group $\mathcal M(\Sigma)$. This group will play the central role in the remainder of this Subsection. There is a natural set of generators of $\mathcal M(\Sigma)$, these are Dehn twists around $3g-1$ cycles. In algebraic form these cycles are the natural homology basis
$a_i, b_i$, $i=1,\ldots,g$ and $g-1$ cycles $c_i = -a_{i} +a_{i+1}$.  We will denote the corresponding Dehn twists as $\mathcal T_{a_i}, \mathcal T_{b_i}$ and $\mathcal T_{c_i}$. In the case of a torus there are only two generators  $\mathcal T_{a}$ and $\mathcal T_{b}$. The homology basis and the basic Dehn twists are illustrated in FIG. \ref{HomB} for the case $g=3$.

It is clear that small coordinate transformations do not change $J$, however the large ones change the homology basis, so the invariance of the intersection numbers will give a relation between the elements of $\mathcal M(\Sigma)$. For example, the Dehn twist $\mathcal T_a$ changes $b$ to $b+a$, while leaving $a$ invariant. A generic large coordinate transformation will induce a change of basis described by an integer valued matrix $M$, demanding that intersection numbers do not change we arrive at
\be
M^T J M = J\,.
\ee
Thus large coordinate transformations represent the action of the symplectic modular group $Sp(2g, \mathbb Z)$. When $g=0$ there are no moduli and when $g=1$ we have an isomorphism $Sp(2,\mathbb Z) \approx SL(2,\mathbb Z)$.  When $g\geq2$ there is a new phenomenon\cite{farb2011primer}: we have a short exact sequence
\be
1\longrightarrow \mathcal I(\Sigma) \longrightarrow \mathcal M(\Sigma) \longrightarrow Sp(2g, \mathbb Z)\longrightarrow 1\,.
\ee
In other words, there are some non-trivial mapping classes that map to the unit matrix. These mapping classes form a normal subgroup known as the Torelli group $\mathcal I(\Sigma)$. The role of the Torelli group in topological phases of matter is not clear.

Next, we discuss forms on $\Sigma$. There is a natural choice of basis in the first cohomology $H^1(\Sigma)$, dual to $(a_i,b_i)$ that we denote $(\alpha_i,\beta_i)$ that satisfies
\be
\int_{a_i}\alpha_j = \int_{b_i} \beta_j = \delta_{ij},\quad \int_{a_i}\beta_j = \int_{b_i} \alpha_j = 0\,.
\ee
There is, however, a more useful basis of abelian differentials of the first kind. To define these we first go to complex, conformal coordinates so that the metric is given by $ds^2 = \sqrt{g} dz d\bar z$. With the notion of complex conjugate at hand we can separate all one-forms into two groups: ones that locally look like $f(z,\bar z) dz$ and ones that locally look like $f(z,\bar z) d\bar z$. In particular, there are forms that look like $\omega = \omega(z)dz$, which are called holomorphic. We can choose a basis in the space of holomorphic forms (there are $g$ of those and $g$ of anti-holomorphic ones) demanding
\be
\int_{a_i} \omega_j = \delta_{ij}
\ee
then the integrals over the $b$-cycles are fixed uniquely
\be
\int_{b_i} \omega_j = \Omega_{ij}\,,
\ee
where the matrix $\Omega_{ij}$ is known as the period matrix and it encodes the moduli. It is not hard to show \cite{alvarez1986theta} that the period matrix is (i) symmetric and (ii) ${\mbox  Im} \,\Omega >0$. The space of matrices satisfying (i) and (ii) is known as Siegel upper half plane. Under the large diffeomorphisms $\Omega_{ij}$ transforms in a non-trivial way. Given a $Sp(2g,\mathbb Z)$ transformation
\be
M = \begin{bmatrix}
    D && C \\
    B && A\\
\end{bmatrix}
\ee
the period matrix transforms as \cite{alvarez1986theta}
\be\la{mpgact}
\Omega^\prime = (A\Omega + B)(C\Omega+D)^{-1}\,.
\ee

With the period matrix at hand one can generalize the notion of $\theta$-functions \cite{fay1973theta}. Generalized $\theta$-function with characteristics is defined as
\be\la{thetagendef}
\theta\begin{bmatrix}
    \boldsymbol a  \\
    \boldsymbol b \\
\end{bmatrix}\Big({\boldsymbol{z}}|\Omega\Big) = \sum_{n\in\mathbb Z^g} e^{\pi i (\boldsymbol n+ \boldsymbol a)\Omega (\boldsymbol n+ \boldsymbol a) + 2\pi i(\boldsymbol n+\boldsymbol a)(\boldsymbol z+\boldsymbol b)}\,,
\ee
where the bolded symbols denote $g$-dimensional vectors. The characteristics themselves have become vectors since on a higher genus there are many cycles through which a flux can be threaded. The Laughlin state will generally be given in terms of the generalized $\theta$-functions. Notice an unpleasant novelty - the generalized $\theta$-functions are functions of $g$ variables instead of one, thus we will need a way to naturally introduce many variables in the trial states.

\subsubsection{Braid group vs. mapping class group}
It turns out that the success of the mapping class group approach to braiding is not accidental. There is a deep and beautiful relation between the mapping class group and various braid groups $\mathcal B_m$. In this Subsection we explore this relation and obtain an explicit geometric representation of the generators of $\mathcal B_m$ (and a slight modification of $\mathcal B_m$) in terms of the generators of $\mathcal M (\Sigma)$. For this Subsection we also fix the following notation: a Riemann surface of genus $g$ with $m$ (indistinguishable) punctures or marked points is denoted as $S_{g,m}$. The surface $S_{0,2M}$ will always be regarded as a $n$ to $1$ projection of $\Sigma_{n,2M}$, given locally by $z(\zeta)$ or, equivalently, as factor of $\Sigma_{n,2M}$ by the action of the $\zn$ automorphism  $\Sigma_{n,2M}/\zn$. All the braiding is done in $z$-variables {\it i.e.} on a sphere with puntctures.

We start by recalling that the (planar) braid group on $m$ strands, $\mathcal B_{m}$, is a group on $m-1$ generators that satisfy relations
\bea\la{braid1}
\sigma_i\sigma_j = \sigma_j \sigma_i\, ,\qquad |i-j|\geq2\\
\la{braid2}
\sigma_i \sigma_{i+1} \sigma_i = \sigma_{i+1} \sigma_{i} \sigma_{i+1}\,.
\eea
The center of the braid group $Z(\mathcal B_{m})$ is spanned by $(\sigma_1\cdot\ldots\cdot\sigma_{m-1})^{m}$. Braid generator $\sigma_i$ acts by intertwining strand $i$ with strand $i+1$. Another way to represent the (planar) braid group is via the mapping class group of a disc with $m$ punctures $\mathcal B_{m} \approx \mathcal M(D_{m})$. To make an explicit map we index the punctures. Then braid generator $\sigma_i$ maps to a Dehn half-twist around a loop that surrounds two punctures $i$ and $i+1$. The center of the braid group is then spanned by Dehn half-twists $\mathcal T_{\p D}$.

Instead of the (planar) braid group $\mathcal M(D_{m})$ it is more convenient to use the {\it spherical} braid group $\mathcal M(S_{0,m})$. To be more precise, the spherical braid group is $\pi_1(S_{0,m})$ and there is a short exact sequence
\be
1\longrightarrow \mathbb Z/2\mathbb Z \longrightarrow \mathcal \pi_1(S_{0,m}) \longrightarrow \mathcal M(S_{0,m})\longrightarrow 1\,,
\ee
but in the following we will disregard the kernel $\mathbb Z/2\mathbb Z$ and will not distinguish the spherical braid group from $\mathcal M(S_{0,m})$.
 The defining relations are slightly different, but conceptually the group is similar. Roughly speaking, the extra relations occur because it is possible to rotate the sphere (see FIG. \ref{Spherical}). Let $\xi_i$ be the generators. Then\footnote{There is a braided category version of  this known as spherical braided category}
\bea\la{sbraid1}
\xi_i\xi_j = \xi_j \xi_i\, ,\qquad |i-j|\geq2\\
\la{sbraid2}
\xi_i \xi_{i+1} \xi_i =\xi_{i+1}\xi_{i} \xi_{i+1}\\
\la{notbraid}
(\xi_1\cdot\ldots\cdot\xi_{m-1})^{m}=1\\ \la{Z2braid}
\xi_1\cdot\ldots\cdot\xi_{m-2}\xi_{m-1}^2\xi_{m-2}\cdot\ldots\cdot\xi_1=1\,.
\eea
The generator $\xi_i$ is a half-twist around any curve that encircles only the points $i$ and $i+1$ \footnote{ For the interested reader we note that relation \eqref{notbraid} is not present in the authentic spherical braid group.}. There is also an obvious homomorphism $\mathcal M(D_m) \rightarrow \mathcal M(S_{0,m+1})$ under which $\mathcal M(D_m)$ maps on the Dehn twists of $\mathcal M(S_{0,m+1})$ that preserve one puncture.

\begin{figure}
  \includegraphics[width=2.5in]{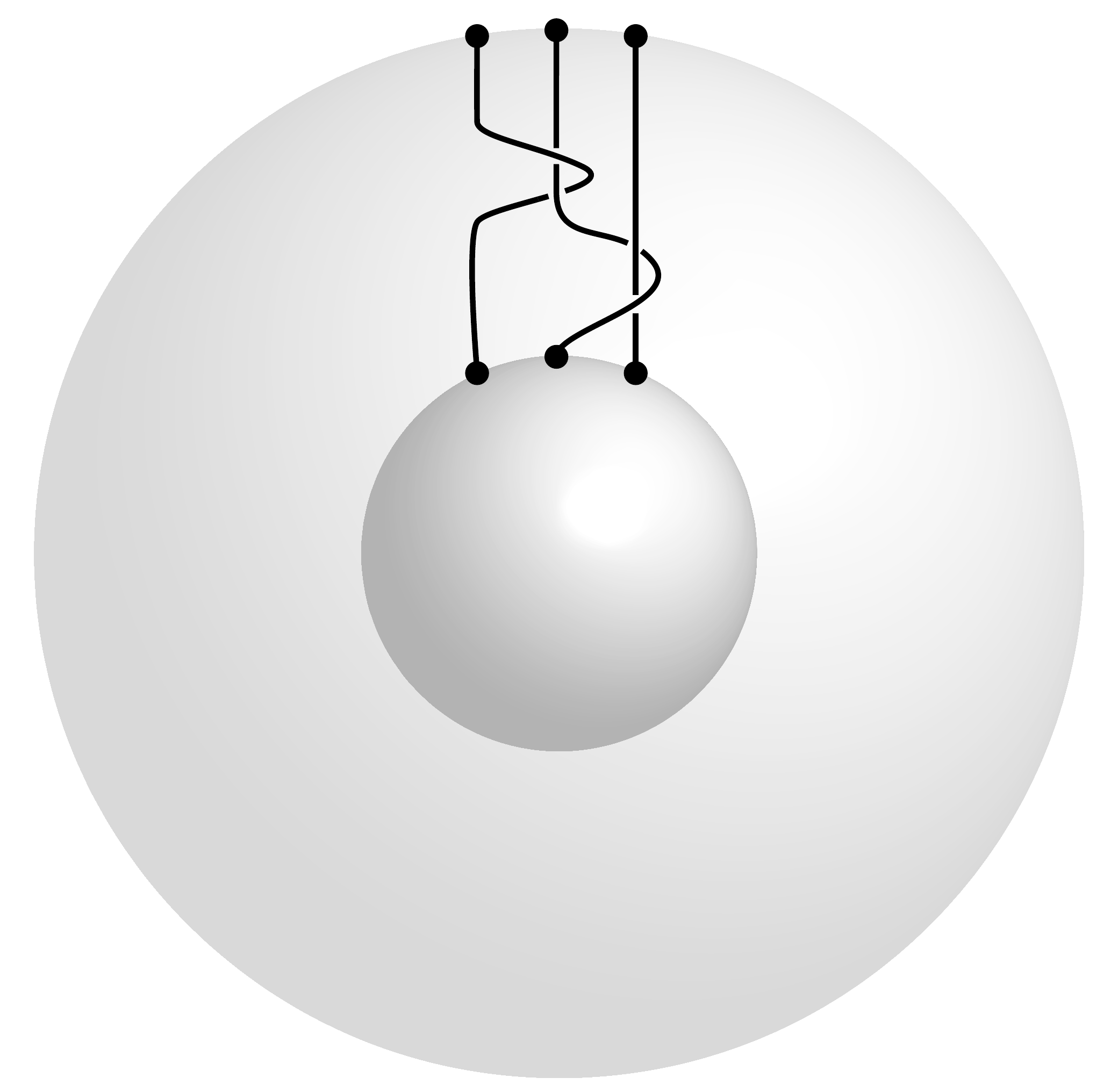}
  \caption{The spherical braid group $\pi_1(S_{0,m})$ consists of the braids defined on the surface of a sphere. The mapping class group $\mathcal M(S_{0,m})$ allows for an extra relation not present in $\pi_1(S_{0,m})$: two elements obtained by a $2\pi$ rotation of the ``inner sphere'' are identified. }
  \label{Spherical}
\end{figure}

The generators of the $\mathcal M(S_{g,0})$ are the Dehn twists $\mathcal T_{a_i},\mathcal T_{b_i},\mathcal T_{c_i}$ which also satisfy the braiding relations \eqref{sbraid1}-\eqref{sbraid2}. Indeed, when two curves do not intersect, the Dehn twists around these curves commute so \eqref{sbraid1} obviously holds. However, when two curves $\gamma_i, \gamma_{i+1} \in \{a_1, c_1, b_1,\ldots,a_g,c_{g-1},b_g\}$ do intersect it is not hard to see that Dehn twists satisfy
\be
\mathcal T_{\gamma_{i+1}}\mathcal T_{\gamma_i}\mathcal T_{\gamma_{i+1}}= \mathcal T_{\gamma_{i}}\mathcal T_{\gamma_{i+1}}\mathcal T_{\gamma_{i}}\,,
\ee 
which is precisely the braiding relation.

Next we are going to explain how the spherical braid group embeds into $\mathcal M(S_{g,0})$. The structure is different for $g=1$, $g=2$ and $g\geq 3$. When $g=1$ the only possible genon is $\Sigma_{2,4}$, which we project to $S_{0,4}$. The mapping class group is a (planar) braid group (divided by its center), {\it i.e.} there is an isomorphism \cite{farb2011primer}
\be
\mathcal M(S_{1,0}) \approx \mathcal B_3/Z(\mathcal B_3) \hookrightarrow \mathcal{M}(S_{0,4}).
\ee
 In fact, we have already constructed this isomorphism explicitly in the previous Subsection. The (planar) braid generators map to Dehn twists $\sigma_1 \mapsto \mathcal T_a$ and $\sigma_2 \mapsto \mathcal T_b= \mathcal S \mathcal T \mathcal S^{-1}$. The spherical braid generators homomorphically map to Dehn twists $\xi_1 , \xi_3\mapsto \mathcal T_a $ and $ \xi_1 \mapsto \mathcal T_b$. For the genus $1$ the spherical braid group is actually richer than the MPG of a closed Riemann surface (see FIG. \ref{gen1}). This is the only case when it is so and this is why a relation to the planar braid group (instead of the spherical one) is present.
 
\begin{figure}
\centerline{
  \includegraphics[width=\linewidth]{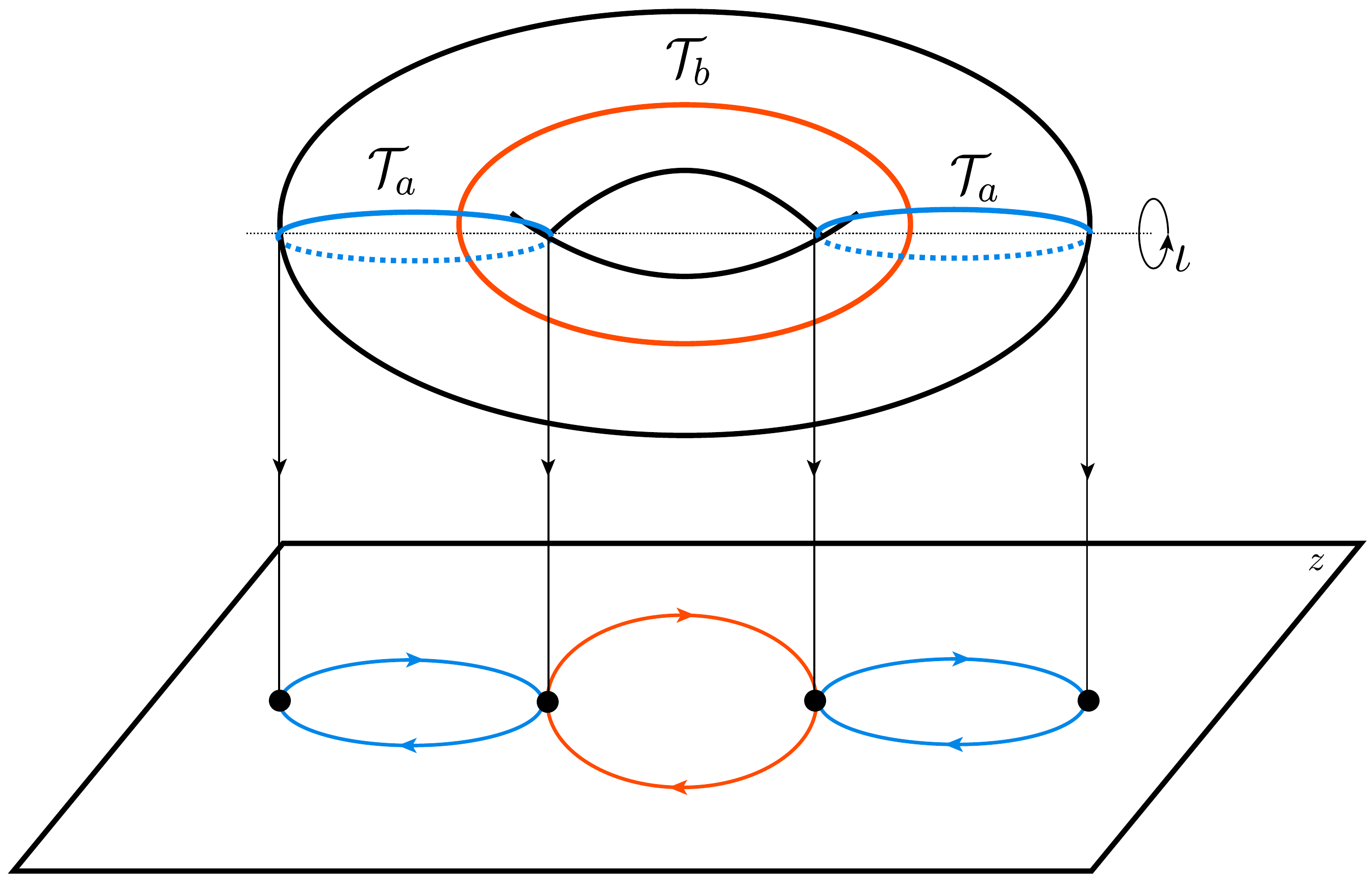}
  }
  \caption{The projection of a torus $S_{1,0}/\iota$. Spherical braid generators $\xi_1$ and $\xi_3$ both map to the Dehn twist $\mathcal T_a$, while $\xi_2$ maps to $\mathcal T_b$. This is the reason for the existence of a homomorphism from $\mathcal M(S_{1,0})$ to the planar braid group $\mathcal B_3$.}
  \label{gen1}
\end{figure}

 When $g=2$ there are two possible genons $\Sigma_{3,4}$ and $\Sigma_{2,6}$. Until specified otherwise we will consider $n=2$ genons. We start by projecting  $\Sigma_{2,6}$ to $S_{0,6}$. Surprisingly there is an isomorphism\cite{birman1973isotopies, birman1974mapping,birman1975braids} $\mathcal M(S_{2,0})/\mathbb Z_2\approx \mathcal M(S_{0,6})$. Under this isomorphism the generators map as (see FIG. \ref{genus2})
\be
 \xi_i \longmapsto\mathcal T_{\gamma_i}\,.
 \ee
This result is known as Birman-Hilden theorem \cite{birman1971mapping,birman1973isotopies}. In the next Subsection we will work out the $\Sigma_{2,6}$ genons in great detail. The reason we restrict to $n=2$ is that Birman-Hilden theorem cleanly works when a surface $S_{g,0}$ has a $\mathbb Z_2$ automorphim (see FIG. \ref{BM}). It did not have to be the case that Birman-Hilden construction corresponds to the computation in the previous Subsection, however it seems to be the general case that it does.

For $g>2$ there are many different genon surfaces, but for now we will consider $\Sigma_{2,2M}$ projected to $S_{0,2M}$. Birman-Hilden theorem then states  $\mathcal {SM}(S_{g,0})\approx \mathcal M(\Sigma_{0,2g+2})$, where $\mathcal {SM}(\Sigma_g)\subset  \mathcal {M}(\Sigma_g)$ is a special subgroup of the mapping class group known as the symmetric mapping class group. It consists of elements of the MPG that commute with the $\mathbb Z_2$ automorphism. It can be explicitly described in terms of generators. The precise map of generators is as follows. We choose a set of curves $\{\{a_i\},\{b_i\}, c_1, c_{g-1}\}$ then
\be\la{BHmap}
\xi_i \longmapsto \mathcal T_{\gamma_i}\,, \qquad \gamma_i \in \{\{b_i\},\{c_i\}, a_1, a_{g}\}\,,
\ee
whereas other generators of MPG do not correspond to braiding. So far we have established that braiding of branch points maps into Dehn twists.

\begin{figure*}
\centerline{
  \includegraphics[width=\linewidth]{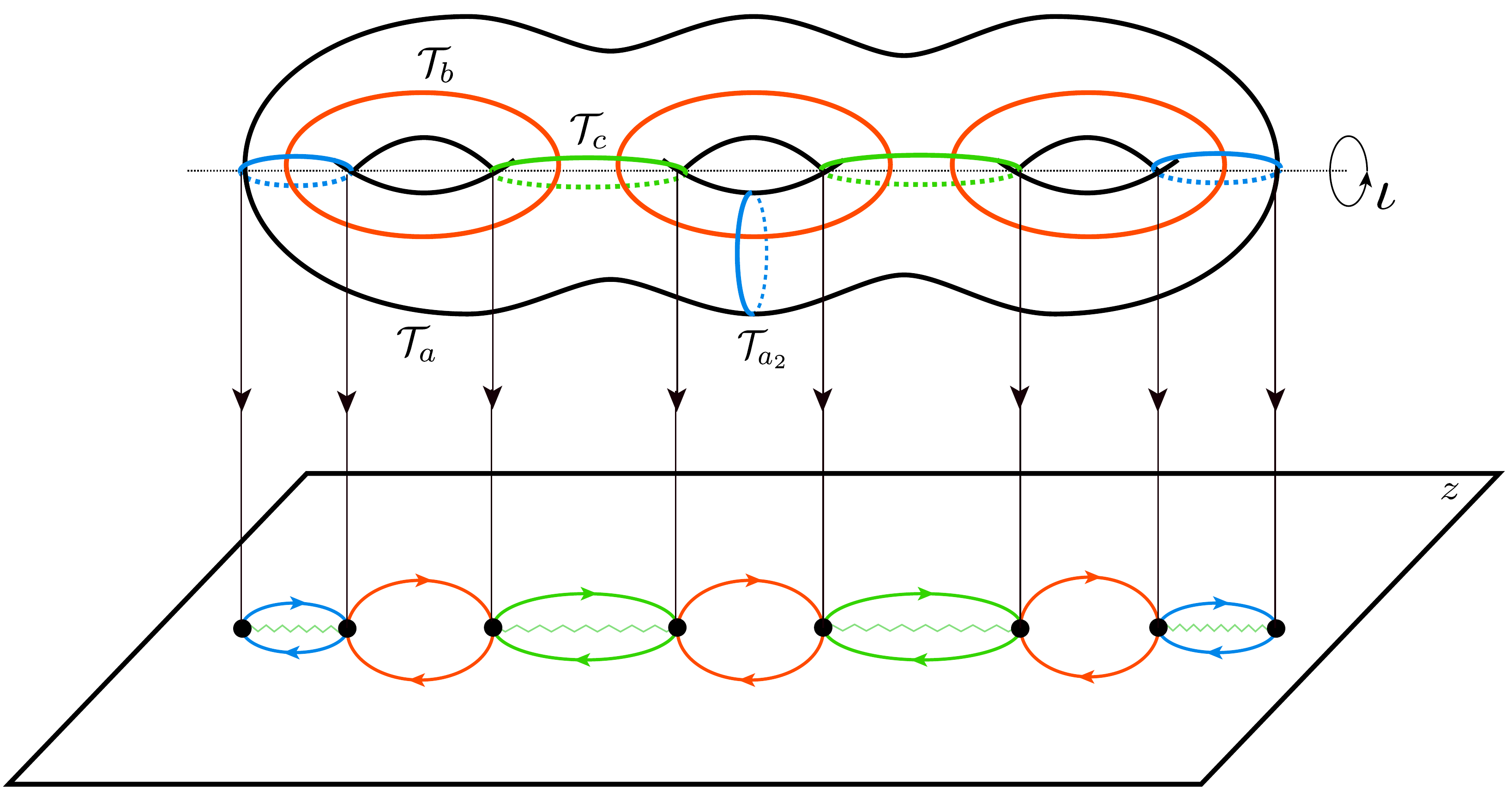}
  }
  \caption{Birman-Hilden representation of a surface with $\mathbb Z_2$ automorphism $\iota$. The automorphism is realized as $\pi$ rotation around a fixed axis. Upon taking a factor by the action of $\iota$ the surface is projected down to $S_{0,2g+2}$ as there are $2g+2$ fixed points of $\iota$. Consequently the elements of the mapping class group $\mathcal M(S_{g,0})$ that commute with the action of $\iota$ are projected to the generators of the spherical braid group of $\mathcal  M(S_{0,2g+2})$. In this example the Dehn twist $\mathcal T_{a_2}$ does not correspond to braiding of genons.}
  \label{BM}
\end{figure*}

That the spherical braid group does not span the entire MPG can be easily seen from FIG. \ref{BM}. Alternatively, we can count the number of generators: there are $2g+1$ generators in the spherical braid group $\mathcal M(\Sigma_{0,2g+2})$ and $3g-1$ generators in the MPG $\mathcal M(\Sigma_{g})$. These numbers are equal to each other only when $g=2$. Introducing extra punctures would, of course, increase the number of spherical braid group generators, but given a genus $g$ surface with a $\mathbb Z_2$ automorphism there is no natural way to introduce more than $2g+2$ special points (see FIG. \ref{BM}).

It is also possible to establish a homomorphism from the planar braid group to $\mathcal B_{2g+1}$ to $\mathcal M(S_{g,1})$. In this case a small circle around the removed point of $S_{g,1}$ becomes the boundary of the disc, when $\mathcal B_{2g+1}$ is realized as $\mathcal M (D_{2g+1})$.

All we need now is the $Sp(2g,\mathbb Z)$ representation of the relevant Dehn twists $\mathcal T_{\gamma_i}$. Fortunately, this problem has been solved long time ago by Birman\cite{birman1971siegel}. She found that 
\bea\la{SP1}
\mathcal T_{a_i}\longmapsto \begin{bmatrix}
    I_g && 0 \\
    A_i && I_g\\
\end{bmatrix},&&\quad
\mathcal T_{b_i}\longmapsto\begin{bmatrix}
    I_g && A_i \\
    0 && I_g\\
\end{bmatrix},\\ \la{SP2}
\mathcal T_{c_i} \longmapsto&& \begin{bmatrix}
    I_g && 0 \\
    B_i && I_g\\
\end{bmatrix}\,,
\eea
where
\be
A_i = E_{ii},\quad B_i = -E_{ii} - E_{i+1i+1}+E_{ii+1}+E_{i+1i}\,,
\ee
where $E_{ij}$ is a matrix with all $0$ entries except $1$ on the intersection of $i$-th row and $j$-th column. 

Thus we have a concrete algorithm for computation of the genon braid matrices. It can be summarized as follows.

(i) Define a state (in $z$-cooridnates) in the presence of genons via
\be\la{Alg1}
\Psi_{gen}(\{z\},\{a\}) = \prod_{i=1}^N\Big( \frac{\p z}{\p\zeta}\Big)^{h}\Big|_{z=z_i}\Big \langle  \prod_{i=1}^N V_q(z_i)\Big\rangle_{z}\,.
\ee

(ii) Use Weyl transformation (accompanied by the change in magnetic field) to remove the geometric singularities
\be \la{Alg2}
\Psi_{gen,p}(\{z\},\{a\})  = e^{ \frac{c}{48\pi} S_L[\sigma]}\prod_{i=1}^N\Big( \frac{\p z}{\p\zeta}\Big)^{h}\Big|_{z=z_i} \Psi_p(\{z\}| \Omega_{ij})
\ee
(iii) The transformation law of $\Psi_{gen,p}(\{z\},\{a\})$ under the action of MPG (the MPG acts on the period matrix $\Omega_{ij}$), combined with the map \eqref{BHmap} generates the braid matrices for $\mathbb Z_2$ genons. The overall phase is determined by the Liouville prefactor.

We note that the relations \eqref{SP1},\,\eqref{SP2} together with \eqref{mpgact} allow, in principle, to calculate the braid matrices of genons on top of any ``parent state'' since the map between the braid generators and the mapping classes is a geometric property of Riemann surfaces with automorphisms and are independent of the physical system placed on the surface. The size of the braid matrices, their explicit form and quantum dimension of genons will, of course, depend on the ``parent'' state. We will calculate the higher genus braid matrices explicitly for the case when the ``parent'' state is the Laughlin state. Before doing that we have to discuss the trial states on higher genus Riemann surfaces.

\subsubsection{FQH states on higher genus Riemann surface}

We will sketch the construction of the Laughlin state (or, any conformal block trial state for that matter). Surprisingly, there is next to nothing said about the FQH states on higher genus surfaces in the literature. Some of the references we could find include Refs.~[\onlinecite{alimohammadi1999coulomb}],[\onlinecite{iengo1994quantum}] where the Laughlin state is guessed on a higher genus surface, but the normalization is not discussed and Ref.~[\onlinecite{bos1989u}], where the wavefunctions of $U(1)_q$ Chern-Simons theory are derived, but again, the normalization is not explicitly calculated. For these reasons we briefly present a CFT construction on a higher genus surface. We refer the interested reader to  [\onlinecite{alvarez1986bosonization, knizhnik1986analytic, eguchi1987chiral,alvarez1987bosonization, dijkgraaf1988c}] and references therein for details about free bosons and fermions on a general Riemann surface and to [\onlinecite{moore1988polynomial}], [\onlinecite{moore1989classical}] for details about rational CFTs on Riemann surfaces.

 We are interested in the correlation function
\be\la{correl}
\Big\langle \prod_{i=1}^N V_q(z_i,\bar z_i)\Big\rangle\,,
\ee
where $V_q$ is the electron operator (in general, including the neutral sector) and we have dropped the background charge. To evaluate this correlator (we focus on the holomorphic sector) we decompose the holomorphic field $\varphi$ as
\be
\varphi(z) = 2\pi \sum_{i=1}^g p_i\int^z \omega_i + \hat \phi (z)\,, 
\ee
where the first term accounts for the zero-modes ($\omega_i$ are the holomorphic differentials) and the second term is orthogonal to the space of zero modes. Then the correlation function reduces to the sum over zero modes $p_i$ and functional integral over $\hat \phi$. We start with the latter. The contribution of $\hat \phi$ works the same way as on a sphere or torus, meaning that we only need to perform the Wick contractions. The propagator is given by
\be
\Big \langle \hat \phi(z) \hat \phi(w) \Big \rangle = \ln E(z,w)\,,
\ee
where $E(z,w)$ is a prime form. Roughly speaking, $E(z,w)$ is a generalization of $\frac{\theta_1(z-w)}{\p\theta_1(0|\tau)}$ and $z-w$ to the higher genus surface in that it is antisymmetric and has first order zero at $z=w$. There is a somewhat explicit form available \cite{fay1973theta}
\be
E(z,w) = \frac{\theta\begin{bmatrix}
    {\bf a_0}  \\
    {\bf b_0} \\
\end{bmatrix}\Big(\int^z_w\boldsymbol \omega\Big|\Omega\Big) }{h(z)h(w)}\,,
\ee
where
\be
 h(z) = \sqrt{\sum_{i=1}^g \omega_i(z) \frac{\p}{\p u_i} \theta\begin{bmatrix}
    {\bf a_0}  \\
    {\bf b_0} \\
\end{bmatrix}\Big({\bf u}\Big|\Omega\Big) }\,,
\ee
where $({\bf  a_0},{\bf b_0})$ is an odd characteristic (in the torus case we had only one odd characteristic $(\1,\1)$, which fixed the choice of $\theta_1$), and $h(z)$ is an analogue of $\p\theta_1(0|\tau)$. On a torus $h(z)$ doesn't actually depend on $z$ since the only holomorphic differential is a constant. The prime form does not depend on the choice of the odd characteristic \cite{fay1973theta} (there is more than one odd $\theta$-function when $g>1$). 

Now we turn to the correlation function \eqref{correl}. Wick contractions will produce the generalization of the Jastrow factor
\be
\prod_{i<j} E(z_i, z_j)^q\,.
\ee
The sum over the zero modes is done over the momentum lattice $p_i \in \mathbb Z^g$ that can be re-written into a {\it finite} sum over the extended conformal blocks. These are given by \cite{dijkgraaf1988c}
\be
\theta\begin{bmatrix}
    \frac{1}{q}{\bf p}  \\
    {\bf 0} \\
\end{bmatrix} \Big(\int^Z \boldsymbol\omega\Big|q\Omega\Big)\prod_{i<j} E(z_i, z_j)^q\,,
\ee
where $Z$ is the center of mass coordinate. The neutralizing background should produce the exponent of $\mathcal K(z_i,\bar z_i)$. 
Thus, putting things together
\be\la{Laughling}
\Psi_{L,\boldsymbol p} = \mathcal N(\Omega) \theta\begin{bmatrix}
    \frac{1}{q}{\bf p}  \\
    {\bf 0} \\
\end{bmatrix} \Big(\int^Z \boldsymbol\omega\Big|q\Omega\Big)\prod_{i<j} E(z_i, z_j)^q e^{- \sum_i \frac{\mathcal K(z_i,\bar z_i)}{4\ell^2}}\,,
\ee
where $N(\Omega)$ is a normalization factor. This factor is not holomorphic in $\Omega$ (it explicitly depends on ${\mbox Im}(\Omega)$) and it ensures that $\Psi_L$ is a modular form of weight $0$ in that it transforms at most by a unitary matrix under any Dehn twist. This factor is calculable from the CFT representation of the wave-function (under the screening assumption), however presently it has never been calculated even for the Laughlin state. Fortunately, we will not need an explicit form of $\mathcal N(\Omega)$, however we point out that is should satisfy
\be
d \bar d \mathcal N(\Omega) = - 2 \eta_H\Omega_{WP}\,,
\ee
where the exterior derivative is taken with respect to the $3g-3$ moduli (encoded in $\Omega$) and $\Omega_{WP}$ is the Weil-Peterson form \cite{klevtsov2015precise}. Notice that $\Psi_L$ is labeled by an integer-valued vector $\boldsymbol p$, thus there are $q^g$ of such states as usual. With \eqref{Laughling} at hand we can find an explicit the representation of the MPG and, consequently, compute the braid matrices of genons.

\subsubsection{Genus $2$}

In this Subsection we will calculate explicitly the braid matrices for $\Sigma_{2,6}$ genons on top of the Laughlin state using all of the ideas discussed in the previous Subsections. We take the Laughlin state to be defined via \eqref{Alg1}. The non-abelian piece of the statistics is fixed by the last factor in \eqref{Alg2} given by \eqref{Laughling}. Thus, to calculate the braid matrices we only need to calculate the action of $Sp(2g,\mathbb Z)$ on \eqref{Laughling} for $g=2$.

We start with the generators of $Sp(2g,\mathbb Z)$. There are $5$ of those
\bea\nonumber
\mathcal T_{a_1}&&= \begin{bmatrix}
    1 && 0&& 0  &&0 \\
     0 && 1 &&0  &&0 \\
     1 && 0 && 1  &&0 \\
       0 && 0&&0  && 1 \\
\end{bmatrix}\,, \quad \mathcal T_{a_2}= \begin{bmatrix}
    1 && 0&&  0 &&0 \\
     0 && 1 &&0  &&0 \\
     0 && 0 && 1  &&0 \\
       0 && 1&&0  && 1 \\
\end{bmatrix} \\ \nonumber
\mathcal T_{b_1}&&= \begin{bmatrix}
    1 && 0&& 1  &&0 \\
     0 && 1 &&0  &&0 \\
     0 && 0 && 1  &&0 \\
       0 && 0&&0  && 1 \\
\end{bmatrix}\,, \quad \mathcal T_{b_2}= \begin{bmatrix}
    1 && 0&&  0 &&0 \\
     0 && 1 &&0  &&1 \\
     0 && 0 && 1  &&0 \\
       0 && 0&&0  && 1 \\
\end{bmatrix}  \\
\mathcal T_{c_1}&&= \begin{bmatrix}
    1 && 0&&  0 &&0 \\
     0 && 1 &&0  &&0 \\
     -1 && 1 && 1  &&0 \\
       1 && -1&&0  && 1 \\
\end{bmatrix}\,.
\eea
Notice that the generator $\mathcal T_{c_1}$ is the real novelty of the higher genus. As a helpful tool we will introduce a matrix suggestively denoted $\mathscr S$
\be
\mathscr S= \begin{bmatrix}
    0 && I_2 \\
     I_2 && 0 \\
\end{bmatrix}\,,
\ee
where $I_2$ is a $2\times 2$ unit matrix. This matrix is a $g=2$ analogue of the modular $\mathcal S$ transformation in that it exchanges the $a$ and $b$ cycles, {\it i.e.}
\be\la{basischangee}
\mathcal T_{b_i} = \mathscr{S} \mathcal T_{a_i} \mathscr S^{-1}\,.
\ee
Now, since the map $\mathcal M(S_{g,0}) \longrightarrow Sp(2g,\mathbb Z)$ is a homomorphism we only need to evaluate the action of $\mathcal T_{a_i}, \mathcal T_{c_i}$ and $\mathscr S$ on the Laughlin state. Then we can obtain $\mathcal T_{b_i}$ by multiplication of matrices just like we did in the case of $g=1$. Indeed, recall that for $g=1$ the action of $\mathcal T_b$ was very complicated, but easily calculable from the action of $\mathcal S$ and $\mathcal T_a$.

\begin{figure*}
  \includegraphics[width=6in]{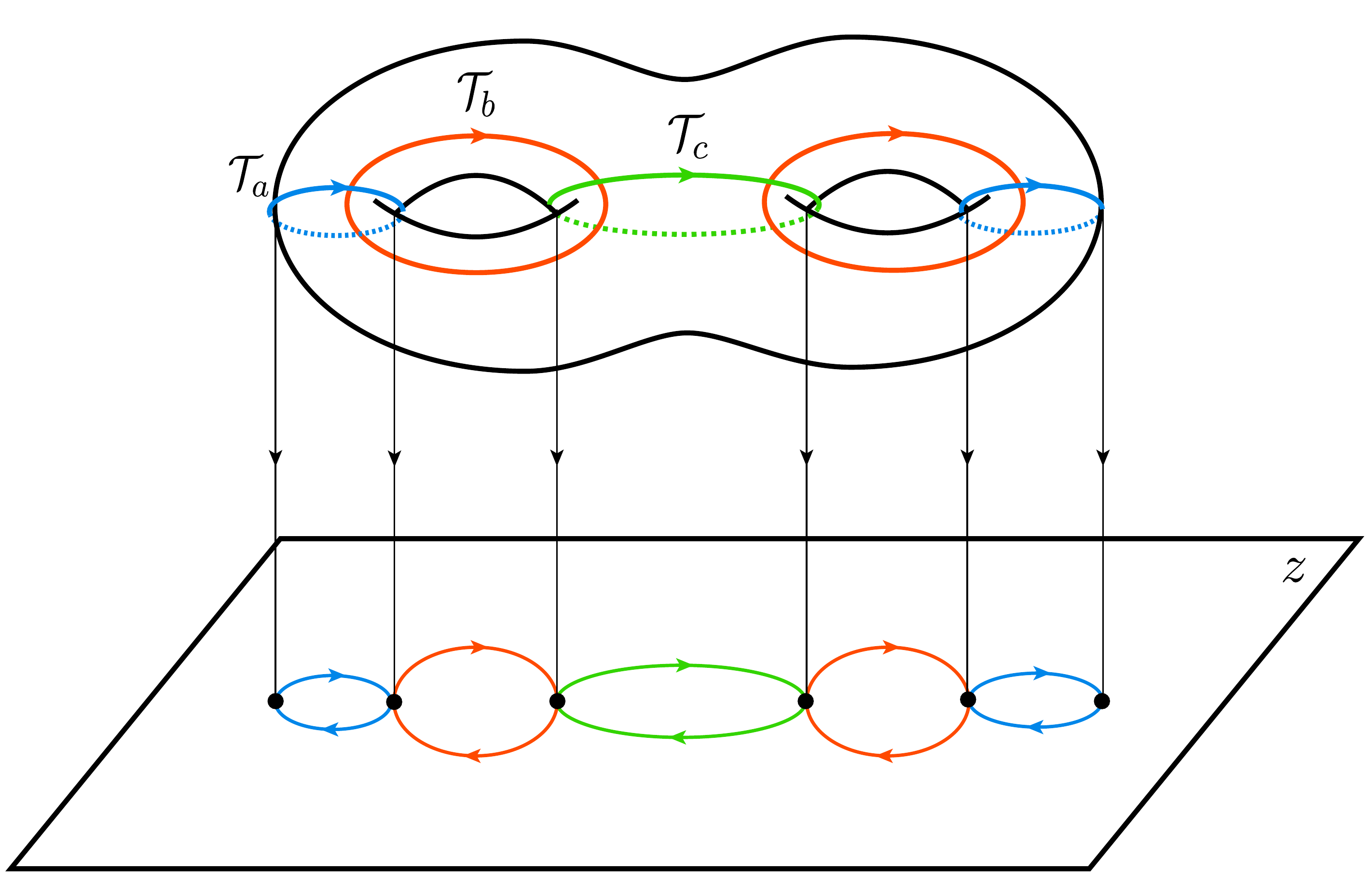}
  \caption{The explicit mapping between the generators of the mapping class group $\mathcal M(S_{2,0})$ and generators of the spherical braid group $\mathcal M(S_{0,6})$ is exhibited. Every Dehn twist corresponds to a braid and vice versa}
  \label{genus2}
\end{figure*}

Next we need the transformation laws of the period matrix $\Omega = \begin{bmatrix}
    \Omega_{11} && \Omega_{12} \\
     \Omega_{21} && \Omega_{22} \\
\end{bmatrix}$. Using \eqref{mpgact} we have
\bea
\mathcal T_{a_i}\Omega = \begin{bmatrix}
    \Omega_{11} && \Omega_{12} \\
     \Omega_{21} && \Omega_{22} \\
\end{bmatrix} + E_{ii}\,, \\
\mathcal T_{c_1}\Omega = \begin{bmatrix}
    \Omega_{11} && \Omega_{12} \\
     \Omega_{21} && \Omega_{22} \\
\end{bmatrix} + B_1\,, \\
\mathscr S \Omega =  \Omega^{-1}\,.
\eea
Notice that the action of $\mathscr S$ is a simple generalization of $\mathcal S$ that sends $\tau \rightarrow - \tau^{-1}$. Also the action of $\mathcal T_{a_i}$ is a simple generalization of $\tau \rightarrow \tau+1$. We will discover the meaning of $\mathcal T_c$ shortly.

We consider the genus $2$ version of the Laughlin wavefunction \eqref{Laughling}. In the calculation we will drop all of the $U(1)$ phases and assume that all of the factors of the type ${\mbox \rm det}\,{\mbox Im}\,\Omega$ are taken care of by the normalization factor $\mathcal N(\Omega)$. The only factor in \eqref{Laughling} that contributes to non-abelian braiding statistics of genons is  
\be
\mathscr X_{\frac{\boldsymbol p}{q}} \equiv \theta\begin{bmatrix}
    \frac{1}{q}{\boldsymbol p}  \\
    {\bf 0} \\
\end{bmatrix} \Big(\int^Z \boldsymbol\omega\Big|q\Omega\Big)
\ee
At the expense of an overall phase in the statistics we can also change the first argument to $\boldsymbol 0$. Thus we are interested in the transformation laws of 
\be
\mathscr X_{\frac{\boldsymbol p}{q}} = \theta\begin{bmatrix}
    \frac{1}{q}{\boldsymbol p}  \\
    {\bf 0} \\
\end{bmatrix} \Big(\boldsymbol 0\Big|q\Omega\Big) = \sum_{\boldsymbol l\in \mathbb Z^g} \exp\pi i \left(\boldsymbol l + \frac{\boldsymbol p}{q}\right)q \Omega \left(\boldsymbol l + \frac{\boldsymbol p}{q}\right) \,.
\ee
There are $q^2$ of such factors and, consequently, the braid matrices are $q^2$ by $q^2$.
It is easy to see that 
\be\la{atwist}
\mathcal T_{a_i} \mathscr X_{\frac{\boldsymbol p}{q}} = e^{2\pi i \frac{p_i^2}{2q}}\mathscr X_{\frac{\boldsymbol p}{q}} \,,
\ee
which is the straight forward generalization of \eqref{Dehntwist} and gives the braid generators $\xi_1$ and $\xi_5$.

The $\mathscr S$ transformation acts as
\be\la{smatrix}
\mathscr S \mathscr X_{\frac{\boldsymbol p}{q}}  = \frac{1}{q} \sum_{\boldsymbol p^\prime} e^{-2\pi i \frac{\boldsymbol p \boldsymbol p^\prime}{2q}} \mathscr X_{\frac{\boldsymbol p^\prime}{q}} = \sum_{\boldsymbol p^\prime}  \mathscr S_{\boldsymbol p \boldsymbol p^\prime}  \mathscr X_{\frac{\boldsymbol p^\prime}{q}}\,.
\ee
Thus the ``$\mathscr S$-matrix'' has two vector indices. The summation goes over values $p^\prime_i = 0,\ldots,q-1$. This can be derived using the multidimensional Poisson resummation formula analogously to the $g=1$ case. Thus, we also know $\mathcal T_{b_i}$ from \eqref{basischangee}.

Finally, we have 
\be\la{ctwist}
\mathcal T_{c_1} \mathscr X_{\frac{\boldsymbol p}{q}}  = e^{2\pi i \frac{(p_1-p_2)^2}{2q}} \mathscr X_{\frac{\boldsymbol p}{q}}\,,
\ee
the phase corresponds to the topological spin of a ``composite'' anyon $(p_1,-p_2)$, which is not too surprising since $c_1= - a_1 +a_2$.

Relations \eqref{atwist} - \eqref{ctwist} together with the correspondence
\bea
\xi_1 \rightarrow \mathcal T_{a_1}\,,\quad \xi_5 \rightarrow \mathcal T_{a_2}\,,&&\quad \xi_2 \rightarrow \mathcal T_{b_1}\,,\quad \xi_4 \rightarrow \mathcal T_{b_2}\,,\\
\xi_3 &&\rightarrow \mathcal T_{c_1}
\eea
 are sufficient to write out the braid matrices for genons. The braid matrices for $\xi_1, \xi_5, \xi_3$ are given by \eqref{atwist} and \eqref{ctwist}, while the braid matrices for $\xi_2, \xi_4$  require multiplying $\mathscr S$ and $\mathcal T_a$, which is not particularly illuminating; the braid matrices have tensor product form as components of $\boldsymbol p$ can be varied independently. 
 
 We note in passing that the Torelli group acts trivially on the Laughlin states (up to an overall phase), however it acts non-trivially if the $\zn$ symmetry is gauged \cite{dijkgraaf1988c}. 
 
 The method described here is applicable to any ``parent'' topological phase as long as the action of the mapping class group of $S_{g,0}$ on the space of ground states is known.

\subsubsection{Generalization to $n>2$}

The case $n=2$ turns out to be simpler since (some) Dehn twists on the surface $S_{g,0}$ map to Dehn half-twists around pairs of genons. With extra difficulties it is possible to move beyond $n=2$. First, we recall how things work for $n=2$ [\onlinecite{birman1971mapping}]. To construct a homomorphism from $\mathcal M(S_{0,2g+2})$ to $\mathcal M(S_{g,0})$ we consider a version of $S_{m,0}$ with $\mathbb Z_2$ automorphism $\iota$ realized as in FIG. \ref{BM}. The symmetric mapping class group $\mathcal {SM}(S_{g,0})$ we mentioned above is a subgroup of $\mathcal M (S_{g,0})$ that commutes with $\iota$. The action of $\iota$ leaves $2g+2$ points fixed. Consider $S_{g,0}/\iota$; it is not hard to see that this factor space is exactly $S_{0,2g+2}$ where each ``puncture'' is a cone with negative curvature $-4\pi$. Now, consider a curve $\gamma$ on $S_{g,0}$ preserved by $\iota$. Dehn twist around $\gamma$ commutes with $\iota$ and thus descends to a Dehn twist on $S_{0,2g+2}$. In fact it descends to a Dehn half-twist around a curve that is an image of $\gamma$ in $S_{0,2g+2}$.
\begin{figure}
  \includegraphics[width=\linewidth]{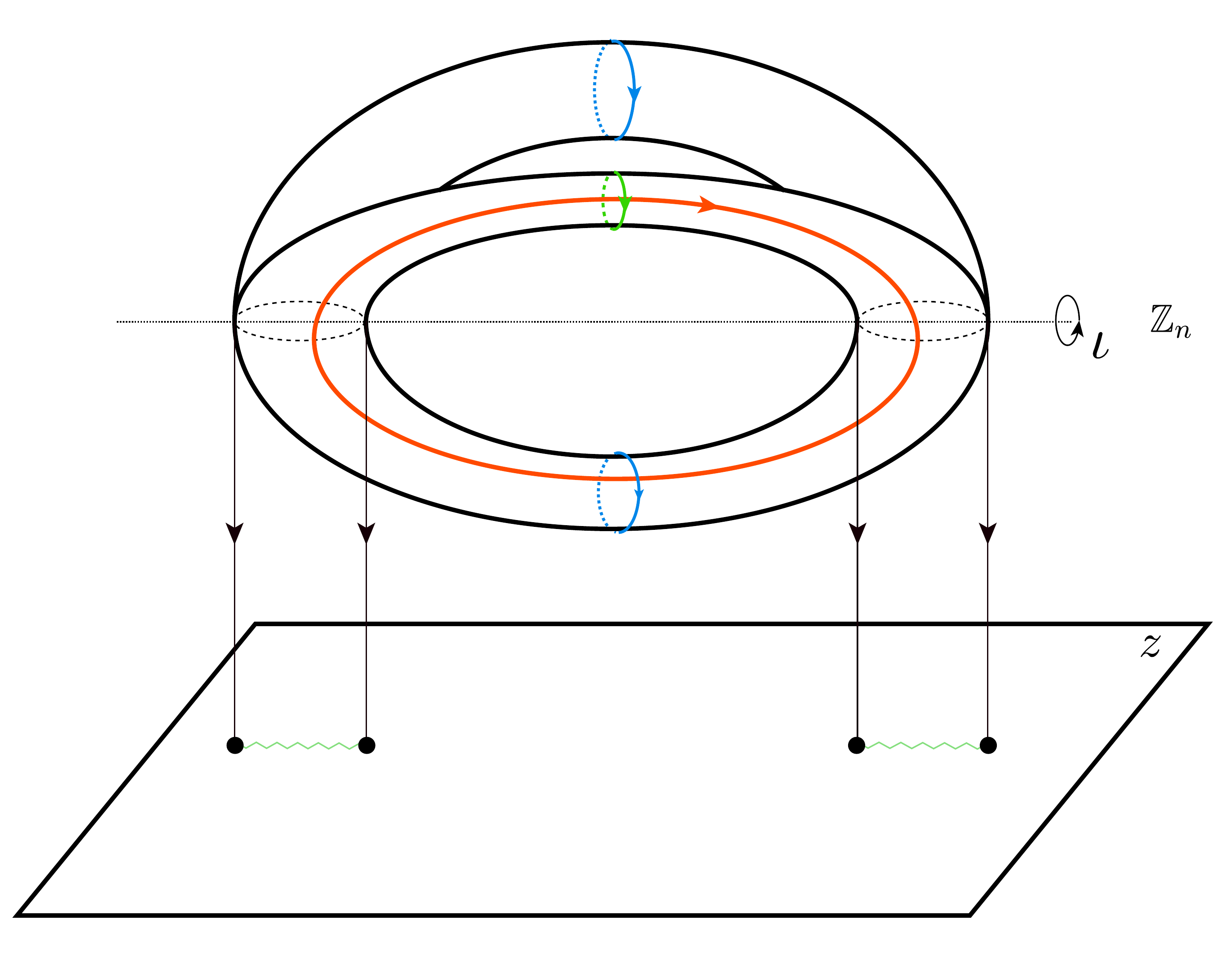}
  \caption{When the automorphism group of a surface is more complicated than $\mathbb Z_2$ the generators of the spherical braid group lift to triple (or, in general, $n$-tuple) products of Dehn twists. The simples case $\Sigma_{3,4}$ with $g=2$ is illustrated.}
  \label{Z3}
\end{figure}

When $n>2$ we have to consider a cyclic $n$-fold covering. See FIG. \ref{Z3} for an example of a smooth surface with a $\mathbb Z_3$ automorphism $\iota$. In this case the same construction goes through, however there is a difficulty. The lift of a braid to the covering surface is not unique. It depends on the explicit representation of the covering. However the following statement is true\cite{birman1973isotopies}. The mapping class group $\mathcal{M}(\Sigma_{n,2M})$ of a $n$-fold cyclic ({\it i.e.} with $\mathbb Z_n$ symmetry) covering of a sphere is isomorphic to the {\it a (non-central) group extension of the cyclic group $\mathbb Z_n$ by the braid group $\mathcal M(S_{0,2M})$}, {\it i.e} $\mathcal M(\Sigma_{n,2M})/ \mathbb Z_n \approx \mathcal M(S_{0,2M})$. More explicitly, let $\xi_i$ be the generators of $\mathcal M(S_{0,2M})$ and let $\mathcal T_{s_i}$ be lifts of $\xi_i$ into $\mathcal M(\Sigma_{n,2M})$. Then\cite{birman1973isotopies} $\mathcal T_{s_i}$ satisfy \eqref{sbraid1},\eqref{sbraid2} and $\eqref{notbraid}$ with $m=2M$ plus the condition
\be
(\mathcal T_{s_1}\cdot\ldots\cdot \mathcal T_{s_{2M-2}} \mathcal T_{s_{2M-1}}^2\mathcal T_{s_{2M-2}}\cdot\ldots\cdot \mathcal T_{s_1})^n=1\,,
\ee
that is $\xi_i$ lifts into a set Dehn twists around mutually intersecting curves. While the relations satisfied by $\mathcal T_{s_i}$ depend only on the the integers $n$ and $2M$, the curves $s_i$ depend on the realization of the covering space, meaning that the curves $s_i$ may transform under the covering transformations.
 This is somewhat similar to what happens in classification of SET phases. In $\mathbb Z_n$ lattice gauge theories one also is forced to consider the (non-central) extensions of the global symmetry group by the gauge group.

The simplest example of a $\mathbb Z_3$ surface is $\Sigma_{3,4}$. There are two braid generators $\xi_1=\xi_3$ and $\xi_2$. The best we can do explicitly is to map the Dehn twist (not half-twist!) $(\xi_1)^2$ to $\mathcal T_{a_1}\cdot\mathcal T_{a_2}\cdot\mathcal T_{c}$ and the Dehn twist $(\xi_2)^2$ maps to $\mathcal T_{b_1}\cdot\mathcal T_{b_2}\mathcal \cdot T_{-b_1+b_2}$. Despite this complication the previous argument guarantees that the half-twists map to Dehn twists.

\section{Discussions and Conclusions}
\subsection{Discussions}
We have provided an algorithm that relates the braid matrices of genons to the elements of the mapping class group of a Riemann surface with automorphism. In the cases when the action of the mapping class group on the space of ground state wave-functions of the ``parent'' topological phase of matter is explicitly known the braid matrices can be readily derived. Such explicit representation is available when there is a representation of the ground states in terms of generalized $\theta$-functions as well as other elliptic functions. This is not always the case. In order to make progress in understanding the genon statistics it is important to understand how the mapping class group acts on the ground state space when such a representation is not available. In fact, the non-abelian statistics of the genons has to be encoded in the RCFT data - braiding and fusion matrices. The action of the mapping class group on the conformal blocks of a RCFT has been worked out in [\onlinecite{moore1989classical}] and, in principle, can be applied to general conformal block trial states. 

In this paper we have focused on the surfaces with the simplest abelian $\mathbb Z_n$ symmetry. It is possible to imagine a generalization to an arbitrary {\it finite} symmetry group $G$. The reason we believe that such generalization is possible is that {\it every finite group can be realized as an automorphism group of a Riemann surface}. Given a Riemann surface $\Sigma$ with an action of $G$ given by an automorphism $\iota_G$ there is always a corresponding factor space $\Sigma/\iota_G$ with singular points. When an anyon, represented by a primary field, is analytically continued around such point it is transformed by the action of the group $G$. The $G$-invariant mapping class group should implement the braiding of the defects, although the details remain to be worked out.

There is another representation for the genons developed in [\onlinecite{barkeshli2013twist}]. Instead of placing a topological phase of matter $\mathfrak C$ on a branched covering one can consider $n$ copies of $\mathfrak C$. Then $\zn$ symmetry is implemented as a symmetry that interchanges the copies, rather than isometry of a Riemann surface. Genons are introduced then as defects of the $\zn$ symmetry. This way of thinking about genons actually suggest a possibility to implement them in a physical setting by allowing quasiparticles to tunnel between different layers of $\mathfrak C$. Detailed discussion of this mechanism for Laughlin states can be found in [\onlinecite{barkeshli2013twist}]. As was emphasized by the authors of  [\onlinecite{barkeshli2013twist}] this opens an exciting opportunity to implement topological phases on higher genus surfaces in a {\it planar} physical sample. The general framework for braiding and fusion of symmetry defects has been developed in [\onlinecite{barkeshli2014symmetry}] and goes under the name of $G$-crossed unitary modular tensor category.

We have not discussed an essential property of quasiholes or genons, namely, fusion. It is easy to visualize the fusion of quasiholes as their charges simply add. In other words, the magnetic fluxes that created quasiholes simply add. Roughly speaking, the fusion rules are abelian 
\be
e^{i\frac{p_1}{\sqrt{q}}\varphi} \times e^{i\frac{p_2}{\sqrt{q}}\varphi} \rightarrow e^{i\frac{p_1 +p_2}{\sqrt{q}}\varphi}\,.
\ee
In the case of genons the fusion rules are somewhat mysterious, since, geometrically speaking, the fusion of genons would correspond to a change in effective topology of the physical space. Moreover, in higher genus cases one can imagine a new type of genons that connect, say, every $m$-th sheet of the surface instead of each sheet. Presumably such genons will have interesting fusion rules. At the same time genons can be understood as twist defect of the $\mathbb Z_n$ symmetry. Fusion rules of the symmetry twist defects have been studied extensively\cite{Bombin2010, barkeshli2014symmetry, teo2015theory, tarantino2015symmetry}and conceptually similar to the fusion of quasiholes, interestingly the fusion rules of $\mathbb Z_n$ twist defects are non-abelian. It would be interesting to understand the fusion of genons from the geometric perspective, which should correspond to topology-changing processes.
\subsection{Conclusions}
We have investigated the properties of the geometric defects created by the curvature fluxes adiabatically threaded through a quantum Hall state. These defects or genons can be assigned an electric charge, a spin and statistics. The charge and spin can be evaluated via the plasma mapping and CFT methods, whereas the non-abelian statistics is determined by a representation of mapping class group of a Riemann surface with an automorphism on a space of groundstates. There is a universal, abelian, part of the statistics that is determined by the gravitational anomaly.  

{\it Acknowledgements:} It is a pleasure to thank D. Aasen, A. Abanov, B. Bradlyn, T. Can, L. Fidkowski, C. Jiang, S. Klevtsov, C. von Keyserlingk, M. Laskin, M. Levin, E. Martinec, M. Roberts, D. T. Son and P. Wiegmann for many stimulating discussions. 

A.G. was supported by the Kadanoff Center for Theoretical Physics Fellowship.

{\it Note added:} While this work was in final stages two preprints appeared on the arXiv where related issues are investigated \cite{laskin2016emergent, barkeshli2016modular}. Ref.~[\onlinecite{laskin2016emergent}] studies Laughlin state on surface with topology of a sphere with and arbitrary number of conical singularities, while Ref.~[\onlinecite{barkeshli2016modular}] studies the action of the mapping class group via a topological charge projection.

\bibliography{Bibliography}

\begin{thebibliography}{107}%
\makeatletter
\providecommand \@ifxundefined [1]{%
 \@ifx{#1\undefined}
}%
\providecommand \@ifnum [1]{%
 \ifnum #1\expandafter \@firstoftwo
 \else \expandafter \@secondoftwo
 \fi
}%
\providecommand \@ifx [1]{%
 \ifx #1\expandafter \@firstoftwo
 \else \expandafter \@secondoftwo
 \fi
}%
\providecommand \natexlab [1]{#1}%
\providecommand \enquote  [1]{``#1''}%
\providecommand \bibnamefont  [1]{#1}%
\providecommand \bibfnamefont [1]{#1}%
\providecommand \citenamefont [1]{#1}%
\providecommand \href@noop [0]{\@secondoftwo}%
\providecommand \href [0]{\begingroup \@sanitize@url \@href}%
\providecommand \@href[1]{\@@startlink{#1}\@@href}%
\providecommand \@@href[1]{\endgroup#1\@@endlink}%
\providecommand \@sanitize@url [0]{\catcode `\\12\catcode `\$12\catcode
  `\&12\catcode `\#12\catcode `\^12\catcode `\_12\catcode `\%12\relax}%
\providecommand \@@startlink[1]{}%
\providecommand \@@endlink[0]{}%
\providecommand \url  [0]{\begingroup\@sanitize@url \@url }%
\providecommand \@url [1]{\endgroup\@href {#1}{\urlprefix }}%
\providecommand \urlprefix  [0]{URL }%
\providecommand \Eprint [0]{\href }%
\providecommand \doibase [0]{http://dx.doi.org/}%
\providecommand \selectlanguage [0]{\@gobble}%
\providecommand \bibinfo  [0]{\@secondoftwo}%
\providecommand \bibfield  [0]{\@secondoftwo}%
\providecommand \translation [1]{[#1]}%
\providecommand \BibitemOpen [0]{}%
\providecommand \bibitemStop [0]{}%
\providecommand \bibitemNoStop [0]{.\EOS\space}%
\providecommand \EOS [0]{\spacefactor3000\relax}%
\providecommand \BibitemShut  [1]{\csname bibitem#1\endcsname}%
\let\auto@bib@innerbib\@empty
\bibitem [{\citenamefont {Moore}\ and\ \citenamefont
  {Read}(1991)}]{moore1991nonabelions}%
  \BibitemOpen
  \bibfield  {author} {\bibinfo {author} {\bibfnamefont {G.}~\bibnamefont
  {Moore}}\ and\ \bibinfo {author} {\bibfnamefont {N.}~\bibnamefont {Read}},\
  }\href@noop {} {\bibfield  {journal} {\bibinfo  {journal} {Nuclear Physics
  B}\ }\textbf {\bibinfo {volume} {360}},\ \bibinfo {pages} {362} (\bibinfo
  {year} {1991})}\BibitemShut {NoStop}%
\bibitem [{\citenamefont {Kitaev}(2006)}]{kitaev2006anyons}%
  \BibitemOpen
  \bibfield  {author} {\bibinfo {author} {\bibfnamefont {A.}~\bibnamefont
  {Kitaev}},\ }\href@noop {} {\bibfield  {journal} {\bibinfo  {journal} {Annals
  of Physics}\ }\textbf {\bibinfo {volume} {321}},\ \bibinfo {pages} {2}
  (\bibinfo {year} {2006})}\BibitemShut {NoStop}%
\bibitem [{\citenamefont {Wen}\ and\ \citenamefont
  {Zee}(1992)}]{WenZeeShiftPaper}%
  \BibitemOpen
  \bibfield  {author} {\bibinfo {author} {\bibfnamefont {X.}~\bibnamefont
  {Wen}}\ and\ \bibinfo {author} {\bibfnamefont {A.}~\bibnamefont {Zee}},\
  }\href@noop {} {\bibfield  {journal} {\bibinfo  {journal} {Phys. Rev. Lett.}\
  }\textbf {\bibinfo {volume} {69}},\ \bibinfo {pages} {953} (\bibinfo {year}
  {1992})}\BibitemShut {NoStop}%
\bibitem [{\citenamefont {Haldane}(2011)}]{2011-Haldane-FQHE}%
  \BibitemOpen
  \bibfield  {author} {\bibinfo {author} {\bibfnamefont {F.~D.~M.}\
  \bibnamefont {Haldane}},\ }\href {http://arxiv.org/abs/1106.3375v2}
  {\bibfield  {journal} {\bibinfo  {journal} {arXiv}\ }\textbf {\bibinfo
  {volume} {cond-mat.mes-hall}} (\bibinfo {year} {2011})},\ \Eprint
  {http://arxiv.org/abs/1106.3375v2} {1106.3375v2} \BibitemShut {NoStop}%
\bibitem [{\citenamefont {Hughes}\ \emph {et~al.}(2011)\citenamefont {Hughes},
  \citenamefont {Leigh},\ and\ \citenamefont {Fradkin}}]{hughes2011torsional}%
  \BibitemOpen
  \bibfield  {author} {\bibinfo {author} {\bibfnamefont {T.~L.}\ \bibnamefont
  {Hughes}}, \bibinfo {author} {\bibfnamefont {R.~G.}\ \bibnamefont {Leigh}}, \
  and\ \bibinfo {author} {\bibfnamefont {E.}~\bibnamefont {Fradkin}},\
  }\href@noop {} {\bibfield  {journal} {\bibinfo  {journal} {Physical Review
  Letters}\ }\textbf {\bibinfo {volume} {107}},\ \bibinfo {pages} {075502}
  (\bibinfo {year} {2011})}\BibitemShut {NoStop}%
\bibitem [{\citenamefont {Hughes}\ \emph {et~al.}(2013)\citenamefont {Hughes},
  \citenamefont {Leigh},\ and\ \citenamefont {Parrikar}}]{hughes2012torsion}%
  \BibitemOpen
  \bibfield  {author} {\bibinfo {author} {\bibfnamefont {T.~L.}\ \bibnamefont
  {Hughes}}, \bibinfo {author} {\bibfnamefont {R.~G.}\ \bibnamefont {Leigh}}, \
  and\ \bibinfo {author} {\bibfnamefont {O.}~\bibnamefont {Parrikar}},\ }\href
  {\doibase 10.1103/PhysRevD.88.025040} {\bibfield  {journal} {\bibinfo
  {journal} {Phys. Rev. D}\ }\textbf {\bibinfo {volume} {88}},\ \bibinfo
  {pages} {025040} (\bibinfo {year} {2013})}\BibitemShut {NoStop}%
\bibitem [{\citenamefont {Abanov}\ and\ \citenamefont
  {Gromov}(2014)}]{Abanov-2014}%
  \BibitemOpen
  \bibfield  {author} {\bibinfo {author} {\bibfnamefont {A.~G.}\ \bibnamefont
  {Abanov}}\ and\ \bibinfo {author} {\bibfnamefont {A.}~\bibnamefont
  {Gromov}},\ }\href {\doibase 10.1103/PhysRevB.90.014435} {\bibfield
  {journal} {\bibinfo  {journal} {Phys. Rev. B}\ }\textbf {\bibinfo {volume}
  {90}},\ \bibinfo {pages} {014435} (\bibinfo {year} {2014})}\BibitemShut
  {NoStop}%
\bibitem [{\citenamefont {Gromov}\ \emph
  {et~al.}(2015{\natexlab{a}})\citenamefont {Gromov}, \citenamefont {Cho},
  \citenamefont {You}, \citenamefont {Abanov},\ and\ \citenamefont
  {Fradkin}}]{GCYFA}%
  \BibitemOpen
  \bibfield  {author} {\bibinfo {author} {\bibfnamefont {A.}~\bibnamefont
  {Gromov}}, \bibinfo {author} {\bibfnamefont {G.~Y.}\ \bibnamefont {Cho}},
  \bibinfo {author} {\bibfnamefont {Y.}~\bibnamefont {You}}, \bibinfo {author}
  {\bibfnamefont {A.~G.}\ \bibnamefont {Abanov}}, \ and\ \bibinfo {author}
  {\bibfnamefont {E.}~\bibnamefont {Fradkin}},\ }\href {\doibase
  10.1103/PhysRevLett.114.016805} {\bibfield  {journal} {\bibinfo  {journal}
  {Phys. Rev. Lett.}\ }\textbf {\bibinfo {volume} {114}},\ \bibinfo {pages}
  {016805} (\bibinfo {year} {2015}{\natexlab{a}})}\BibitemShut {NoStop}%
\bibitem [{\citenamefont {Klevtsov}(2014)}]{klevtsov2014random}%
  \BibitemOpen
  \bibfield  {author} {\bibinfo {author} {\bibfnamefont {S.}~\bibnamefont
  {Klevtsov}},\ }\href@noop {} {\bibfield  {journal} {\bibinfo  {journal}
  {Journal of High Energy Physics}\ }\textbf {\bibinfo {volume} {2014}},\
  \bibinfo {pages} {1} (\bibinfo {year} {2014})}\BibitemShut {NoStop}%
\bibitem [{\citenamefont {Can}\ \emph {et~al.}(2015)\citenamefont {Can},
  \citenamefont {Laskin},\ and\ \citenamefont {Wiegmann}}]{can2014geometry}%
  \BibitemOpen
  \bibfield  {author} {\bibinfo {author} {\bibfnamefont {T.}~\bibnamefont
  {Can}}, \bibinfo {author} {\bibfnamefont {M.}~\bibnamefont {Laskin}}, \ and\
  \bibinfo {author} {\bibfnamefont {P.~B.}\ \bibnamefont {Wiegmann}},\
  }\href@noop {} {\bibfield  {journal} {\bibinfo  {journal} {Annals of
  Physics}\ }\textbf {\bibinfo {volume} {362}},\ \bibinfo {pages} {752}
  (\bibinfo {year} {2015})}\BibitemShut {NoStop}%
\bibitem [{\citenamefont {Can}\ \emph {et~al.}(2014{\natexlab{a}})\citenamefont
  {Can}, \citenamefont {Laskin},\ and\ \citenamefont {Wiegmann}}]{CLW}%
  \BibitemOpen
  \bibfield  {author} {\bibinfo {author} {\bibfnamefont {T.}~\bibnamefont
  {Can}}, \bibinfo {author} {\bibfnamefont {M.}~\bibnamefont {Laskin}}, \ and\
  \bibinfo {author} {\bibfnamefont {P.}~\bibnamefont {Wiegmann}},\ }\href
  {\doibase 10.1103/PhysRevLett.113.046803} {\bibfield  {journal} {\bibinfo
  {journal} {Phys. Rev. Lett.}\ }\textbf {\bibinfo {volume} {113}},\ \bibinfo
  {pages} {046803} (\bibinfo {year} {2014}{\natexlab{a}})}\BibitemShut
  {NoStop}%
\bibitem [{\citenamefont {Laskin}\ \emph {et~al.}(2015)\citenamefont {Laskin},
  \citenamefont {Can},\ and\ \citenamefont {Wiegmann}}]{laskin2015collective}%
  \BibitemOpen
  \bibfield  {author} {\bibinfo {author} {\bibfnamefont {M.}~\bibnamefont
  {Laskin}}, \bibinfo {author} {\bibfnamefont {T.}~\bibnamefont {Can}}, \ and\
  \bibinfo {author} {\bibfnamefont {P.}~\bibnamefont {Wiegmann}},\ }\href@noop
  {} {\bibfield  {journal} {\bibinfo  {journal} {Physical Review B}\ }\textbf
  {\bibinfo {volume} {92}},\ \bibinfo {pages} {235141} (\bibinfo {year}
  {2015})}\BibitemShut {NoStop}%
\bibitem [{\citenamefont {Ferrari}\ and\ \citenamefont
  {Klevtsov}(2014)}]{Klevtsov-fields}%
  \BibitemOpen
  \bibfield  {author} {\bibinfo {author} {\bibfnamefont {F.}~\bibnamefont
  {Ferrari}}\ and\ \bibinfo {author} {\bibfnamefont {S.}~\bibnamefont
  {Klevtsov}},\ }\href {http://dx.doi.org/10.1007/JHEP12%282014%29086}
  {\bibfield  {journal} {\bibinfo  {journal} {Journal of High Energy Physics 
  86}\ }\textbf {\bibinfo {volume} {2014}} (\bibinfo {year}
  {2014})}\BibitemShut {NoStop}%
\bibitem [{\citenamefont {Klevtsov}\ \emph {et~al.}(2015)\citenamefont
  {Klevtsov}, \citenamefont {Ma}, \citenamefont {Marinescu},\ and\
  \citenamefont {Wiegmann}}]{klevtsov2015quantum}%
  \BibitemOpen
  \bibfield  {author} {\bibinfo {author} {\bibfnamefont {S.}~\bibnamefont
  {Klevtsov}}, \bibinfo {author} {\bibfnamefont {X.}~\bibnamefont {Ma}},
  \bibinfo {author} {\bibfnamefont {G.}~\bibnamefont {Marinescu}}, \ and\
  \bibinfo {author} {\bibfnamefont {P.}~\bibnamefont {Wiegmann}},\ }\href@noop
  {} {\bibfield  {journal} {\bibinfo  {journal} {arXiv preprint
  arXiv:1510.06720}\ } (\bibinfo {year} {2015})}\BibitemShut {NoStop}%
\bibitem [{\citenamefont {Gromov}\ \emph
  {et~al.}(2015{\natexlab{b}})\citenamefont {Gromov}, \citenamefont {Jensen},\
  and\ \citenamefont {Abanov}}]{gromov2015boundary}%
  \BibitemOpen
  \bibfield  {author} {\bibinfo {author} {\bibfnamefont {A.}~\bibnamefont
  {Gromov}}, \bibinfo {author} {\bibfnamefont {K.}~\bibnamefont {Jensen}}, \
  and\ \bibinfo {author} {\bibfnamefont {A.~G.}\ \bibnamefont {Abanov}},\
  }\href@noop {} {\bibfield  {journal} {\bibinfo  {journal} {arXiv preprint
  arXiv:1506.07171}\ } (\bibinfo {year} {2015}{\natexlab{b}})}\BibitemShut
  {NoStop}%
\bibitem [{\citenamefont {Cappelli}\ and\ \citenamefont
  {Randellini}(2015)}]{cappelli2015multipole}%
  \BibitemOpen
  \bibfield  {author} {\bibinfo {author} {\bibfnamefont {A.}~\bibnamefont
  {Cappelli}}\ and\ \bibinfo {author} {\bibfnamefont {E.}~\bibnamefont
  {Randellini}},\ }\href@noop {} {\bibfield  {journal} {\bibinfo  {journal}
  {arXiv preprint arXiv:1512.02147}\ } (\bibinfo {year} {2015})}\BibitemShut
  {NoStop}%
\bibitem [{\citenamefont {Avron}\ \emph {et~al.}(1995)\citenamefont {Avron},
  \citenamefont {Seiler},\ and\ \citenamefont
  {Zograf}}]{1995-AvronSeilerZograf}%
  \BibitemOpen
  \bibfield  {author} {\bibinfo {author} {\bibfnamefont {J.~E.}\ \bibnamefont
  {Avron}}, \bibinfo {author} {\bibfnamefont {R.}~\bibnamefont {Seiler}}, \
  and\ \bibinfo {author} {\bibfnamefont {P.~G.}\ \bibnamefont {Zograf}},\
  }\href
  {http://www.google.com/search?client=safari&rls=en-us&q=VISCOSITY+OF+QUANTUM+HALL+FLUIDS&ie=UTF-8&oe=UTF-8}
  {\bibfield  {journal} {\bibinfo  {journal} {Phys Rev Lett}\ }\textbf
  {\bibinfo {volume} {75}},\ \bibinfo {pages} {697} (\bibinfo {year}
  {1995})}\BibitemShut {NoStop}%
\bibitem [{\citenamefont {Landau}\ \emph {et~al.}(1986)\citenamefont {Landau},
  \citenamefont {Lifshitz}, \citenamefont {Kosevich},\ and\ \citenamefont
  {Pitaevski{\u\i}}}]{LandauLifshitz-7}%
  \BibitemOpen
  \bibfield  {author} {\bibinfo {author} {\bibfnamefont {L.}~\bibnamefont
  {Landau}}, \bibinfo {author} {\bibfnamefont {E.}~\bibnamefont {Lifshitz}},
  \bibinfo {author} {\bibfnamefont {A.}~\bibnamefont {Kosevich}}, \ and\
  \bibinfo {author} {\bibfnamefont {L.}~\bibnamefont {Pitaevski{\u\i}}},\
  }\href {http://books.google.com/books?id=tpY-VkwCkAIC} {\emph {\bibinfo
  {title} {Theory of elasticity}}},\ Theoretical Physics\ (\bibinfo
  {publisher} {Butterworth-Heinemann},\ \bibinfo {year} {1986})\BibitemShut
  {NoStop}%
\bibitem [{\citenamefont {Johri}\ \emph {et~al.}(2015)\citenamefont {Johri},
  \citenamefont {Papic}, \citenamefont {Schmitteckert}, \citenamefont {Bhatt},\
  and\ \citenamefont {Haldane}}]{johri2015probing}%
  \BibitemOpen
  \bibfield  {author} {\bibinfo {author} {\bibfnamefont {S.}~\bibnamefont
  {Johri}}, \bibinfo {author} {\bibfnamefont {Z.}~\bibnamefont {Papic}},
  \bibinfo {author} {\bibfnamefont {P.}~\bibnamefont {Schmitteckert}}, \bibinfo
  {author} {\bibfnamefont {R.}~\bibnamefont {Bhatt}}, \ and\ \bibinfo {author}
  {\bibfnamefont {F.}~\bibnamefont {Haldane}},\ }\href@noop {} {\bibfield
  {journal} {\bibinfo  {journal} {arXiv preprint arXiv:1512.08698}\ } (\bibinfo
  {year} {2015})}\BibitemShut {NoStop}%
\bibitem [{\citenamefont {Barkeshli}\ and\ \citenamefont
  {Qi}(2012)}]{barkeshli2012topological}%
  \BibitemOpen
  \bibfield  {author} {\bibinfo {author} {\bibfnamefont {M.}~\bibnamefont
  {Barkeshli}}\ and\ \bibinfo {author} {\bibfnamefont {X.-L.}\ \bibnamefont
  {Qi}},\ }\href@noop {} {\bibfield  {journal} {\bibinfo  {journal} {Physical
  Review X}\ }\textbf {\bibinfo {volume} {2}},\ \bibinfo {pages} {031013}
  (\bibinfo {year} {2012})}\BibitemShut {NoStop}%
\bibitem [{\citenamefont {Cho}\ \emph {et~al.}(2015)\citenamefont {Cho},
  \citenamefont {Parrikar}, \citenamefont {You}, \citenamefont {Leigh},\ and\
  \citenamefont {Hughes}}]{cho2015condensation}%
  \BibitemOpen
  \bibfield  {author} {\bibinfo {author} {\bibfnamefont {G.~Y.}\ \bibnamefont
  {Cho}}, \bibinfo {author} {\bibfnamefont {O.}~\bibnamefont {Parrikar}},
  \bibinfo {author} {\bibfnamefont {Y.}~\bibnamefont {You}}, \bibinfo {author}
  {\bibfnamefont {R.~G.}\ \bibnamefont {Leigh}}, \ and\ \bibinfo {author}
  {\bibfnamefont {T.~L.}\ \bibnamefont {Hughes}},\ }\href@noop {} {\bibfield
  {journal} {\bibinfo  {journal} {Physical Review B}\ }\textbf {\bibinfo
  {volume} {91}},\ \bibinfo {pages} {035122} (\bibinfo {year}
  {2015})}\BibitemShut {NoStop}%
\bibitem [{\citenamefont {Laskin}\ \emph {et~al.}(2016)\citenamefont {Laskin},
  \citenamefont {Chiu}, \citenamefont {Can},\ and\ \citenamefont
  {Wiegmann}}]{laskin2016emergent}%
  \BibitemOpen
  \bibfield  {author} {\bibinfo {author} {\bibfnamefont {M.}~\bibnamefont
  {Laskin}}, \bibinfo {author} {\bibfnamefont {Y.}~\bibnamefont {Chiu}},
  \bibinfo {author} {\bibfnamefont {T.}~\bibnamefont {Can}}, \ and\ \bibinfo
  {author} {\bibfnamefont {P.}~\bibnamefont {Wiegmann}},\ }\href@noop {}
  {\bibfield  {journal} {\bibinfo  {journal} {arXiv preprint arXiv:1602.04802}\
  } (\bibinfo {year} {2016})}\BibitemShut {NoStop}%
\bibitem [{\citenamefont {Read}\ and\ \citenamefont
  {Green}(2000)}]{2000-ReadGreen}%
  \BibitemOpen
  \bibfield  {author} {\bibinfo {author} {\bibfnamefont {N.}~\bibnamefont
  {Read}}\ and\ \bibinfo {author} {\bibfnamefont {D.}~\bibnamefont {Green}},\
  }\href@noop {} {\bibfield  {journal} {\bibinfo  {journal} {Physical Review
  B}\ }\textbf {\bibinfo {volume} {61}},\ \bibinfo {pages} {10267} (\bibinfo
  {year} {2000})}\BibitemShut {NoStop}%
\bibitem [{\citenamefont {Stone}(2012)}]{Stone-Gravitational}%
  \BibitemOpen
  \bibfield  {author} {\bibinfo {author} {\bibfnamefont {M.}~\bibnamefont
  {Stone}},\ }\href@noop {} {\bibfield  {journal} {\bibinfo  {journal}
  {Physical Review B}\ }\textbf {\bibinfo {volume} {85}},\ \bibinfo {pages}
  {184503} (\bibinfo {year} {2012})}\BibitemShut {NoStop}%
\bibitem [{\citenamefont {Gromov}\ and\ \citenamefont
  {Abanov}(2015)}]{gromov-thermal}%
  \BibitemOpen
  \bibfield  {author} {\bibinfo {author} {\bibfnamefont {A.}~\bibnamefont
  {Gromov}}\ and\ \bibinfo {author} {\bibfnamefont {A.~G.}\ \bibnamefont
  {Abanov}},\ }\href {\doibase 10.1103/PhysRevLett.114.016802} {\bibfield
  {journal} {\bibinfo  {journal} {Phys. Rev. Lett.}\ }\textbf {\bibinfo
  {volume} {114}},\ \bibinfo {pages} {016802} (\bibinfo {year}
  {2015})}\BibitemShut {NoStop}%
\bibitem [{\citenamefont {Bradlyn}\ and\ \citenamefont
  {Read}(2015{\natexlab{a}})}]{bradlyn2014low}%
  \BibitemOpen
  \bibfield  {author} {\bibinfo {author} {\bibfnamefont {B.}~\bibnamefont
  {Bradlyn}}\ and\ \bibinfo {author} {\bibfnamefont {N.}~\bibnamefont {Read}},\
  }\href@noop {} {\bibfield  {journal} {\bibinfo  {journal} {Phys. Rev. B}\
  }\textbf {\bibinfo {volume} {91}},\ \bibinfo {pages} {125303} (\bibinfo
  {year} {2015}{\natexlab{a}})}\BibitemShut {NoStop}%
\bibitem [{\citenamefont {Bradlyn}\ and\ \citenamefont
  {Read}(2015{\natexlab{b}})}]{bradlyn2015topological}%
  \BibitemOpen
  \bibfield  {author} {\bibinfo {author} {\bibfnamefont {B.}~\bibnamefont
  {Bradlyn}}\ and\ \bibinfo {author} {\bibfnamefont {N.}~\bibnamefont {Read}},\
  }\href@noop {} {\bibfield  {journal} {\bibinfo  {journal} {Phys. Rev. B}\
  }\textbf {\bibinfo {volume} {91}},\ \bibinfo {pages} {165306} (\bibinfo
  {year} {2015}{\natexlab{b}})}\BibitemShut {NoStop}%
\bibitem [{\citenamefont {Klevtsov}\ and\ \citenamefont
  {Wiegmann}(2015)}]{klevtsov2015precise}%
  \BibitemOpen
  \bibfield  {author} {\bibinfo {author} {\bibfnamefont {S.}~\bibnamefont
  {Klevtsov}}\ and\ \bibinfo {author} {\bibfnamefont {P.}~\bibnamefont
  {Wiegmann}},\ }\href {\doibase 10.1103/PhysRevLett.115.086801} {\bibfield
  {journal} {\bibinfo  {journal} {Phys. Rev. Lett.}\ }\textbf {\bibinfo
  {volume} {115}},\ \bibinfo {pages} {086801} (\bibinfo {year}
  {2015})}\BibitemShut {NoStop}%
\bibitem [{\citenamefont {Bradlyn}\ \emph {et~al.}(2012)\citenamefont
  {Bradlyn}, \citenamefont {Goldstein},\ and\ \citenamefont
  {Read}}]{bradlyn-read-2012kubo}%
  \BibitemOpen
  \bibfield  {author} {\bibinfo {author} {\bibfnamefont {B.}~\bibnamefont
  {Bradlyn}}, \bibinfo {author} {\bibfnamefont {M.}~\bibnamefont {Goldstein}},
  \ and\ \bibinfo {author} {\bibfnamefont {N.}~\bibnamefont {Read}},\
  }\href@noop {} {\bibfield  {journal} {\bibinfo  {journal} {Physical Review
  B}\ }\textbf {\bibinfo {volume} {86}},\ \bibinfo {pages} {245309} (\bibinfo
  {year} {2012})}\BibitemShut {NoStop}%
\bibitem [{\citenamefont {Einarsson}(1991)}]{einarsson1991fractional}%
  \BibitemOpen
  \bibfield  {author} {\bibinfo {author} {\bibfnamefont {T.}~\bibnamefont
  {Einarsson}},\ }\href@noop {} {\bibfield  {journal} {\bibinfo  {journal}
  {Modern Physics Letters B}\ }\textbf {\bibinfo {volume} {5}},\ \bibinfo
  {pages} {675} (\bibinfo {year} {1991})}\BibitemShut {NoStop}%
\bibitem [{\citenamefont {Einarsson}\ \emph {et~al.}(1995)\citenamefont
  {Einarsson}, \citenamefont {Sondhi}, \citenamefont {Girvin},\ and\
  \citenamefont {Arovas}}]{einarsson1995fractional}%
  \BibitemOpen
  \bibfield  {author} {\bibinfo {author} {\bibfnamefont {T.}~\bibnamefont
  {Einarsson}}, \bibinfo {author} {\bibfnamefont {S.}~\bibnamefont {Sondhi}},
  \bibinfo {author} {\bibfnamefont {S.}~\bibnamefont {Girvin}}, \ and\ \bibinfo
  {author} {\bibfnamefont {D.}~\bibnamefont {Arovas}},\ }\href@noop {}
  {\bibfield  {journal} {\bibinfo  {journal} {Nuclear Physics B}\ }\textbf
  {\bibinfo {volume} {441}},\ \bibinfo {pages} {515} (\bibinfo {year}
  {1995})}\BibitemShut {NoStop}%
\bibitem [{\citenamefont {Hoyos}\ and\ \citenamefont
  {Son}(2012)}]{2012-HoyosSon}%
  \BibitemOpen
  \bibfield  {author} {\bibinfo {author} {\bibfnamefont {C.}~\bibnamefont
  {Hoyos}}\ and\ \bibinfo {author} {\bibfnamefont {D.~T.}\ \bibnamefont
  {Son}},\ }\href {\doibase 10.1103/PhysRevLett.108.066805} {\bibfield
  {journal} {\bibinfo  {journal} {Phys. Rev. Lett.}\ }\textbf {\bibinfo
  {volume} {108}},\ \bibinfo {pages} {066805} (\bibinfo {year}
  {2012})}\BibitemShut {NoStop}%
\bibitem [{\citenamefont {Son}(2013)}]{son2013newton}%
  \BibitemOpen
  \bibfield  {author} {\bibinfo {author} {\bibfnamefont {D.~T.}\ \bibnamefont
  {Son}},\ }\href@noop {} {\bibfield  {journal} {\bibinfo  {journal}
  {arXiv:1306.0638}\ } (\bibinfo {year} {2013})}\BibitemShut {NoStop}%
\bibitem [{\citenamefont {Gromov}\ and\ \citenamefont
  {Abanov}(2014)}]{Gromov-galilean}%
  \BibitemOpen
  \bibfield  {author} {\bibinfo {author} {\bibfnamefont {A.}~\bibnamefont
  {Gromov}}\ and\ \bibinfo {author} {\bibfnamefont {A.~G.}\ \bibnamefont
  {Abanov}},\ }\href {\doibase 10.1103/PhysRevLett.113.266802} {\bibfield
  {journal} {\bibinfo  {journal} {Phys. Rev. Lett.}\ }\textbf {\bibinfo
  {volume} {113}},\ \bibinfo {pages} {266802} (\bibinfo {year}
  {2014})}\BibitemShut {NoStop}%
\bibitem [{\citenamefont {Jensen}(2014)}]{jensen2014coupling}%
  \BibitemOpen
  \bibfield  {author} {\bibinfo {author} {\bibfnamefont {K.}~\bibnamefont
  {Jensen}},\ }\href@noop {} {\bibfield  {journal} {\bibinfo  {journal} {arXiv
  preprint arXiv:1408.6855}\ } (\bibinfo {year} {2014})}\BibitemShut {NoStop}%
\bibitem [{\citenamefont {Geracie}\ \emph {et~al.}(2015)\citenamefont
  {Geracie}, \citenamefont {Prabhu},\ and\ \citenamefont
  {Roberts}}]{geracie2015fields}%
  \BibitemOpen
  \bibfield  {author} {\bibinfo {author} {\bibfnamefont {M.}~\bibnamefont
  {Geracie}}, \bibinfo {author} {\bibfnamefont {K.}~\bibnamefont {Prabhu}}, \
  and\ \bibinfo {author} {\bibfnamefont {M.~M.}\ \bibnamefont {Roberts}},\
  }\href@noop {} {\bibfield  {journal} {\bibinfo  {journal} {arXiv preprint
  arXiv:1503.02680}\ } (\bibinfo {year} {2015})}\BibitemShut {NoStop}%
\bibitem [{\citenamefont {Maciejko}\ \emph {et~al.}(2013)\citenamefont
  {Maciejko}, \citenamefont {Hsu}, \citenamefont {Kivelson}, \citenamefont
  {Park},\ and\ \citenamefont {Sondhi}}]{maciejko2013field}%
  \BibitemOpen
  \bibfield  {author} {\bibinfo {author} {\bibfnamefont {J.}~\bibnamefont
  {Maciejko}}, \bibinfo {author} {\bibfnamefont {B.}~\bibnamefont {Hsu}},
  \bibinfo {author} {\bibfnamefont {S.}~\bibnamefont {Kivelson}}, \bibinfo
  {author} {\bibfnamefont {Y.}~\bibnamefont {Park}}, \ and\ \bibinfo {author}
  {\bibfnamefont {S.}~\bibnamefont {Sondhi}},\ }\href@noop {} {\bibfield
  {journal} {\bibinfo  {journal} {Physical Review B}\ }\textbf {\bibinfo
  {volume} {88}},\ \bibinfo {pages} {125137} (\bibinfo {year}
  {2013})}\BibitemShut {NoStop}%
\bibitem [{\citenamefont {You}\ and\ \citenamefont
  {Fradkin}(2013)}]{you2013field}%
  \BibitemOpen
  \bibfield  {author} {\bibinfo {author} {\bibfnamefont {Y.}~\bibnamefont
  {You}}\ and\ \bibinfo {author} {\bibfnamefont {E.}~\bibnamefont {Fradkin}},\
  }\href@noop {} {\bibfield  {journal} {\bibinfo  {journal} {Physical Review
  B}\ }\textbf {\bibinfo {volume} {88}},\ \bibinfo {pages} {235124} (\bibinfo
  {year} {2013})}\BibitemShut {NoStop}%
\bibitem [{\citenamefont {Laughlin}(1983)}]{1983-Laughlin}%
  \BibitemOpen
  \bibfield  {author} {\bibinfo {author} {\bibfnamefont {R.~B.}\ \bibnamefont
  {Laughlin}},\ }\href@noop {} {\bibfield  {journal} {\bibinfo  {journal} {Phys
  Rev Lett}\ }\textbf {\bibinfo {volume} {50}},\ \bibinfo {pages} {1395}
  (\bibinfo {year} {1983})}\BibitemShut {NoStop}%
\bibitem [{\citenamefont {Barkeshli}\ \emph {et~al.}(2013)\citenamefont
  {Barkeshli}, \citenamefont {Jian},\ and\ \citenamefont
  {Qi}}]{barkeshli2013twist}%
  \BibitemOpen
  \bibfield  {author} {\bibinfo {author} {\bibfnamefont {M.}~\bibnamefont
  {Barkeshli}}, \bibinfo {author} {\bibfnamefont {C.-M.}\ \bibnamefont {Jian}},
  \ and\ \bibinfo {author} {\bibfnamefont {X.-L.}\ \bibnamefont {Qi}},\
  }\href@noop {} {\bibfield  {journal} {\bibinfo  {journal} {Physical Review
  B}\ }\textbf {\bibinfo {volume} {87}},\ \bibinfo {pages} {045130} (\bibinfo
  {year} {2013})}\BibitemShut {NoStop}%
\bibitem [{\citenamefont {Witten}(1989)}]{witten1989quantum}%
  \BibitemOpen
  \bibfield  {author} {\bibinfo {author} {\bibfnamefont {E.}~\bibnamefont
  {Witten}},\ }\href@noop {} {\bibfield  {journal} {\bibinfo  {journal}
  {Communications in Mathematical Physics}\ }\textbf {\bibinfo {volume}
  {121}},\ \bibinfo {pages} {351} (\bibinfo {year} {1989})}\BibitemShut
  {NoStop}%
\bibitem [{\citenamefont {Gurarie}\ and\ \citenamefont
  {Nayak}(1997)}]{gurarie1997plasma}%
  \BibitemOpen
  \bibfield  {author} {\bibinfo {author} {\bibfnamefont {V.}~\bibnamefont
  {Gurarie}}\ and\ \bibinfo {author} {\bibfnamefont {C.}~\bibnamefont
  {Nayak}},\ }\href@noop {} {\bibfield  {journal} {\bibinfo  {journal} {Nuclear
  Physics B}\ }\textbf {\bibinfo {volume} {506}},\ \bibinfo {pages} {685}
  (\bibinfo {year} {1997})}\BibitemShut {NoStop}%
\bibitem [{\citenamefont {Bonderson}\ \emph {et~al.}(2011)\citenamefont
  {Bonderson}, \citenamefont {Gurarie},\ and\ \citenamefont
  {Nayak}}]{bonderson2011plasma}%
  \BibitemOpen
  \bibfield  {author} {\bibinfo {author} {\bibfnamefont {P.}~\bibnamefont
  {Bonderson}}, \bibinfo {author} {\bibfnamefont {V.}~\bibnamefont {Gurarie}},
  \ and\ \bibinfo {author} {\bibfnamefont {C.}~\bibnamefont {Nayak}},\
  }\href@noop {} {\bibfield  {journal} {\bibinfo  {journal} {Physical Review
  B}\ }\textbf {\bibinfo {volume} {83}},\ \bibinfo {pages} {075303} (\bibinfo
  {year} {2011})}\BibitemShut {NoStop}%
\bibitem [{\citenamefont {Read}(2009{\natexlab{a}})}]{2009-Read-HallViscosity}%
  \BibitemOpen
  \bibfield  {author} {\bibinfo {author} {\bibfnamefont {N.}~\bibnamefont
  {Read}},\ }\href {\doibase 10.1103/PhysRevB.79.045308} {\bibfield  {journal}
  {\bibinfo  {journal} {Phys Rev B}\ }\textbf {\bibinfo {volume} {79}},\
  \bibinfo {pages} {045308} (\bibinfo {year} {2009}{\natexlab{a}})}\BibitemShut
  {NoStop}%
\bibitem [{Note1()}]{Note1}%
  \BibitemOpen
  \bibinfo {note} {In fact, there is extra freedom in introducing the coupling
  to the curved space. One can always multiply the wavefunction by a factor
  $\DOTSB \prod@ \slimits@ _{i} \protect \sqrt {g(z_i)}^j$. In this paper we
  take $j=0$, but in general $j$ will change the geometric spin.}\BibitemShut
  {Stop}%
\bibitem [{\citenamefont {Douglas}\ and\ \citenamefont
  {Klevtsov}(2010)}]{douglas2010bergman}%
  \BibitemOpen
  \bibfield  {author} {\bibinfo {author} {\bibfnamefont {M.~R.}\ \bibnamefont
  {Douglas}}\ and\ \bibinfo {author} {\bibfnamefont {S.}~\bibnamefont
  {Klevtsov}},\ }\href@noop {} {\bibfield  {journal} {\bibinfo  {journal}
  {Communications in Mathematical Physics}\ }\textbf {\bibinfo {volume}
  {293}},\ \bibinfo {pages} {205} (\bibinfo {year} {2010})}\BibitemShut
  {NoStop}%
\bibitem [{\citenamefont {Hoyos}\ and\ \citenamefont
  {Son}(2011)}]{2011-HoyosSon}%
  \BibitemOpen
  \bibfield  {author} {\bibinfo {author} {\bibfnamefont {C.}~\bibnamefont
  {Hoyos}}\ and\ \bibinfo {author} {\bibfnamefont {D.~T.}\ \bibnamefont
  {Son}},\ }\href {http://arxiv.org/abs/1109.2651v1} {\bibfield  {journal}
  {\bibinfo  {journal} {arXiv: 1109.2651}\ } (\bibinfo {year} {2011})},\
  \Eprint {http://arxiv.org/abs/1109.2651v1} {1109.2651v1} \BibitemShut
  {NoStop}%
\bibitem [{\citenamefont {Haldane}(1983)}]{1983-Haldane-hierarchy}%
  \BibitemOpen
  \bibfield  {author} {\bibinfo {author} {\bibfnamefont {F.~D.~M.}\
  \bibnamefont {Haldane}},\ }\href@noop {} {\bibfield  {journal} {\bibinfo
  {journal} {Phys Rev Lett}\ }\textbf {\bibinfo {volume} {51}},\ \bibinfo
  {pages} {605} (\bibinfo {year} {1983})}\BibitemShut {NoStop}%
\bibitem [{\citenamefont {Fr\"ohlich}\ and\ \citenamefont
  {Studer}(1993)}]{1993-frohlich}%
  \BibitemOpen
  \bibfield  {author} {\bibinfo {author} {\bibfnamefont {J.}~\bibnamefont
  {Fr\"ohlich}}\ and\ \bibinfo {author} {\bibfnamefont {U.~M.}\ \bibnamefont
  {Studer}},\ }\href {\doibase 10.1103/RevModPhys.65.733} {\bibfield  {journal}
  {\bibinfo  {journal} {Rev. Mod. Phys.}\ }\textbf {\bibinfo {volume} {65}},\
  \bibinfo {pages} {733} (\bibinfo {year} {1993})}\BibitemShut {NoStop}%
\bibitem [{\citenamefont {Schine}\ \emph {et~al.}(2015)\citenamefont {Schine},
  \citenamefont {Ryou}, \citenamefont {Gromov}, \citenamefont {Sommer},\ and\
  \citenamefont {Simon}}]{schine2015synthetic}%
  \BibitemOpen
  \bibfield  {author} {\bibinfo {author} {\bibfnamefont {N.}~\bibnamefont
  {Schine}}, \bibinfo {author} {\bibfnamefont {A.}~\bibnamefont {Ryou}},
  \bibinfo {author} {\bibfnamefont {A.}~\bibnamefont {Gromov}}, \bibinfo
  {author} {\bibfnamefont {A.}~\bibnamefont {Sommer}}, \ and\ \bibinfo {author}
  {\bibfnamefont {J.}~\bibnamefont {Simon}},\ }\href@noop {} {\bibfield
  {journal} {\bibinfo  {journal} {arXiv preprint arXiv:1511.07381}\ } (\bibinfo
  {year} {2015})}\BibitemShut {NoStop}%
\bibitem [{\citenamefont {Arovas}\ \emph {et~al.}(1984)\citenamefont {Arovas},
  \citenamefont {Schrieffer},\ and\ \citenamefont
  {Wilczek}}]{arovas1984fractional}%
  \BibitemOpen
  \bibfield  {author} {\bibinfo {author} {\bibfnamefont {D.}~\bibnamefont
  {Arovas}}, \bibinfo {author} {\bibfnamefont {J.~R.}\ \bibnamefont
  {Schrieffer}}, \ and\ \bibinfo {author} {\bibfnamefont {F.}~\bibnamefont
  {Wilczek}},\ }\href@noop {} {\bibfield  {journal} {\bibinfo  {journal}
  {Physical review letters}\ }\textbf {\bibinfo {volume} {53}},\ \bibinfo
  {pages} {722} (\bibinfo {year} {1984})}\BibitemShut {NoStop}%
\bibitem [{\citenamefont {Read}(2008)}]{read2008quasiparticle}%
  \BibitemOpen
  \bibfield  {author} {\bibinfo {author} {\bibfnamefont {N.}~\bibnamefont
  {Read}},\ }\href@noop {} {\bibfield  {journal} {\bibinfo  {journal} {arXiv
  preprint arXiv:0807.3107}\ } (\bibinfo {year} {2008})}\BibitemShut {NoStop}%
\bibitem [{\citenamefont {Tokatly}\ and\ \citenamefont
  {Vignale}(2007)}]{2007-TokatlyVignale}%
  \BibitemOpen
  \bibfield  {author} {\bibinfo {author} {\bibfnamefont {I.~V.}\ \bibnamefont
  {Tokatly}}\ and\ \bibinfo {author} {\bibfnamefont {G.}~\bibnamefont
  {Vignale}},\ }\href {\doibase 10.1103/PhysRevB.76.161305} {\bibfield
  {journal} {\bibinfo  {journal} {Phys Rev B}\ }\textbf {\bibinfo {volume}
  {76}},\ \bibinfo {pages} {161305} (\bibinfo {year} {2007})}\BibitemShut
  {NoStop}%
\bibitem [{\citenamefont {Read}(2009{\natexlab{b}})}]{2009-Read}%
  \BibitemOpen
  \bibfield  {author} {\bibinfo {author} {\bibfnamefont {N.}~\bibnamefont
  {Read}},\ }\href {\doibase 10.1103/PhysRevB.79.245304} {\bibfield  {journal}
  {\bibinfo  {journal} {Phys Rev B}\ }\textbf {\bibinfo {volume} {79}},\
  \bibinfo {pages} {245304} (\bibinfo {year} {2009}{\natexlab{b}})}\BibitemShut
  {NoStop}%
\bibitem [{\citenamefont {Read}\ and\ \citenamefont {Rezayi}(1999)}]{Read1999}%
  \BibitemOpen
  \bibfield  {author} {\bibinfo {author} {\bibfnamefont {N.}~\bibnamefont
  {Read}}\ and\ \bibinfo {author} {\bibfnamefont {E.}~\bibnamefont {Rezayi}},\
  }\href@noop {} {\bibfield  {journal} {\bibinfo  {journal} {Phys. Rev. B}\
  }\textbf {\bibinfo {volume} {59}},\ \bibinfo {pages} {8084} (\bibinfo {year}
  {1999})}\BibitemShut {NoStop}%
\bibitem [{\citenamefont {Read}\ and\ \citenamefont
  {Rezayi}(2011)}]{2011-ReadRezayi}%
  \BibitemOpen
  \bibfield  {author} {\bibinfo {author} {\bibfnamefont {N.}~\bibnamefont
  {Read}}\ and\ \bibinfo {author} {\bibfnamefont {E.~H.}\ \bibnamefont
  {Rezayi}},\ }\href {\doibase 10.1103/PhysRevB.84.085316} {\bibfield
  {journal} {\bibinfo  {journal} {Phys Rev B}\ }\textbf {\bibinfo {volume}
  {84}},\ \bibinfo {pages} {085316} (\bibinfo {year} {2011})}\BibitemShut
  {NoStop}%
\bibitem [{\citenamefont {Sondhi}\ and\ \citenamefont
  {Kivelson}(1992)}]{sondhi1992long}%
  \BibitemOpen
  \bibfield  {author} {\bibinfo {author} {\bibfnamefont {S.}~\bibnamefont
  {Sondhi}}\ and\ \bibinfo {author} {\bibfnamefont {S.}~\bibnamefont
  {Kivelson}},\ }\href@noop {} {\bibfield  {journal} {\bibinfo  {journal}
  {Physical Review B}\ }\textbf {\bibinfo {volume} {46}},\ \bibinfo {pages}
  {13319} (\bibinfo {year} {1992})}\BibitemShut {NoStop}%
\bibitem [{\citenamefont {Can}\ \emph {et~al.}(2014{\natexlab{b}})\citenamefont
  {Can}, \citenamefont {Laskin},\ and\ \citenamefont
  {Wiegmann}}]{can2014field}%
  \BibitemOpen
  \bibfield  {author} {\bibinfo {author} {\bibfnamefont {T.}~\bibnamefont
  {Can}}, \bibinfo {author} {\bibfnamefont {M.}~\bibnamefont {Laskin}}, \ and\
  \bibinfo {author} {\bibfnamefont {P.}~\bibnamefont {Wiegmann}},\ }\href@noop
  {} {\bibfield  {journal} {\bibinfo  {journal} {arXiv:1412.8716}\ } (\bibinfo
  {year} {2014}{\natexlab{b}})}\BibitemShut {NoStop}%
\bibitem [{\citenamefont {Kvorning}(2013)}]{kvorning2013quantum}%
  \BibitemOpen
  \bibfield  {author} {\bibinfo {author} {\bibfnamefont {T.}~\bibnamefont
  {Kvorning}},\ }\href@noop {} {\bibfield  {journal} {\bibinfo  {journal}
  {Physical Review B}\ }\textbf {\bibinfo {volume} {87}},\ \bibinfo {pages}
  {195131} (\bibinfo {year} {2013})}\BibitemShut {NoStop}%
\bibitem [{Note2()}]{Note2}%
  \BibitemOpen
  \bibinfo {note} {Notice that it happens to be Sugawara stress tensor of
  simple currents $J$.}\BibitemShut {Stop}%
\bibitem [{\citenamefont {Francesco}\ \emph {et~al.}(1997)\citenamefont
  {Francesco}, \citenamefont {Mathieu},\ and\ \citenamefont
  {S{\'e}n{\'e}chal}}]{CFT-book}%
  \BibitemOpen
  \bibfield  {author} {\bibinfo {author} {\bibfnamefont {P.~D.}\ \bibnamefont
  {Francesco}}, \bibinfo {author} {\bibfnamefont {P.}~\bibnamefont {Mathieu}},
  \ and\ \bibinfo {author} {\bibfnamefont {D.}~\bibnamefont
  {S{\'e}n{\'e}chal}},\ }\href {http://books.google.com/books?id=keUrdME5rhIC}
  {\emph {\bibinfo {title} {Conformal field theory}}},\ Graduate texts in
  contemporary physics\ (\bibinfo  {publisher} {Springer},\ \bibinfo {year}
  {1997})\BibitemShut {NoStop}%
\bibitem [{\citenamefont {Thouless}\ \emph {et~al.}(1982)\citenamefont
  {Thouless}, \citenamefont {Kohmoto}, \citenamefont {Nightingale},\ and\
  \citenamefont {den Nijs}}]{TKNN}%
  \BibitemOpen
  \bibfield  {author} {\bibinfo {author} {\bibfnamefont {D.~J.}\ \bibnamefont
  {Thouless}}, \bibinfo {author} {\bibfnamefont {M.}~\bibnamefont {Kohmoto}},
  \bibinfo {author} {\bibfnamefont {M.~P.}\ \bibnamefont {Nightingale}}, \ and\
  \bibinfo {author} {\bibfnamefont {M.}~\bibnamefont {den Nijs}},\ }\href
  {\doibase 10.1103/PhysRevLett.49.405} {\bibfield  {journal} {\bibinfo
  {journal} {Phys. Rev. Lett.}\ }\textbf {\bibinfo {volume} {49}},\ \bibinfo
  {pages} {405} (\bibinfo {year} {1982})}\BibitemShut {NoStop}%
\bibitem [{\citenamefont {L{\'e}vay}(1995)}]{levay1995berry}%
  \BibitemOpen
  \bibfield  {author} {\bibinfo {author} {\bibfnamefont {P.}~\bibnamefont
  {L{\'e}vay}},\ }\href@noop {} {\bibfield  {journal} {\bibinfo  {journal}
  {Journal of Mathematical Physics}\ }\textbf {\bibinfo {volume} {36}},\
  \bibinfo {pages} {2792} (\bibinfo {year} {1995})}\BibitemShut {NoStop}%
\bibitem [{\citenamefont {Verlinde}(1988)}]{verlinde1988fusion}%
  \BibitemOpen
  \bibfield  {author} {\bibinfo {author} {\bibfnamefont {E.}~\bibnamefont
  {Verlinde}},\ }\href@noop {} {\bibfield  {journal} {\bibinfo  {journal}
  {Nuclear Physics B}\ }\textbf {\bibinfo {volume} {300}},\ \bibinfo {pages}
  {360} (\bibinfo {year} {1988})}\BibitemShut {NoStop}%
\bibitem [{\citenamefont {Wen}\ and\ \citenamefont
  {Niu}(1990)}]{wen1990ground}%
  \BibitemOpen
  \bibfield  {author} {\bibinfo {author} {\bibfnamefont {X.-G.}\ \bibnamefont
  {Wen}}\ and\ \bibinfo {author} {\bibfnamefont {Q.}~\bibnamefont {Niu}},\
  }\href@noop {} {\bibfield  {journal} {\bibinfo  {journal} {Physical Review
  B}\ }\textbf {\bibinfo {volume} {41}},\ \bibinfo {pages} {9377} (\bibinfo
  {year} {1990})}\BibitemShut {NoStop}%
\bibitem [{\citenamefont {Haldane}\ and\ \citenamefont
  {Rezayi}(1985)}]{haldane1985periodic}%
  \BibitemOpen
  \bibfield  {author} {\bibinfo {author} {\bibfnamefont {F.~D.~M.}\
  \bibnamefont {Haldane}}\ and\ \bibinfo {author} {\bibfnamefont {E.~H.}\
  \bibnamefont {Rezayi}},\ }\href@noop {} {\bibfield  {journal} {\bibinfo
  {journal} {Physical Review B}\ }\textbf {\bibinfo {volume} {31}},\ \bibinfo
  {pages} {2529} (\bibinfo {year} {1985})}\BibitemShut {NoStop}%
\bibitem [{\citenamefont {Fay}(1973)}]{fay1973theta}%
  \BibitemOpen
  \bibfield  {author} {\bibinfo {author} {\bibfnamefont {J.~D.}\ \bibnamefont
  {Fay}},\ }\href@noop {} {\  (\bibinfo {year} {1973})}\BibitemShut {NoStop}%
\bibitem [{\citenamefont {Cappelli}\ and\ \citenamefont
  {Zemba}(1997)}]{cappelli1997modular}%
  \BibitemOpen
  \bibfield  {author} {\bibinfo {author} {\bibfnamefont {A.}~\bibnamefont
  {Cappelli}}\ and\ \bibinfo {author} {\bibfnamefont {G.~R.}\ \bibnamefont
  {Zemba}},\ }\href@noop {} {\bibfield  {journal} {\bibinfo  {journal} {Nuclear
  Physics B}\ }\textbf {\bibinfo {volume} {490}},\ \bibinfo {pages} {595}
  (\bibinfo {year} {1997})}\BibitemShut {NoStop}%
\bibitem [{\citenamefont {Polyakov}(1988)}]{polyakov1988fermi}%
  \BibitemOpen
  \bibfield  {author} {\bibinfo {author} {\bibfnamefont {A.~M.}\ \bibnamefont
  {Polyakov}},\ }\href@noop {} {\bibfield  {journal} {\bibinfo  {journal}
  {Modern Physics Letters A}\ }\textbf {\bibinfo {volume} {3}},\ \bibinfo
  {pages} {325} (\bibinfo {year} {1988})}\BibitemShut {NoStop}%
\bibitem [{\citenamefont {Tze}(1988)}]{tze1988manifold}%
  \BibitemOpen
  \bibfield  {author} {\bibinfo {author} {\bibfnamefont {C.-H.}\ \bibnamefont
  {Tze}},\ }\href@noop {} {\bibfield  {journal} {\bibinfo  {journal}
  {International Journal of Modern Physics A}\ }\textbf {\bibinfo {volume}
  {3}},\ \bibinfo {pages} {1959} (\bibinfo {year} {1988})}\BibitemShut
  {NoStop}%
\bibitem [{\citenamefont {Lee}\ and\ \citenamefont
  {Wen}(1994)}]{lee1994orbital}%
  \BibitemOpen
  \bibfield  {author} {\bibinfo {author} {\bibfnamefont {D.-H.}\ \bibnamefont
  {Lee}}\ and\ \bibinfo {author} {\bibfnamefont {X.-G.}\ \bibnamefont {Wen}},\
  }\href@noop {} {\bibfield  {journal} {\bibinfo  {journal} {Physical Review
  B}\ }\textbf {\bibinfo {volume} {49}},\ \bibinfo {pages} {11066} (\bibinfo
  {year} {1994})}\BibitemShut {NoStop}%
\bibitem [{\citenamefont {Cho}\ \emph {et~al.}(2014)\citenamefont {Cho},
  \citenamefont {You},\ and\ \citenamefont {Fradkin}}]{2014-ChoYouFradkin}%
  \BibitemOpen
  \bibfield  {author} {\bibinfo {author} {\bibfnamefont {G.~Y.}\ \bibnamefont
  {Cho}}, \bibinfo {author} {\bibfnamefont {Y.}~\bibnamefont {You}}, \ and\
  \bibinfo {author} {\bibfnamefont {E.}~\bibnamefont {Fradkin}},\ }\href
  {\doibase 10.1103/PhysRevB.90.115139} {\bibfield  {journal} {\bibinfo
  {journal} {Phys. Rev. B}\ }\textbf {\bibinfo {volume} {90}},\ \bibinfo
  {pages} {115139} (\bibinfo {year} {2014})}\BibitemShut {NoStop}%
\bibitem [{\citenamefont {Leinaas}(2002)}]{leinaas2002spin}%
  \BibitemOpen
  \bibfield  {author} {\bibinfo {author} {\bibfnamefont {J.~M.}\ \bibnamefont
  {Leinaas}},\ }in\ \href@noop {} {\emph {\bibinfo {booktitle} {Confluence of
  Cosmology, Massive Neutrinos, Elementary Particles, and Gravitation}}}\
  (\bibinfo  {publisher} {Springer},\ \bibinfo {year} {2002})\ pp.\ \bibinfo
  {pages} {149--161}\BibitemShut {NoStop}%
\bibitem [{\citenamefont {Li}(1993{\natexlab{a}})}]{li1993intrinsic}%
  \BibitemOpen
  \bibfield  {author} {\bibinfo {author} {\bibfnamefont {D.}~\bibnamefont
  {Li}},\ }\href@noop {} {\bibfield  {journal} {\bibinfo  {journal} {Modern
  Physics Letters B}\ }\textbf {\bibinfo {volume} {7}},\ \bibinfo {pages}
  {1103} (\bibinfo {year} {1993}{\natexlab{a}})}\BibitemShut {NoStop}%
\bibitem [{\citenamefont {Li}(1993{\natexlab{b}})}]{li1993anyons}%
  \BibitemOpen
  \bibfield  {author} {\bibinfo {author} {\bibfnamefont {D.}~\bibnamefont
  {Li}},\ }\href@noop {} {\bibfield  {journal} {\bibinfo  {journal} {Nuclear
  Physics B}\ }\textbf {\bibinfo {volume} {396}},\ \bibinfo {pages} {411}
  (\bibinfo {year} {1993}{\natexlab{b}})}\BibitemShut {NoStop}%
\bibitem [{Note3()}]{Note3}%
  \BibitemOpen
  \bibinfo {note} {To be more precise Eq. \protect \textup {\hbox
  {\mathsurround \z@ \protect \normalfont (\ignorespaces \ref {2gr}\unskip
  \@@italiccorr )}} is somewhat symbolic as every term in the sum is written in
  a chart that covers vicinity of $z=a$. Indeed when singularities $a_i$ move
  around they induce the perturbations of metric outside of their
  location}\BibitemShut {NoStop}%
\bibitem [{\citenamefont {Lunin}\ and\ \citenamefont
  {Mathur}(2001)}]{lunin2001correlation}%
  \BibitemOpen
  \bibfield  {author} {\bibinfo {author} {\bibfnamefont {O.}~\bibnamefont
  {Lunin}}\ and\ \bibinfo {author} {\bibfnamefont {S.~D.}\ \bibnamefont
  {Mathur}},\ }\href@noop {} {\bibfield  {journal} {\bibinfo  {journal}
  {Communications in Mathematical Physics}\ }\textbf {\bibinfo {volume}
  {219}},\ \bibinfo {pages} {399} (\bibinfo {year} {2001})}\BibitemShut
  {NoStop}%
\bibitem [{\citenamefont {Knizhnik}(1987)}]{knizhnik1987analytic}%
  \BibitemOpen
  \bibfield  {author} {\bibinfo {author} {\bibfnamefont {V.}~\bibnamefont
  {Knizhnik}},\ }\href@noop {} {\bibfield  {journal} {\bibinfo  {journal}
  {Communications in Mathematical Physics}\ }\textbf {\bibinfo {volume}
  {112}},\ \bibinfo {pages} {567} (\bibinfo {year} {1987})}\BibitemShut
  {NoStop}%
\bibitem [{\citenamefont {Bershadsky}\ and\ \citenamefont
  {Radul}(1987{\natexlab{a}})}]{bershadsky1987conformal}%
  \BibitemOpen
  \bibfield  {author} {\bibinfo {author} {\bibfnamefont {M.}~\bibnamefont
  {Bershadsky}}\ and\ \bibinfo {author} {\bibfnamefont {A.}~\bibnamefont
  {Radul}},\ }\href@noop {} {\bibfield  {journal} {\bibinfo  {journal}
  {International Journal of Modern Physics A}\ }\textbf {\bibinfo {volume}
  {2}},\ \bibinfo {pages} {165} (\bibinfo {year}
  {1987}{\natexlab{a}})}\BibitemShut {NoStop}%
\bibitem [{\citenamefont {Bershadsky}\ and\ \citenamefont
  {Radul}(1987{\natexlab{b}})}]{bershadsky1987g}%
  \BibitemOpen
  \bibfield  {author} {\bibinfo {author} {\bibfnamefont {M.}~\bibnamefont
  {Bershadsky}}\ and\ \bibinfo {author} {\bibfnamefont {A.}~\bibnamefont
  {Radul}},\ }\href@noop {} {\bibfield  {journal} {\bibinfo  {journal} {Physics
  Letters B}\ }\textbf {\bibinfo {volume} {193}},\ \bibinfo {pages} {213}
  (\bibinfo {year} {1987}{\natexlab{b}})}\BibitemShut {NoStop}%
\bibitem [{\citenamefont {Bershadsky}\ and\ \citenamefont
  {Radul}(1988)}]{bershadsky1988fermionic}%
  \BibitemOpen
  \bibfield  {author} {\bibinfo {author} {\bibfnamefont {M.}~\bibnamefont
  {Bershadsky}}\ and\ \bibinfo {author} {\bibfnamefont {A.}~\bibnamefont
  {Radul}},\ }\href@noop {} {\bibfield  {journal} {\bibinfo  {journal}
  {Communications in mathematical physics}\ }\textbf {\bibinfo {volume}
  {116}},\ \bibinfo {pages} {689} (\bibinfo {year} {1988})}\BibitemShut
  {NoStop}%
\bibitem [{\citenamefont {Dixon}\ \emph {et~al.}(1987)\citenamefont {Dixon},
  \citenamefont {Friedan}, \citenamefont {Martinec},\ and\ \citenamefont
  {Shenker}}]{dixon1987conformal}%
  \BibitemOpen
  \bibfield  {author} {\bibinfo {author} {\bibfnamefont {L.}~\bibnamefont
  {Dixon}}, \bibinfo {author} {\bibfnamefont {D.}~\bibnamefont {Friedan}},
  \bibinfo {author} {\bibfnamefont {E.}~\bibnamefont {Martinec}}, \ and\
  \bibinfo {author} {\bibfnamefont {S.}~\bibnamefont {Shenker}},\ }\href@noop
  {} {\bibfield  {journal} {\bibinfo  {journal} {Nuclear Physics B}\ }\textbf
  {\bibinfo {volume} {282}},\ \bibinfo {pages} {13} (\bibinfo {year}
  {1987})}\BibitemShut {NoStop}%
\bibitem [{\citenamefont {Alvarez-Gaume}\ \emph
  {et~al.}(1986{\natexlab{a}})\citenamefont {Alvarez-Gaume}, \citenamefont
  {Moore},\ and\ \citenamefont {Vafa}}]{alvarez1986theta}%
  \BibitemOpen
  \bibfield  {author} {\bibinfo {author} {\bibfnamefont {L.}~\bibnamefont
  {Alvarez-Gaume}}, \bibinfo {author} {\bibfnamefont {G.}~\bibnamefont
  {Moore}}, \ and\ \bibinfo {author} {\bibfnamefont {C.}~\bibnamefont {Vafa}},\
  }\href@noop {} {\bibfield  {journal} {\bibinfo  {journal} {Communications in
  Mathematical Physics}\ }\textbf {\bibinfo {volume} {106}},\ \bibinfo {pages}
  {1} (\bibinfo {year} {1986}{\natexlab{a}})}\BibitemShut {NoStop}%
\bibitem [{\citenamefont {Farb}\ and\ \citenamefont
  {Margalit}(2011)}]{farb2011primer}%
  \BibitemOpen
  \bibfield  {author} {\bibinfo {author} {\bibfnamefont {B.}~\bibnamefont
  {Farb}}\ and\ \bibinfo {author} {\bibfnamefont {D.}~\bibnamefont
  {Margalit}},\ }\href@noop {} {\emph {\bibinfo {title} {A Primer on Mapping
  Class Groups (PMS-49)}}}\ (\bibinfo  {publisher} {Princeton University
  Press},\ \bibinfo {year} {2011})\BibitemShut {NoStop}%
\bibitem [{Note4()}]{Note4}%
  \BibitemOpen
  \bibinfo {note} {There is a braided category version of this known as
  spherical braided category}\BibitemShut {NoStop}%
\bibitem [{Note5()}]{Note5}%
  \BibitemOpen
  \bibinfo {note} {For the interested reader we note that relation \protect
  \textup {\hbox {\mathsurround \z@ \protect \normalfont (\ignorespaces \ref
  {notbraid}\unskip \@@italiccorr )}} is not present in the authentic spherical
  braid group.}\BibitemShut {Stop}%
\bibitem [{\citenamefont {Birman}\ and\ \citenamefont
  {Hilden}(1973)}]{birman1973isotopies}%
  \BibitemOpen
  \bibfield  {author} {\bibinfo {author} {\bibfnamefont {J.~S.}\ \bibnamefont
  {Birman}}\ and\ \bibinfo {author} {\bibfnamefont {H.~M.}\ \bibnamefont
  {Hilden}},\ }\href@noop {} {\bibfield  {journal} {\bibinfo  {journal} {Annals
  of Mathematics}\ ,\ \bibinfo {pages} {424}} (\bibinfo {year}
  {1973})}\BibitemShut {NoStop}%
\bibitem [{\citenamefont {Birman}(1974)}]{birman1974mapping}%
  \BibitemOpen
  \bibfield  {author} {\bibinfo {author} {\bibfnamefont {J.~S.}\ \bibnamefont
  {Birman}},\ }\href@noop {} {\bibfield  {journal} {\bibinfo  {journal}
  {Discontinuous Groups and Riemann Surfaces (College Park 1973)}\ ,\ \bibinfo
  {pages} {57}} (\bibinfo {year} {1974})}\BibitemShut {NoStop}%
\bibitem [{\citenamefont {Birman}(1975)}]{birman1975braids}%
  \BibitemOpen
  \bibfield  {author} {\bibinfo {author} {\bibfnamefont {J.~S.}\ \bibnamefont
  {Birman}},\ }\href@noop {} {\emph {\bibinfo {title} {Braids, links, and
  mapping class groups}}},\ \bibinfo {number} {82}\ (\bibinfo  {publisher}
  {Princeton University Press},\ \bibinfo {year} {1975})\BibitemShut {NoStop}%
\bibitem [{\citenamefont {Birman}\ and\ \citenamefont
  {Hilden}(1971)}]{birman1971mapping}%
  \BibitemOpen
  \bibfield  {author} {\bibinfo {author} {\bibfnamefont {J.~S.}\ \bibnamefont
  {Birman}}\ and\ \bibinfo {author} {\bibfnamefont {H.~M.}\ \bibnamefont
  {Hilden}},\ }in\ \href@noop {} {\emph {\bibinfo {booktitle} {Advances in the
  theory of Riemann surfaces (Proc. Conf., Stony Brook, NY, 1969)}}}\ (\bibinfo
  {year} {1971})\ pp.\ \bibinfo {pages} {81--115}\BibitemShut {NoStop}%
\bibitem [{\citenamefont {Birman}(1971)}]{birman1971siegel}%
  \BibitemOpen
  \bibfield  {author} {\bibinfo {author} {\bibfnamefont {J.~S.}\ \bibnamefont
  {Birman}},\ }\href@noop {} {\bibfield  {journal} {\bibinfo  {journal}
  {Mathematische Annalen}\ }\textbf {\bibinfo {volume} {191}},\ \bibinfo
  {pages} {59} (\bibinfo {year} {1971})}\BibitemShut {NoStop}%
\bibitem [{\citenamefont {Alimohammadi}\ and\ \citenamefont
  {Sadjadi}(1999)}]{alimohammadi1999coulomb}%
  \BibitemOpen
  \bibfield  {author} {\bibinfo {author} {\bibfnamefont {M.}~\bibnamefont
  {Alimohammadi}}\ and\ \bibinfo {author} {\bibfnamefont {H.~M.}\ \bibnamefont
  {Sadjadi}},\ }\href@noop {} {\bibfield  {journal} {\bibinfo  {journal}
  {Journal of Physics A: Mathematical and General}\ }\textbf {\bibinfo {volume}
  {32}},\ \bibinfo {pages} {4433} (\bibinfo {year} {1999})}\BibitemShut
  {NoStop}%
\bibitem [{\citenamefont {Iengo}\ and\ \citenamefont
  {Li}(1994)}]{iengo1994quantum}%
  \BibitemOpen
  \bibfield  {author} {\bibinfo {author} {\bibfnamefont {R.}~\bibnamefont
  {Iengo}}\ and\ \bibinfo {author} {\bibfnamefont {D.}~\bibnamefont {Li}},\
  }\href@noop {} {\bibfield  {journal} {\bibinfo  {journal} {Nuclear Physics
  B}\ }\textbf {\bibinfo {volume} {413}},\ \bibinfo {pages} {735} (\bibinfo
  {year} {1994})}\BibitemShut {NoStop}%
\bibitem [{\citenamefont {Bos}\ and\ \citenamefont {Nair}(1989)}]{bos1989u}%
  \BibitemOpen
  \bibfield  {author} {\bibinfo {author} {\bibfnamefont {M.}~\bibnamefont
  {Bos}}\ and\ \bibinfo {author} {\bibfnamefont {V.}~\bibnamefont {Nair}},\
  }\href@noop {} {\bibfield  {journal} {\bibinfo  {journal} {Physics Letters
  B}\ }\textbf {\bibinfo {volume} {223}},\ \bibinfo {pages} {61} (\bibinfo
  {year} {1989})}\BibitemShut {NoStop}%
\bibitem [{\citenamefont {Alvarez-Gaume}\ \emph
  {et~al.}(1986{\natexlab{b}})\citenamefont {Alvarez-Gaume}, \citenamefont
  {Moore}, \citenamefont {Nelson}, \citenamefont {Vafa},\ and\ \citenamefont
  {Bost}}]{alvarez1986bosonization}%
  \BibitemOpen
  \bibfield  {author} {\bibinfo {author} {\bibfnamefont {L.}~\bibnamefont
  {Alvarez-Gaume}}, \bibinfo {author} {\bibfnamefont {G.}~\bibnamefont
  {Moore}}, \bibinfo {author} {\bibfnamefont {P.}~\bibnamefont {Nelson}},
  \bibinfo {author} {\bibfnamefont {C.}~\bibnamefont {Vafa}}, \ and\ \bibinfo
  {author} {\bibfnamefont {J.}~\bibnamefont {Bost}},\ }\href@noop {} {\bibfield
   {journal} {\bibinfo  {journal} {Physics Letters B}\ }\textbf {\bibinfo
  {volume} {178}},\ \bibinfo {pages} {41} (\bibinfo {year}
  {1986}{\natexlab{b}})}\BibitemShut {NoStop}%
\bibitem [{\citenamefont {Knizhnik}(1986)}]{knizhnik1986analytic}%
  \BibitemOpen
  \bibfield  {author} {\bibinfo {author} {\bibfnamefont {V.}~\bibnamefont
  {Knizhnik}},\ }\href@noop {} {\bibfield  {journal} {\bibinfo  {journal}
  {Physics Letters B}\ }\textbf {\bibinfo {volume} {180}},\ \bibinfo {pages}
  {247} (\bibinfo {year} {1986})}\BibitemShut {NoStop}%
\bibitem [{\citenamefont {Eguchi}\ and\ \citenamefont
  {Ooguri}(1987)}]{eguchi1987chiral}%
  \BibitemOpen
  \bibfield  {author} {\bibinfo {author} {\bibfnamefont {T.}~\bibnamefont
  {Eguchi}}\ and\ \bibinfo {author} {\bibfnamefont {H.}~\bibnamefont
  {Ooguri}},\ }\href@noop {} {\bibfield  {journal} {\bibinfo  {journal}
  {Physics Letters B}\ }\textbf {\bibinfo {volume} {187}},\ \bibinfo {pages}
  {127} (\bibinfo {year} {1987})}\BibitemShut {NoStop}%
\bibitem [{\citenamefont {Alvarez-Gaume}\ \emph {et~al.}(1987)\citenamefont
  {Alvarez-Gaume}, \citenamefont {Bost}, \citenamefont {Moore}, \citenamefont
  {Nelson},\ and\ \citenamefont {Vafa}}]{alvarez1987bosonization}%
  \BibitemOpen
  \bibfield  {author} {\bibinfo {author} {\bibfnamefont {L.}~\bibnamefont
  {Alvarez-Gaume}}, \bibinfo {author} {\bibfnamefont {J.-B.}\ \bibnamefont
  {Bost}}, \bibinfo {author} {\bibfnamefont {G.}~\bibnamefont {Moore}},
  \bibinfo {author} {\bibfnamefont {P.}~\bibnamefont {Nelson}}, \ and\ \bibinfo
  {author} {\bibfnamefont {C.}~\bibnamefont {Vafa}},\ }\href@noop {} {\bibfield
   {journal} {\bibinfo  {journal} {Communications in Mathematical Physics}\
  }\textbf {\bibinfo {volume} {112}},\ \bibinfo {pages} {503} (\bibinfo {year}
  {1987})}\BibitemShut {NoStop}%
\bibitem [{\citenamefont {Dijkgraaf}\ \emph {et~al.}(1988)\citenamefont
  {Dijkgraaf}, \citenamefont {Verlinde},\ and\ \citenamefont
  {Verlinde}}]{dijkgraaf1988c}%
  \BibitemOpen
  \bibfield  {author} {\bibinfo {author} {\bibfnamefont {R.}~\bibnamefont
  {Dijkgraaf}}, \bibinfo {author} {\bibfnamefont {E.}~\bibnamefont {Verlinde}},
  \ and\ \bibinfo {author} {\bibfnamefont {H.}~\bibnamefont {Verlinde}},\
  }\href@noop {} {\bibfield  {journal} {\bibinfo  {journal} {Communications in
  Mathematical Physics}\ }\textbf {\bibinfo {volume} {115}},\ \bibinfo {pages}
  {649} (\bibinfo {year} {1988})}\BibitemShut {NoStop}%
\bibitem [{\citenamefont {Moore}\ and\ \citenamefont
  {Seiberg}(1988)}]{moore1988polynomial}%
  \BibitemOpen
  \bibfield  {author} {\bibinfo {author} {\bibfnamefont {G.}~\bibnamefont
  {Moore}}\ and\ \bibinfo {author} {\bibfnamefont {N.}~\bibnamefont
  {Seiberg}},\ }\href@noop {} {\bibfield  {journal} {\bibinfo  {journal}
  {Physics Letters B}\ }\textbf {\bibinfo {volume} {212}},\ \bibinfo {pages}
  {451} (\bibinfo {year} {1988})}\BibitemShut {NoStop}%
\bibitem [{\citenamefont {Moore}\ and\ \citenamefont
  {Seiberg}(1989)}]{moore1989classical}%
  \BibitemOpen
  \bibfield  {author} {\bibinfo {author} {\bibfnamefont {G.}~\bibnamefont
  {Moore}}\ and\ \bibinfo {author} {\bibfnamefont {N.}~\bibnamefont
  {Seiberg}},\ }\href@noop {} {\bibfield  {journal} {\bibinfo  {journal}
  {Communications in Mathematical Physics}\ }\textbf {\bibinfo {volume}
  {123}},\ \bibinfo {pages} {177} (\bibinfo {year} {1989})}\BibitemShut
  {NoStop}%
\bibitem [{\citenamefont {Barkeshli}\ \emph {et~al.}(2014)\citenamefont
  {Barkeshli}, \citenamefont {Bonderson}, \citenamefont {Cheng},\ and\
  \citenamefont {Wang}}]{barkeshli2014symmetry}%
  \BibitemOpen
  \bibfield  {author} {\bibinfo {author} {\bibfnamefont {M.}~\bibnamefont
  {Barkeshli}}, \bibinfo {author} {\bibfnamefont {P.}~\bibnamefont
  {Bonderson}}, \bibinfo {author} {\bibfnamefont {M.}~\bibnamefont {Cheng}}, \
  and\ \bibinfo {author} {\bibfnamefont {Z.}~\bibnamefont {Wang}},\ }\href@noop
  {} {\bibfield  {journal} {\bibinfo  {journal} {arXiv preprint
  arXiv:1410.4540}\ } (\bibinfo {year} {2014})}\BibitemShut {NoStop}%
\bibitem [{\citenamefont {Bombin}(2010)}]{Bombin2010}%
  \BibitemOpen
  \bibfield  {author} {\bibinfo {author} {\bibfnamefont {H.}~\bibnamefont
  {Bombin}},\ }\href {\doibase 10.1103/PhysRevLett.105.030403} {\bibfield
  {journal} {\bibinfo  {journal} {Phys. Rev. Lett.}\ }\textbf {\bibinfo
  {volume} {105}},\ \bibinfo {pages} {030403} (\bibinfo {year}
  {2010})}\BibitemShut {NoStop}%
\bibitem [{\citenamefont {Teo}\ \emph {et~al.}(2015)\citenamefont {Teo},
  \citenamefont {Hughes},\ and\ \citenamefont {Fradkin}}]{teo2015theory}%
  \BibitemOpen
  \bibfield  {author} {\bibinfo {author} {\bibfnamefont {J.~C.}\ \bibnamefont
  {Teo}}, \bibinfo {author} {\bibfnamefont {T.~L.}\ \bibnamefont {Hughes}}, \
  and\ \bibinfo {author} {\bibfnamefont {E.}~\bibnamefont {Fradkin}},\
  }\href@noop {} {\bibfield  {journal} {\bibinfo  {journal} {Annals of
  Physics}\ }\textbf {\bibinfo {volume} {360}},\ \bibinfo {pages} {349}
  (\bibinfo {year} {2015})}\BibitemShut {NoStop}%
\bibitem [{\citenamefont {Tarantino}\ \emph {et~al.}(2015)\citenamefont
  {Tarantino}, \citenamefont {Lindner},\ and\ \citenamefont
  {Fidkowski}}]{tarantino2015symmetry}%
  \BibitemOpen
  \bibfield  {author} {\bibinfo {author} {\bibfnamefont {N.}~\bibnamefont
  {Tarantino}}, \bibinfo {author} {\bibfnamefont {N.}~\bibnamefont {Lindner}},
  \ and\ \bibinfo {author} {\bibfnamefont {L.}~\bibnamefont {Fidkowski}},\
  }\href@noop {} {\bibfield  {journal} {\bibinfo  {journal} {arXiv preprint
  arXiv:1506.06754}\ } (\bibinfo {year} {2015})}\BibitemShut {NoStop}%
\bibitem [{\citenamefont {Barkeshli}\ and\ \citenamefont
  {Freedman}(2016)}]{barkeshli2016modular}%
  \BibitemOpen
  \bibfield  {author} {\bibinfo {author} {\bibfnamefont {M.}~\bibnamefont
  {Barkeshli}}\ and\ \bibinfo {author} {\bibfnamefont {M.}~\bibnamefont
  {Freedman}},\ }\href@noop {} {\bibfield  {journal} {\bibinfo  {journal}
  {arXiv preprint arXiv:1602.01093}\ } (\bibinfo {year} {2016})}\BibitemShut
  {NoStop}%
\bibitem [{\citenamefont {Arnol'd}(1968)}]{arnol1968remark}%
  \BibitemOpen
  \bibfield  {author} {\bibinfo {author} {\bibfnamefont {V.~I.}\ \bibnamefont
  {Arnol'd}},\ }\href@noop {} {\bibfield  {journal} {\bibinfo  {journal}
  {Functional Analysis and Its Applications}\ }\textbf {\bibinfo {volume}
  {2}},\ \bibinfo {pages} {187} (\bibinfo {year} {1968})}\BibitemShut {NoStop}%
\end{thebibliography}%

\appendix

\section{Statistics from the Induced Action}
\subsection{Quasiholes}
In this appendix we will calculated the statistics of the quasi-holes in Laughlin state using the induced action, meaning we will show that 
\be
\gamma_{{\mbox \rm stat}} = 2\pi i \frac{p_1 p_2}{q}
\ee
follows from the induced action
\be
S_{ind} = \frac{1}{4\pi q}\int A d A 
\ee
We consider a gauge field configuration that corresponds to moving a flux $2\pi p_1$ around a flux $2\pi p_2$.
\be
A = p_1 \left( \frac{dz}{z-z_1(t)} - \frac{d\bar z}{\bar z - \bar z_1(t)}\right) +  p_2 \left( \frac{dz}{z-z_2} - \frac{d\bar z}{\bar z - \bar z_2}\right)\,.
\ee
It is clear that with the help of 
\be
\tilde d\frac{ dz}{z} = -\left(\bar \p \frac{1}{z}\right)  dz   d\bar z = - \pi \delta(z)  dz   d\bar z\,,
\ee
where $\tilde d$ is spatial exterior derivative $\tilde d = dz \p + d\bar z \bar \p$\,. Next we evaluate the Chern-Simons action on this field configuration. We have (we write out only the relevant terms)
\be
d A \approx p_1 \left( \frac{\dot z_1}{(z-z_1)^2}dt  dz - \frac{\dot{\bar z}_1}{(\bar z-\bar z_1)^2}dt  d\bar z  \right)   
\ee
Then (again, only cross terms are relevant)
\bea\nonumber 
\frac{1}{4\pi q}\int AdA&& \approx \frac{ p_1 p_2 }{4\pi q} \int \left(\frac{\dot z_1(t)}{(z-z_1(t))^2} \frac{1}{(\bar z - \bar z_2)} - {\mbox \rm c.c.}\right) dt  dz  d\bar z \\\nonumber
&&= \frac{ p_1 p_2 }{4q} \int \left(\frac{\dot z_1(t)}{(z-z_1(t))} \delta(z-z_2)- {\mbox \rm c.c.}\right) dt  dz  d\bar z \\ \nonumber
&&= \frac{ p_1 p_2 }{4q} \int \left(\frac{\dot z_1(t)}{(z_2-z_1(t))} - {\mbox \rm c.c.}\right) dt \\
&&=\frac{ p_1 p_2 }{4q} \oint \left(\frac{dz_1}{(z_2-z_1)} - {\mbox \rm c.c.}\right) = 2\pi i \frac{p_1p_2}{2q}
\eea
Strictly speaking this computation holds for general $p_1$ and $p_2$, however it only corresponds to quasiholes when fluxes are integer. We expect that the adiabatic phase makes sense and is universal even for a generic magnetic flux, despite the fact that the computation is outside of applicability of the induced action.
\subsection{Genons}
\la{cone}
It is possible to compute the statistics of generic curvature fluxes using the same method. The result, however, does not agree with monodromy computations of Section 3. The difference in the computation of the statistics of curvature fluxes is that singularities may change the topology of the space. To avoid this we place extra curvature flux at infinity. We consider the spin connection configuration of the form
\be
\omega = \omega_1 + \omega_2 + \omega_\infty\,,
\ee
where 
\bea
\omega_1&&= \alpha_1\left(\frac{dz}{z-z_1(t)} -\frac{d\bar z}{\bar z-\bar z_1(t)} \right)\,,\\
\omega_2&&= \alpha_2\left(\frac{dz}{z-z_2} -\frac{d\bar z}{\bar z-\bar z_2} \right)\,,\\
\omega_\infty&&= (2-\alpha_1-\alpha_2)\left(\frac{dz}{z-z_\infty} -\frac{d\bar z}{\bar z-\bar z_\infty} \right)\,.
\eea
The spatial curvature $2$-form is
\bea\nonumber
 R_1 &&= 2\tilde d \omega_1 = -4\pi\alpha_1\Big(\delta(z-z_1(t))+ \delta(z-z_2)\Big)dz d\bar z\,,\\ \nonumber
 R_\infty && = 2\tilde d \omega_\infty = 4\pi(2- \alpha_1-\alpha_2)\delta(z-z_\infty)dz d\bar z\,,
\eea
and the Euler characteristic comes out correctly $\chi = 2$. Evaluation of the gravitational Chern-Simons action proceeds the same as in the quasihole case
\bea
\frac{c_\mathrm{w}}{48\pi} \int \omega d \omega &&= 2\pi i \frac{c_\mathrm{w}}{24} \alpha_1\alpha_2\,.
\eea
We assumed that infinity point is not encircled by the path $z_1(t)$. This result agrees with \cite{laskin2016emergent}, however it disagrees with the monodromy computation. It is amusing to note that if the contour happens to include $z_\infty$ then the braiding phase does not depend on $\alpha_2$ (or, more generally, on any cone that is enclosed by the contour)
\be
\gamma_{stat} = \frac{c_\mathrm{w}}{24}\alpha_1(\alpha_1-2) =  \frac{c_\mathrm{w}}{24} (n^2-1)\,,
\ee
in the last equality we took $\alpha_1 = -(n-1)$, which is the case discussed in the main text. In this case the statistics comes out without a factor of $\frac{1}{n} = \frac{1}{\alpha_1+1}$. This statistics does not satisfy the spin-statistics theorem. It appears that the induced action computation only agrees with monodromy computation when $0<\alpha_1<<1$.
\section{Explicit evaluation of the Liouville action}
\la{Laction}
In this Appendix we are going to elaborate on the derivation of \eqref{2gfactor}. We need to evaluate the Liouville action on a singular geometry described by the metric
\be
ds^2 = n^2 |z-a_1|^{2n-2} |z-a_2|^{2n-2}dz d\bar z\,.
\ee
To do so we cut a s small disc of radius $\epsilon$ around the points where curvature becomes singular and replace it with a flat patch with regular metric
\be
ds^2 = 4 \epsilon dz d\bar z
\ee
The Liouville action is evaluated in $\zeta$ coordinates.  We consider the Liouville action in $\zeta$-space, with discs removed, $\Sigma^\prime$. First, since there is no curvature in $\zeta$ coordinates the term $R\sigma$ does not contribute. Second, we integrate by parts in the kinetic energy term
\be
\frac{c}{96\pi} \int_{\Sigma^\prime} -\sigma \Delta \sigma + \frac{c}{96\pi}\int_{\p{\Sigma^\prime}} \sigma n^\mu\p_\mu \sigma\,,
\ee
where the first term vanishes because $\Delta \sigma$ is a sum of $\delta$-functions with support outside of $\Sigma^\prime$. Thus
\be\la{Lboundary}
S_L[\sigma] = \frac{c}{96\pi}\int_{\p{\Sigma^\prime}} \sigma n^\mu\p_\mu \sigma
\ee
There are two types of points that we have removed from $\Sigma$. First type is the points $a_i$. These points contained negative curvature. Second type is the points that map to $z=\infty$. In order to study these points we need to specify the covering map $\zeta(z)$. A convenient choice is 
\be
\zeta = a \frac{z^n}{z^n - (z-s)^n}\,,
\ee
where the branch points in $\zeta$-space are located in $\zeta=0$ and $\zeta=a$.
The zeroes of denominator are
\be
z_k = \frac{s}{1-\alpha_k}\,
\ee
and lie in a strip parallel to the imaginary axis.
We can consider the parameter $s$ as regulator, that upon the limit $s\rightarrow \infty$ places all of the zeroes to infinity. The curvature at finite $s$ is
\bea\nonumber
\sqrt{g} R &&= -4\pi(n-1) \delta(z) -4\pi(n-1) \delta(z-1) \\
&&+ 4\pi \sum_{k=1}^{n-1}  \delta(z-z_k) + 4\pi \delta(z-z_\infty)
\eea
Thus we need to evaluate the boundary integral around $z=0$,$z=1$,$z=z_k$,$z=\infty$. Integrals around $0$ and $1$ are equal to each other as are the integrals around $z_k$'s; the integral around $z=\infty$ vanishes. We start with the integral around $0$.
\bea\nonumber
\frac{c}{96\pi}\int_{\p{\Sigma^\prime}}&& \sigma n^\mu\p_\mu \sigma = \frac{c}{96\pi} \int d\theta \epsilon^{\frac{1}{n}} \frac{n-1}{\epsilon^\frac{1}{n}} \ln \left|a^2 n^2 z^{2n-2}\right| + {\mbox \rm c.c.}\\
&&\frac{c}{48}\epsilon^{\frac{1}{n}} \frac{n-1}{\epsilon^\frac{1}{n}} \ln \left|a^{-\frac{2}{n}} n^{-\frac{2}{n}} \epsilon^{\frac{2n-2}{n}}\right| + {\mbox \rm c.c.}\\
&&=- \frac{c}{6} \frac{n-1}{n} \ln |a| + \ldots\,,
\eea
where we kept only the dependence on $\ln a$. When it comes to points $z=z_k$ each of these points has degree $1$ instead of $n$, but computation goes in an identical way. We have
\be
\frac{c}{96\pi}\int_{\p{\Sigma^\prime}} \sigma n^\mu\p_\mu \sigma = - \frac{c}{6}(n-1) \ln |a| +\ldots
\ee
Adding the contributions together we have 
\be
S_L[\sigma] = - \frac{c}{6} \frac{n^2-1}{n} \ln |a|\,.
\ee
The result does not depend on $s$ and upon replacing $a \rightarrow a_1 - a_2$ we recover  \eqref{2gfactor}.

\section{Singular surfaces with $\zn$-symmetry}
\la{ZnSect}
The genon surfaces $\Sigma_{n,2M}$ can be described yet in another way. Consider a set o points $(\zeta,z)\in \mathbb C^2$ that satisfies equation
\be\la{Zncurve}
z^n =\prod_{i=1}^{2M}(\zeta-a_{i})\,.
\ee
This curve is exactly the same curve we have studied before. To see this it is sufficient to assume that all $a_i$ are sufficiently far apart and consider a neighborhood of a point $\zeta=a_i$. In this neighborhood the Eq.\eqref{Zncurve}
\be
z^n = C (\zeta-a_i)\,,
\ee
where $C$ is an overall constant $C = \prod_{i\neq j} (a_i-a_j)$. This is precisely the coordinates we have used in Eq.\eqref{zeta}. In this neighborhood $\zeta$ is the multivalued coordinate we used to set up the computation of the correlation functions. Globally, Riemann-Roch theorem ensures that genus comes out correctly.

There is a natural, albeit not canonical set of holomorphic differentials $f_i$ on the algebraic curve \eqref{Zncurve}. These are\cite{bershadsky1987conformal}
\be
f_{i,l} = \frac{\zeta^{i-1}}{z^l}d\zeta\,,
\ee
where $l=1,\ldots, n-1$ and $i =1,\ldots, M-1$, so there are precisely $g(\Sigma_{n,2M})$ of them. This basis can be transformed back to the canonical basis of holomorphic differentials $\omega_i$
\be\la{basischange}
\omega_i = L_{ij}f_{j}\,,
\ee
where we have denoted $f_j$ a $g$-dimensional vector made from the matrix $f_{j,l}$. We also define
\be
A_{ij} =  \oint_{a_i} f_j\,,\qquad B_{ij} =  \oint_{b_i} f_j
\ee
 An explicit expression for $L_{ij}$ is derived by applying $\oint_{a_k}$ to both sides of \eqref{basischange}
\be
 L_{ij} = A^{-1}_{ij}\,,
\ee

Now, the period matrix can be expressed in terms of $f_j$ and, consequently, in terms of $a_i$. We have
\be\la{ubereq}
\Omega_{ij} = \oint_{b_i} \omega_j = A^{-1}_{jk}\oint_{b_i}f_k = A^{-1}_{jk} B_{ki}\
\ee
To illustrate this abstract approach we consider case of torus $\Sigma_{2,4}$. Then \eqref{ubereq} reduces to
\be
\tau = \frac{\oint_b f}{\oint_a f}\,, \quad f = \frac{d\zeta}{\sqrt{(\zeta-a_1)(\zeta-a_2)(\zeta-a_3)(\zeta-a_4)}}\,.
\ee
These integrals are contour integral representations of hypergeometric function $F(x) =\mathstrut_2 F_1(1/2,1/2,1;x)$, with $x$ being the anharmonic ratio, provided we fix three of $a_i$ to be at $0, 1$ and $\infty$. Thus we get
\be\la{supereq2}
\tau = i \frac{F\left(\frac{1}{x}\right)}{F\left(1-\frac{1}{x}\right)}\,,
\ee
which is inverse of \eqref{supereq}. While it is far from obvious that \eqref{supereq} and \eqref{supereq2} are inverse of each other, it can be checked on Mathematica.

Despite the apparent simplicity the Eq.\eqref{ubereq} is not very friendly to work with since each entry in $A$ and $B$ matrices is an elliptic integral of the form
\be
\oint_{c_i} \frac{\zeta^{i-1}}{(\zeta-a_1)^{\frac{l}{n}}(\zeta-a_2)^{\frac{l}{n}}\cdot\ldots\cdot(\zeta-a_{2M})^{\frac{l}{n}}}d\zeta\,,
\ee
where $c_i$ is either $a_i$ or $b_i$. Arnold derived an explicit representation for braid matrices in the case $\Sigma_{2,6}$ using Pickard-Lefshetz theory \cite{arnol1968remark}. We will refrain from the detailed analysis of matrices $A$ and $B$ and their monodromies, instead we discuss some general geometric features of $\Sigma_{n,2M}$. 

There are two extreme cases we can consider. The first case is $\Sigma_{n,4}$. This surface has genus $n-1$ and there are $3g-3 = 3n-6 $ moduli, however there is only one anharmonic ratio $x$ and there are only two independent braids. In this case the relevant elliptic integrals are of the type
\be
\oint_{c_i} \frac{d\zeta}{\zeta^{\frac{l}{n}}(\zeta-1)^{\frac{l}{n}}(\zeta-x)^{\frac{l}{n}}}
\ee

These integrals can be expressed in terms of the hypergeometric function $\mathstrut_2 F_1\left(\frac{l}{n},1-\frac{l}{n},2(1-\frac{l}{n});x\right)$.

The opposite extreme case is the surface $\Sigma_{2,2M}$. In this case the genus is $M-1$ and there are $3M-6$ moduli. At the same time there are $2M-3$ independent anharmonic ratios. This is the case with the biggest number of anharmonic ratios available and the surface is still not generic. The reason is that even $\mathbb Z_2$ symmetry conflicts with deformations in some directions in the moduli space and even more so for $\mathbb Z_n$ case. Interestingly, in the genus $2$ case, when $M=3$ the surface is generic as there are as many anharmonic ratios as moduli. For all such surfaces the scaling dimension  of a branch point is $1/16$ and when $q=2$ they still should describe Moore-Read quasi-holes.

In principle, the route to calculation of the braiding matrices is clear. Eq.\eqref{ubereq} translates braiding of $a_i$ to modular symplectic $Sp(2g,\mathbb Z)$ transformations. The latter act on the Laughlin state on the genus $g$ surface and are known explicitly.

\end{document}